\documentclass[universe,review,accept,pdftex]{Definitions/mdpi} 

\usepackage{subfigure}
\usepackage{graphicx}
\usepackage{amssymb}
\usepackage{url}
\usepackage[italicdiff]{physics}
\usepackage{natbib}
\usepackage{verbatim}
\setcitestyle{authoryear}
\usepackage{longtable}

\firstpage{1} 
\pubvolume{xx}
\issuenum{1}
\articlenumber{5}
\pubyear{2021}
\copyrightyear{2021}
\history{Received: 31 March 2021; Accepted: 29 April 2021; Published: May 11, 2021}

\newcommand*{\MG}[1]{\textcolor{magenta}{\textsf{[MG: #1]}}}

\def\stacksymbols #1#2#3#4{\def\theguybelow{#2}
        \def\verticalposition{\lower#3pt}
        \def\spacingwithinsymbol{\baselineskip0pt\lineskip#4pt}
        \mathrel{\mathpalette\intermediary#1}}
\def\intermediary #1#2{\verticalposition\vbox{\spacingwithinsymbol
        \everycr={}\tabskip0pt
        \halign{$\mathsurround0pt#1\hfil##\hfil$\crcr#2\crcr
                \theguybelow\crcr}}}
\def\lta{\stacksymbols{<}{\sim}{2.5}{.2}}

\def\kms{{\rm km\,s^{-1}}}

\def\Tx{$T_{\rm x}$}
\def\tx{T_{\rm x}}
\def\Lx{$L_{\rm x}$}
\def\lx{L_{\rm x}}
\def\R500{$R_{500}$}
\def\r500{R_{500}}

\def\sige{\sigma_{\rm e}}
\def\Re{$R_{\rm e}$}

\newcommand{\msun}{M_{\odot}}

\newcommand{\es}{{\rm erg\,s^{-1}}}

\newcommand{\Chandra}{\textit{Chandra}}

\newcommand{\XMM}{XMM-\textit{Newton}}
\newcommand{\ROSAT}{\textit{ROSAT}}
\newcommand{\ASCA}{\textit{ASCA}}

\Title{Feedback from Active Galactic Nuclei in Galaxy Groups}

\Author{Dominique~Eckert  
$^{1,\star,\dagger}$\orcidD{}, Massimo~Gaspari $^{2,3,\dagger}$\orcidM{}, Fabio~Gastaldello $^{4,\dagger}$\orcidF{},  Amandine~M.~C.~Le Brun $^{5,\dagger}$\orcidA{}, and  Ewan~O'Sullivan $^{6,\dagger}$\orcidE{} }

\AuthorNames{Dominique Eckert, Fabio Gastaldello, Massimo Gaspari, Amandine Le Brun, Ewan O'Sullivan}

\corres{Correspondence: Dominique.Eckert@unige.ch}

\address{%
$^{1}$ \quad Department of Astronomy, University of Geneva, Ch. d'Ecogia 16, CH-1290 Versoix, Switzerland \\
$^{2}$ \quad INAF---Osservatorio di Astrofisica e Scienza dello Spazio di Bologna, via Piero Gobetti 93/3, I-40129 Bologna, Italy\\
$^{3}$ \quad Department of Astrophysical Sciences, Princeton University, 4 Ivy Lane, Princeton, NJ 08544, USA\\
$^{4}$ \quad IASF---Milano, INAF, Via A. Corti 12, I-20133 Milano, Italy\\
$^{5}$ \quad LUTh, UMR 8102 CNRS, Observatoire de Paris, PSL Research University, Université Paris Diderot, 5 Place Jules Janssen, 92190 Meudon, France\\
$^{6}$ \quad Center for Astrophysics $|$ Harvard \& Smithsonian, 60 Garden Street, Cambridge, MA 02138, USA\\
$^\star$ \quad Correspondence: Dominique.Eckert@unige.ch\\
}

\firstnote{All authors contributed equally to this work.}

\abstract{The co-evolution between supermassive black holes and their environment is most directly traced by the hot atmospheres of dark matter halos. Cooling of the hot atmosphere supplies the central regions with fresh gas, igniting active galactic nuclei (AGN) with long duty cycles. Outflows from the central engine tightly couple with the surrounding gaseous medium and provide the dominant heating source preventing runaway cooling by carving cavities and driving shocks across the medium. The AGN feedback loop is a key feature of all modern galaxy evolution models. Here we review our knowledge of the AGN feedback process in the specific context of galaxy groups. Galaxy groups are uniquely suited to constrain the mechanisms governing the cooling-heating balance. Unlike in more massive halos, the energy supplied by the central AGN to the hot intragroup medium can exceed the gravitational binding energy of halo gas particles. We report on the state-of-the-art in observations of the feedback phenomenon and in theoretical models of the heating-cooling balance in galaxy groups. We also describe how our knowledge of the AGN feedback process impacts on galaxy evolution models and on large-scale baryon distributions. Finally, we discuss how new instrumentation will answer key open questions on the topic. }


\begin{document}


\section{Introduction} \label{s:intro}

Structure formation in the Universe operates as a bottom-up process in which small halos formed at high redshift progressively merge and accrete the surrounding material to form the massive halos we see today \citep{Springel2005b}. Given the evolution of the halo mass function, the peak of the mass density in the current Universe occurs in halos of $\sim10^{13}M_\odot$ - the \emph{galaxy group} regime. At the current epoch, galaxy groups are the building blocks of the structure formation process and thus they occupy a key regime in the evolution of galaxies. Typical $L_\star$ galaxies exist within groups rather than within isolated halos \citep{Robotham:2011}. The galaxy stellar mass function exhibits a cut-off at $M_\star\sim10^{11}M_\odot$ \citep{Davidzon:2017}, corresponding to the central galaxies of galaxy groups, brightest group galaxies (BGGs). Abundance matching studies show that the star formation efficiency reaches a maximum at $M_h \sim 10^{12}M_\odot$ and decreases both at higher and lower masses \citep{Behroozi:2013,Leauthaud:2012,Coupon:2015}. At the high-mass end, non-gravitational energy input is needed to quench star formation and reproduce the shape of the stellar mass function \citep{Silk:1998}. 

Feedback from active supermassive black holes (SMBH) is currently the favored mechanism to regulate the star formation activity in massive galaxies, explain the observed shape of the stellar mass function, and quench catastrophic cooling flows. Outflows and jets from the central active galactic nuclei (AGN) interact with the surrounding hot intragroup medium (IGrM) and release a large amount of energy, which prevents the gas reservoir from cooling efficiently and fueling star formation (for previous reviews see \citealt{McNamara:2007}; \citealt{Fabian:2012}; \citealt{Gitti:2012}; \citealt{Gaspari:2020}). All modern galaxy evolution models include a prescription for AGN feedback to reproduce the shape of the galaxy luminosity function and the halo baryon fraction \citep[see the companion review by][]{Oppenheimer:2021}. Earlier attempts at reproducing these observables with supernova feedback resulted in catastrophic cooling and largely overestimated the stellar content of groups \citep[e.g.,][]{LeBrun:2014}. State-of-the-art cosmological simulations all implement sub-grid prescriptions for AGN feedback, either in the form of thermal, isotropic feedback or in the form of mechanical, directional feedback. The adopted feedback scheme strongly affects the gas properties of galaxy groups. Indeed, strong feedback raises the entropy level of the surrounding gas particles, which can lead to a global depletion of baryons in group-scale halos. Cosmological simulations are now facing the important challenge of reproducing at the same time realistic galaxy populations and gas properties.

The imprint of AGN feedback is most easily observed through high-resolution X-ray observations of nearby galaxy groups and clusters. Bubbles of expanding energetic material carve cavities in the gas distribution, spatially coinciding with energetic AGN outflows traced by their radio emission. Supersonic outflows also drive shock fronts permeating the surrounding IGrM and distributing heat across the environment. In parallel, extended H$\alpha$ nebulae demonstrate the existence of efficient gas cooling from the hot phase, thereby feeding the central SMBH. Recent ALMA observations at millimetric wavelengths also provide evidence for large amounts of cool gas at the vicinity of the SMBH. Multi-wavelength observations of the cores of nearby massive structures thus allow us to investigate in detail the balance between gas cooling and AGN heating.

While a great deal of attention has been devoted to studying these phenomena in the most massive nearby clusters such as Perseus \citep[see][for a review]{McNamara:2007}, observations of similar quality only exist for a handful of groups, such as NGC 5044 \citep{Gastaldello:2009} and NGC 5813 \citep{Randall:2015}. Detailed observations of galaxy groups are crucial for our understanding of feedback processes, as the physical conditions differ from those of galaxy clusters in several important ways. First and foremost, the ratio of feedback energy to gravitational energy is different from that of clusters. The AGN energy input is sometimes parameterized as $\dot E_{\rm feed} \approx \epsilon_f \epsilon_r \dot M_{\rm BH} c^2$ with $\epsilon_r\sim10\%$ the energy output of the BH and $\epsilon_f$ the coupling efficiency between the BH outflows and the surrounding medium. On the other hand, the gravitational binding energy is a strong function of halo mass, $E_{\rm bind} \propto M_h^2$. If the coupling efficiency only depends on the physical properties of the gas and the feedback loop has a long duty cycle, the total integrated feedback energy $E_{\rm feed}$ becomes comparable to $E_{\rm bind}$ or even exceeds it in group-scale halos, whereas it remains substantially lower in the most massive systems. Similarly, the radius inside which non-gravitational energy dominates over gravitational energy is comparably larger in group-scale halos. The impact of AGN in groups is spread over much larger volumes and can even lead to a depletion of baryons within the virial radius. On top of that, the radiative cooling function experiences a transition of regime between the temperature range of clusters and that of groups. For temperatures greater than $\sim 3$ keV, the plasma is almost completely ionized and the Bremsstrahlung process dominates. For temperatures of $\sim 1$ keV, line cooling dominates, which makes radiative losses comparably more important. Therefore, the radiative cooling time can become much shorter than the Hubble time even at relatively low gas densities, and the supply of gas to the SMBH can be sustained more easily. For all these reasons, studying the feedback loop across a wide range of halo masses is necessary to inform our theoretical models.

Here we review the current state of the art in our knowledge of the AGN feedback process in the specific case of galaxy groups. For the purpose of this review, we define galaxy groups as galaxy concentrations with halo masses in the range $10^{13}-10^{14}M_\odot$ and with an X-ray bright intragroup medium (IGrM). Such masses correspond to virial temperatures of $\sim0.5-2$\ keV. Most of the processes discussed in this review are also relevant in the case of X-ray bright isolated elliptical galaxies and massive spirals with $kT\sim0.3-0.5$ keV. Whenever it is appropriate, we will discuss halos of lower masses as well. The paper is organized as follows. In \S\ref{s:gal}, we describe why AGN feedback is presently thought to be a key ingredient in the evolution of galaxy groups and how invoking AGN feedback can solve a number of overarching issues in galaxy evolution. In \S\ref{s:obs}, we review the current observational evidence for AGN feedback in nearby galaxy groups, with our main focus on observations at X-ray and radio wavelengths and additional information coming from millimeter and H$\alpha$ observations. \S\ref{s:theory} summarizes the theoretical framework put into place to interpret the heating/cooling cycle and the main physical processes involved, with a specific focus on the galaxy group regime. \S\ref{s:large_scale} discusses how AGN feedback is implemented in modern cosmological simulations and its impact on the evolution and large-scale distribution of the baryonic component of the Universe. We conclude our review in \S\ref{s:future} with a short presentation of the most relevant upcoming experiments and their expected contribution to the field.


\section{The need for AGN feedback in galaxy evolution} \label{s:gal}

Even though the first indications that AGN feedback could be one of the missing elements in theories of galaxy formation and evolution are already nearly half a century old \citep{Larson1974,Rees1977,White1978}, theoretical and observational studies of the process are still in their infancy. In the last fifteen years, AGN feedback has emerged as the most promising solution to a number of overarching problems in galaxy formation and evolution both in semi-analytic models and cosmological hydrodynamical simulations \citep[e.g.][]{Granato:2004,Bower2006,Croton2006,DiMatteo2008,Puchwein:2008}. In short, the main issues are: cosmic `downsizing' \citep{Cowie:1996}, i.e. the observation that the majority of both star formation and AGN activity took place before redshift $z\sim1$ \citep[e.g.][]{Shaver:1996,Madau:1996}, the shape of the galaxy stellar mass function at the high-mass end, the gas content of massive galaxies, groups and clusters, the absence of a cooling flow, the deviations from self-similarity of gas scaling relations, and the entropy floor (the latter three are discussed in detail in \S\ref{ss:X-ray}). AGN feedback progressively appeared to be a promising solution to solve each of these problems separately, eventually leading to the realization that these issues could in fact be seen as different facets of the same problem \citep[e.g.][]{Benson2003,Bower2006,Bower2008}. In this section, we highlight the main reasons why AGN feedback plays a central role in galaxy evolution models, specifically at the scale of galaxy groups.

\subsection{\textbf{The shape of the galaxy stellar mass function}}
\label{s:galaxy_evol}

\begin{figure}[t]
 \centering
     \includegraphics[width=1.0\textwidth]{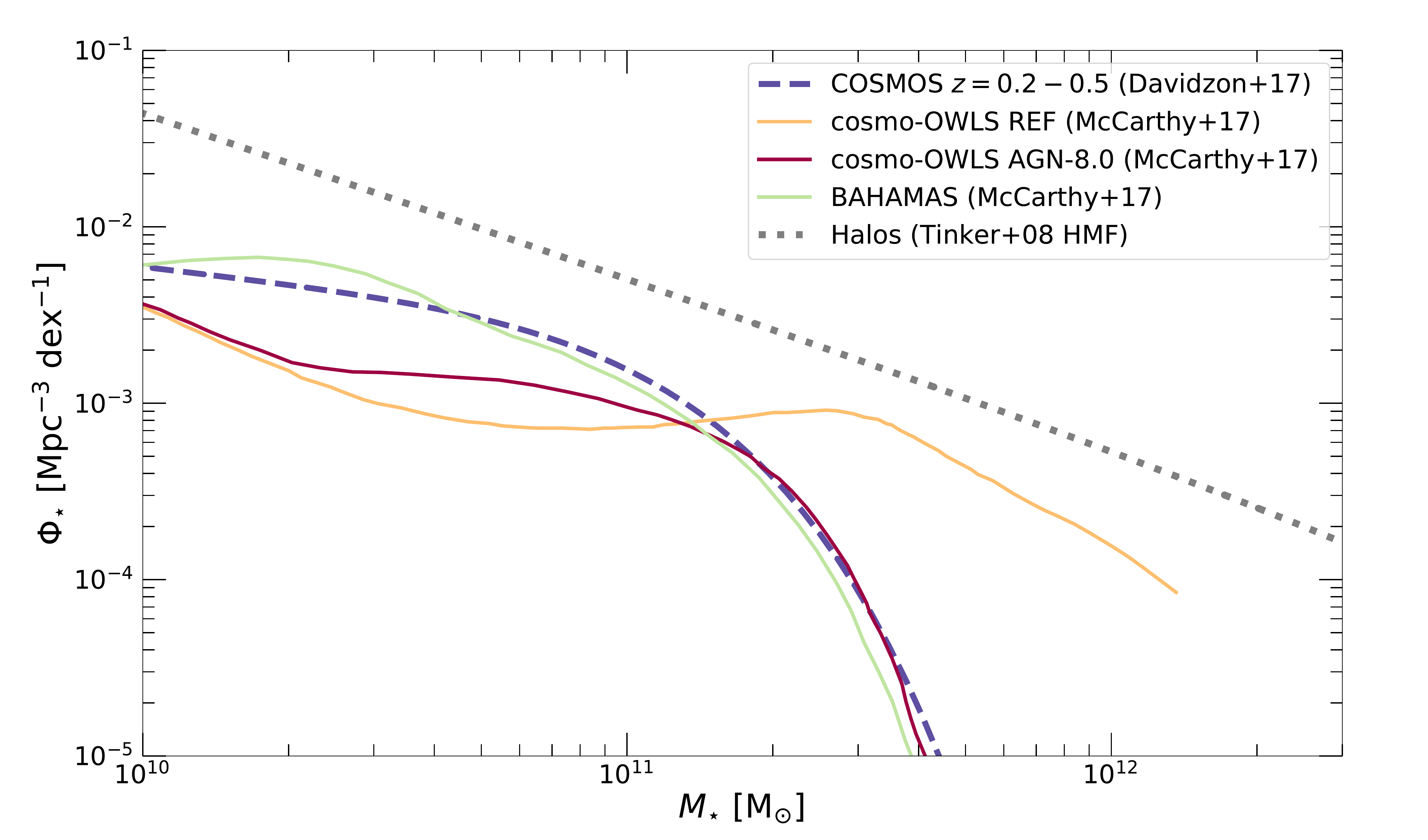}
     \caption{Galaxy stellar mass function (GSMF) $\Phi_\star(M_\star)=\frac{dN}{dM_{*}dV}$ in observations and simulations. The dashed purple curve shows the double-Schechter fit to the local GSMF measured in the COSMOS survey \citep{Davidzon:2017}. The solid curves show the predictions of the cosmo-OWLS/BAHAMAS model \citep{McCarthy:2017,LeBrun:2014} in the case with only stellar feedback (REF, yellow), cosmo-OWLS with AGN feedback (AGN-8.0, maroon) and the latest BAHAMAS model (green), in which the feedback model was tuned to reproduce jointly the GSMF and the gas fraction. For comparison, the dotted gray curve shows the \citet{Tinker:2008} halo mass function in \emph{Planck} cosmology scaled by $f_b = \Omega_b/\Omega_m$, which highlights what one expects to see in the case in which each halo is populated by a galaxy with $M_\star = f_b M_h$. }
     \label{fig:GSMF}
 \end{figure}

\citet{White1978} presented one of the first models of galaxy formation in a cosmological context \citep[see also][]{Rees1977,Silk1977}. The authors proposed a two-stage process in which galaxies form via radiative cooling of the baryons within halos that had already formed via gravitational collapse of the collisionless dark matter. The authors also argued that an additional non-gravitational process, called feedback, was needed to avoid the overproduction of faint galaxies compared to observations. Indeed, in that model, the galaxy stellar mass function would simply be a scaled-down version of the dark matter halo mass function \citep[see discussion in e.g.][illustrated by their figure 1 and Fig.~\ref{fig:GSMF} of this review]{Benson2003}. Later studies pointed out that the scaled halo mass function model overpredicts the abundance of the most massive galaxies compared to observations \citep[e.g.][]{Cole1991,White1991,Blanchard1992,Katz1992,Kauffmann1999,vanKampen1999,Cole2000,Benson2003}. In short, galaxy formation is fundamentally an inefficient process as only a small fraction of the Universe's baryons are in the form of stars \citep[about 10 per cent; e.g.][]{Cole2001,Balogh2001,Lin2003} and the star formation efficiency strongly depends on halo mass. The mass of the dark matter halo thus plays a fundamental role in shaping the galaxies that it contains. Observational evidence for this halo-mass dependency of the efficiency of galaxy formation was first obtained using galaxy group catalogues \citep[][]{Yang2005,Eke2006}. 

To illustrate the first point, in Fig.~\ref{fig:GSMF}, we show the local galaxy stellar mass function (GSMF) $\Phi_\star(M_\star)$ measured within the COSMOS survey \citep{Davidzon:2017} modeled using a double Schechter function. The dashed line shows the \citet{Tinker:2008} halo mass function with the masses scaled by the universal baryon fraction $f_b=\Omega_b/\Omega_m$, which produces the GSMF one would expect to see if every halo was populated by a single galaxy and all the available baryon content had been converted into stars. The observed GSMF vastly differs from the scaled halo mass function both in shape and normalization. The lower normalization implies that the star formation efficiency is much less than 100\%, whereas the steep decline at high masses shows that the growth of galaxies does not follow the structure formation process. Attempts at reproducing the shape of the GSMF with feedback from supernovae and star formation were unsuccessful, as the injected energy was insufficient to offset cooling and regulate the star formation efficiency \citep{Benson2003}. In Fig.~\ref{fig:GSMF}, we compare the observed GSMF with the predictions of hydrodynamical cosmological simulations from the cosmo-OWLS and BAHAMAS suites \citep{LeBrun:2014,McCarthy:2017} implementing several prescriptions for baryonic physics. 
The cosmo-OWLS run including cooling, star formation and stellar feedback (labelled REF) suffers from overcooling, and thus it overpredicts the observed abundance of massive galaxies ($M_\star\gtrsim10^{12}M_\odot$) by several orders of magnitude. Conversely, including AGN feedback allows the model to reproduce the abundance of massive galaxies closely. AGN feedback is implemented by releasing a fraction of the rest-mass energy of cooling gas particles within the surrounding environment, which reheats the gas and regulates the star formation efficiency (see \S\ref{s:cosmo_sims} for details). The effect of AGN feedback kicks in around $M_\star\sim2-3\times10^{11}M_\odot$ corresponding to the stellar masses of BGGs, which highlights the crucial role played by the galaxy group regime.

Following the early works highlighting the discrepancy between the observed GSMF and the structure formation theory, it took over 30 years to pinpoint the most likely source of feedback. Early models of non-gravitational heating invoked a `pre-heating' of the baryonic content before the epoch of formation of massive haloes, but did not specify the source of the energy injection \citep[e.g.][]{Evrard:1991,Kaiser:1991,Blanchard1992}. As these models cannot predict the impact on the galaxy population but only on the intra-group medium, they are not discussed any further here (but see \S\ref{ss:X-ray} for a detailed discussion). Models implementing gas cooling without feedback were able to reproduce the breakdown of self-similarity observed in X-ray selected samples, but this came at the price of greatly overpredicting the abundance of galaxies \citep[e.g.][and \cite{Balogh2008} for a review]{Bryan:2000,Muanwong:2002,Voit:2001,Voit:2002,Wu:2002}. In the late 1990s and early 2000s, teams working with both semi-analytic models and hydrodynamical simulations, who had started to model the effects of cooling, star formation and stellar feedback, came to the realization that supernova heating, while being a suitable explanation for the inefficiency of galaxy formation at the low-mass end, could neither solve the remaining problem with the high-mass end of the galaxy stellar mass function (if anything it made it worse; see e.g. \citealt{Menci:2000};\citealt{Bower:2001};\citealt{Benson2003}) nor reproduce the properties of the gas in massive galaxies, groups and clusters \citep[e.g.][]{Kay:2003,Valdarnini:2003, Nagai:2007}. Indeed, the amount of energy injected by supernovae is insufficient to eject gas from the potential wells of groups and clusters even if one assumes that the feedback is one hundred per cent efficient \citep[e.g.][]{Valageas:1999,Kravtsov:2000,Wu:2000}, and such efficiencies were in conflict with contemporary observations of galactic outflows \citep[e.g.][]{Martin:1999}. Heating from thermal conduction was also investigated \citep[e.g.][]{Fabian:2002,Benson2003,Dolag:2004,Pope:2005} and eventually ruled out as it required that the conduction coefficient should exceed the Spitzer rate expected for a fully ionized plasma. By contrast, the energy injected by the supermassive black holes at the center of galaxies is in principle sufficient to eject gas from the potential wells of groups and clusters \citep[e.g.][]{Kauffmann:2000,Benson2003,Granato:2004}. Additionally, the existence of a feedback loop between supermassive black holes and galaxy formation provides an attractive solution to explain the observed correlation between galaxy and black hole properties (see \S\ref{s:coevol} and~\ref{s:Mbh} for details). Galaxy groups are the best astrophysical laboratories for studying the impact of various feedback mechanisms, since they have managed to retain enough hot gas to allow for a study of the impact of feedback on the IGrM while at the same time representing a transitional regime for the stellar properties of galaxies. We will discuss this point in more detail in the remainder of this review.

\subsection{\textbf{Co-evolution between black hole mass and galaxy properties}}
\label{s:coevol}

Since the late 1990s, it has become clear that central SMBH co-evolve with the properties of their host galaxies. Thanks to increasingly precise measurements of SMBH masses obtained through spatially resolved dynamical measurements of stars and gas at the BH's vicinity, it is now well established that the vast majority of galaxies host a central SMBH with a mass that correlates with the properties of its host galaxy \citep[see][for an extensive review]{Kormendy:2013}. The SMBH mass was found to correlate with a galaxy's near-infrared luminosity $L_K$, i.e. with the galaxy's integrated stellar content, and with the velocity dispersion $\sigma_v$ of the stars in the bulge, which is often used as a proxy for the halo mass \citep{Magorrian:1998,Ferrarese:2000,Gultekin:2009, Kormendy:2013, McConnell:2011, McConnell:2013, vBosch:2016}. SMBH masses scale with galaxy properties as $M_{\rm BH} \propto \sige^{4.5}$ and $M_{\rm BH} \propto L_K^{1.1}$. While it is still unclear whether the relations between SMBH mass and galaxy properties are fundamental or derive from correlations between hidden variables, the existence of these scaling relations implies that the processes leading to the growth of BH are tightly linked with the evolution of the host halo. Widespread AGN feedback provides a natural explanation for the existence of SMBH scaling relations \citep{Silk:1998,Fabian:1999,Granato:2004}. The radiative and mechanical energy output of AGN outbursts can in principle be sufficient to expel cold gas away from galaxy bulges, thereby regulating at the same time the AGN accretion rate and the star formation activity of the host galaxy. Feedback from AGN has the potential to directly link the properties of supermassive black holes and their host galaxies. The coupling of the energy released by the formation of the supermassive black hole to the surrounding forming galaxy should lead to a relationship close to the observed $M_{\rm BH}-\sigma_{\rm e}$ relation \citep{Cattaneo:1999,Kauffmann:2000}. The generality of these arguments imply that this result may be reasonably independent of the specific details of the feedback model. 

While the $M_{\rm BH}-L_K$ and $M_{\rm BH}-\sige$ relations are now well established, the typical intrinsic scatter of optical/stellar relations remains substantial (even in early-type galaxies), $\epsilon \sim 0.4-0.5$ dex, especially for stellar luminosities and masses (e.g., \citealt{Saglia:2016}). Moreover, optical extraction radii are necessarily limited to galactic half-light radii \Re\ ($\sim2-5$\,kpc), neglecting the key role of the host halo of the group. It has been suggested that SMBH properties are fundamentally linked with the mass of the host halo \citep{Ferrarese:2002,Booth:2010} and that the $M_{\rm BH}-L_K$ and $M_{\rm BH}-\sigma_v$ relations arise as a byproduct from the scaling relations between halo mass and optical galaxy properties. Several recent studies seem to confirm that the mass of the central SMBH is more tightly related to the temperature of the host gaseous halo, i.e., the global gravitational potential and hot-halo processes \citep{Bogdan:2018,Gaspari:2019,Lakhchaura:2019}. We discuss this point in detail in \S\ref{s:Mbh}.

\section{Observational evidence} \label{s:obs}

\subsection{\textbf{X-ray observations}} \label{ss:X-ray}

\subsubsection{Feedback-induced hydrodynamical features}

The X-ray observatories in orbit for the past 20 years, \Chandra\ and \XMM, have revolutionized our understanding of the cores of relaxed galaxies, groups, and clusters, which show a highly peaked X-ray emission from a hot interstellar medium whose radiative cooling time is often less than 1~Gyr. Soon after the launch of \XMM\ and \Chandra\ it was realized that the gas in the central regions of nearby groups and clusters does not efficiently cool from the X-ray phase, condense, and flow toward the center, as expected from the original `cooling flow' model \citep{Fabian:1994}. Spectroscopic observations with \Chandra\ and \XMM\ have established that there is little evidence for emission from gas cooling below $\sim T_{vir}/3$ \citep{Kaastra:2001,Peterson:2001,Molendi:2001}. Precisely where the gas should be cooling most rapidly, it appears not to be cooling at all. This effect is known as the `cooling flow problem' \citep[e.g,][]{Peterson:2006}.

A compensating heat source must therefore resupply the radiative losses, and many possibilities have been proposed, including thermal conduction \citep[e.g.,][]{Narayan:2001}, energy released by mergers \citep[e.g.,][]{Motl:2004,ZuHone:2010}, or by supernovae \citep[e.g.,][]{Silk:1986}; see \S\ref{s:galaxy_evol}. However, feedback from the central AGN was rapidly established as the most appealing solution to the problem. There is, in fact, clear observational evidence for AGN heating as the majority of brightest cluster galaxies of cool-core clusters and groups host a radio loud AGN \citep[e.g.,][]{Burns:1990,Best:2007} and, following the launch of \Chandra, disturbances such as shocks, ripples, and cavities have been found in the central atmospheres of many clusters, groups, and elliptical galaxies \citep[e.g.,][]{Finoguenov:2001,Vrtilek:2002,Birzan:2004,Forman:2005,Allen:2006,Fabian:2006,Dunn:2006,Jetha:2007,Croston:2008}. The cavities, which appear as X-ray surface brightness depressions, have been interpreted as bubbles of low-density relativistic plasma inflated by radio jets, displacing the thermal gas and causing $PdV$ heating \citep[e.g.,][]{Churazov:2002}. Weak shocks associated with outbursts, long expected in models of jet-fed radio lobes \citep[][]{Scheuer:74} were also finally detected in deep \Chandra~ observations, for example in 
M87 \citep{Forman:2005}, Hydra A \citep{Nulsen:2005_Hydra} and MS 0735+7241 \citep{McNamara:2005}. The energies available from the AGN were found to be not only comparable to those needed to stop gas from cooling, but the mean power of the outbursts was well correlated with the radiative losses from the IGrM \citep{Nulsen:07}. 

\begin{figure}[t]
\includegraphics[width=0.49\textwidth,bb=36 126 577 667]{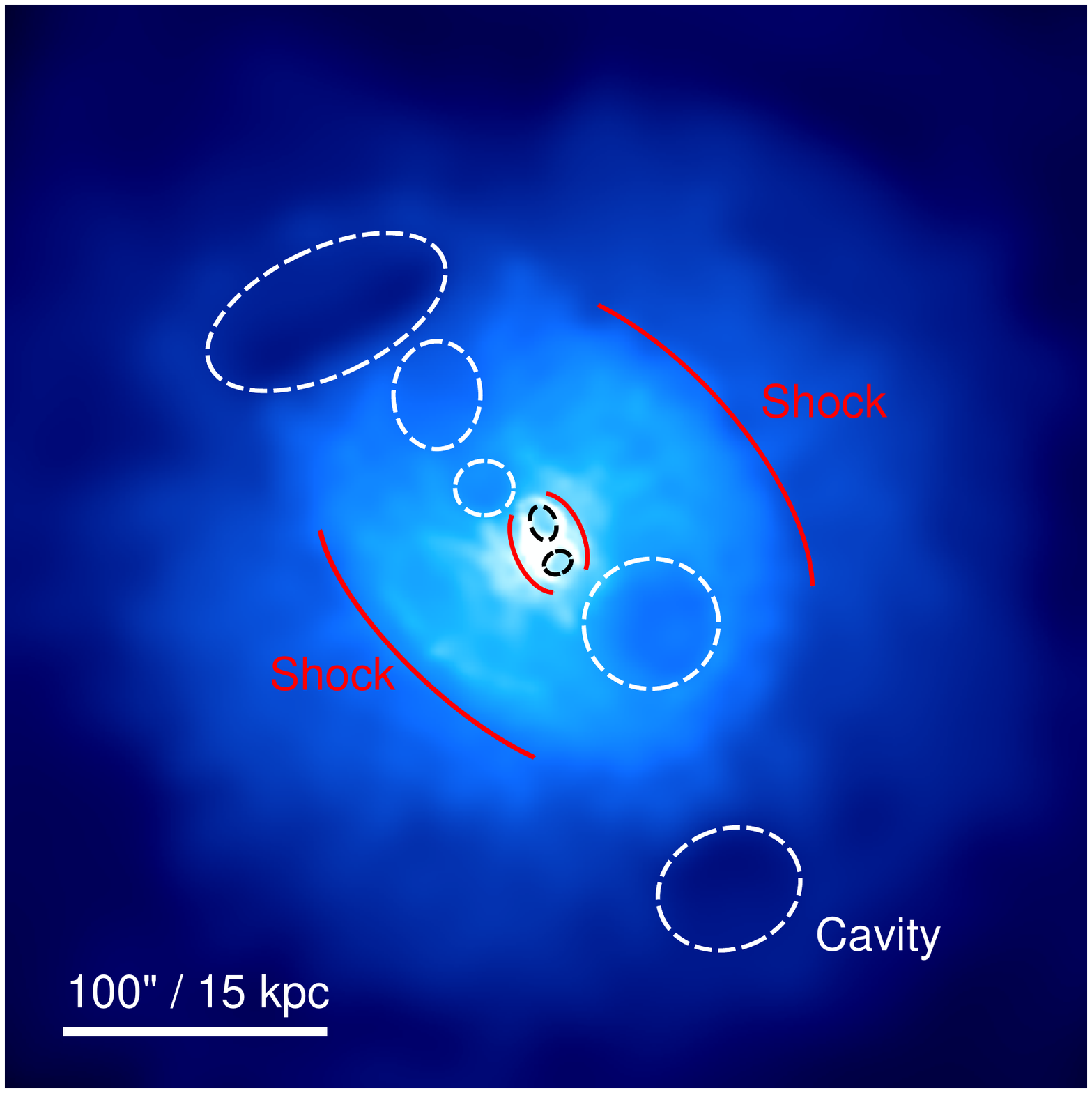}
\includegraphics[width=0.502\textwidth,bb=36 133 577 660]{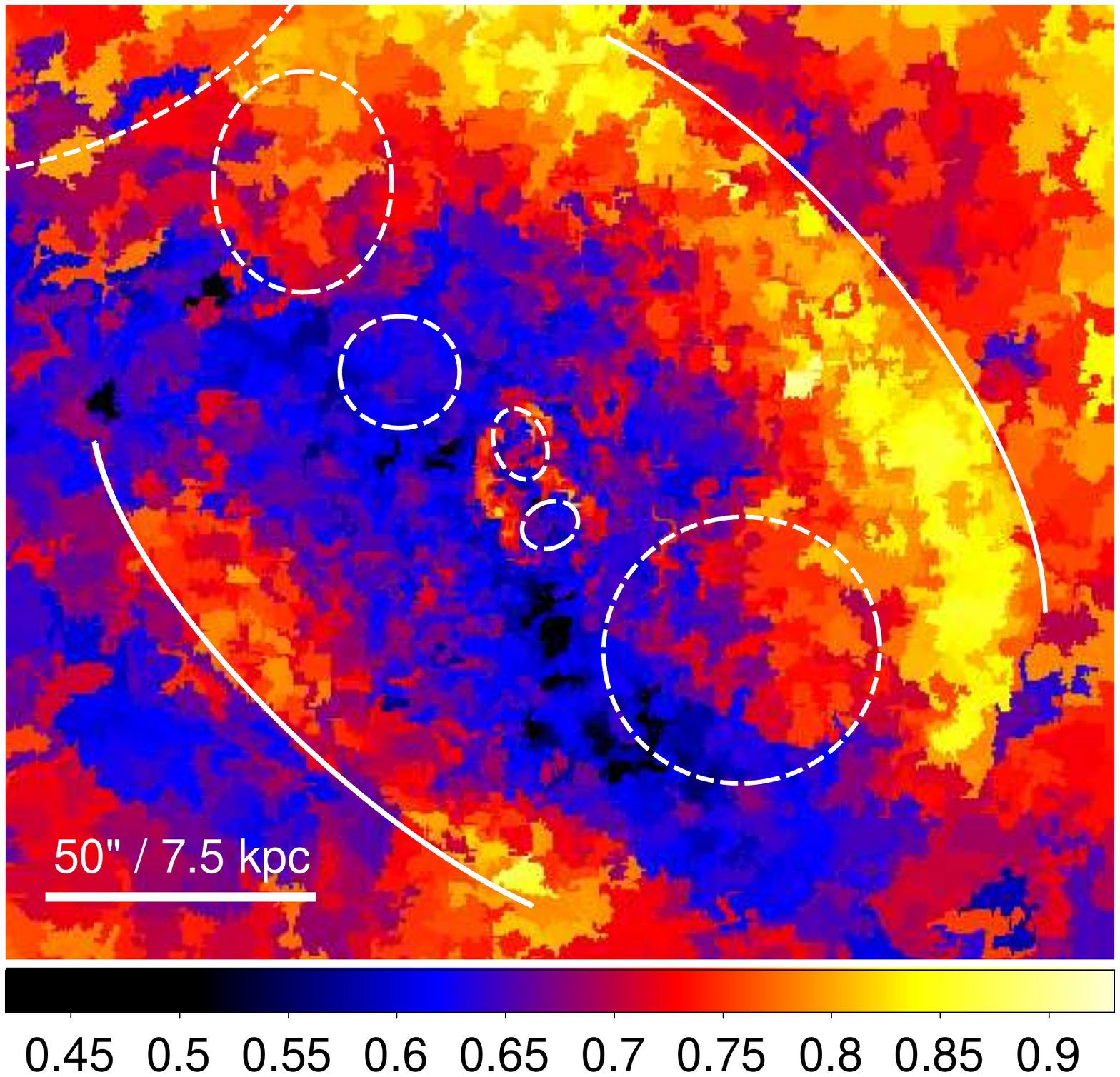}
\caption{
\label{fig:N5813}
Cavities and shocks in the NGC~5813 galaxy group. The \textit{left panel} shows an adaptively smoothed \Chandra\ 0.5-2~keV image of the group, with cavities marked by dashed ellipses and two pairs of shock fronts by solid curved lines. The \textit{right panel} shows a temperature map (in units of keV) with the cavities and outer shock fronts marked. Note the shock-heated gas (red and yellow) behind the outer shock fronts and in the shocked rims of the innermost set of cavities. Images drawn from the \Chandra\ Early-type Galaxy Atlas
\protect\citep{Kimetal19}
}
\end{figure}

Given the lower surface brightness of groups, in the early days of the \Chandra\ era, the study of AGN feedback in these systems did not progress at the same pace as for more massive clusters \citep{McNamara:2007}. However, the situation has improved in more recent times, with a number of studies addressing sizable samples of groups and characterizing the cavities in their IGrM \citep{Cavagnoloetal10,Dong:2010,OSullivan:2017,Shin:2016}. Table \ref{tab:cavity_properties} presents a list of groups with known cavities detected using high-resolution X-ray observations. Deep \Chandra\ X-ray data are now available for a number of galaxy groups (HCG 62, NGC 5044, NGC 5813). In particular, \citet{Randall:2015} presented the results of a very deep (650 ks) observation of NGC 5813, the longest such observation available to date. In Fig. \ref{fig:N5813}, we show the \Chandra\ image and temperature map of NGC 5813, revealing an impressive number of feedback-induced features. Multiple pairs of X-ray cavities can be observed on both sides of the nucleus, indicating the system has undergone several consecutive AGN outbursts inflating powerful expanding bubbles. Two pairs of concentric shock fronts were discovered perpendicular to the jet axis. The passage of the shock fronts reheats the IGrM, as evidenced by the higher temperatures measured in the post-shock regions (see the right-hand panel of Fig. \ref{fig:N5813}). 


Since cavities are the most commonly detected feedback-related structures, AGN energy input is usually gauged from their properties, as briefly summarized below. The energy required to inflate radio bubbles creating cavities in the X-ray emitting gas is usually expressed as the enthalpy, i.e. the sum of the work done to carve out the cavity and the internal energy of the radio lobes:
\begin{equation}
H = E_{\rm int} + pV = \frac{\gamma}{\gamma-1} pV
\end{equation}

\noindent where $p$ is the pressure of the surrounding IGrM, $V$ is the volume of the cavity and $\gamma$ is the ratio of the specific heats of the plasma filling the cavities. If the plasma is relativistic, $\gamma=4/3$ and $H=4 pV$; if it is non-relativistic, $\gamma=5/3$ and $H=2.5 pV$.
The exact composition of the cavities is still unknown, even though X-ray, radio \citep[][and references therein]{McNamara:2012} and SZ observations \citep{Abdulla:2019} suggest that it is likely a mixture of the two species. The other key physical quantity which can be estimated from observations is the age of the cavity. 
As summarized by \citet{Birzan:2004}, several age estimators have been suggested:
\begin{enumerate}[label=(\alph*)]
\item the \emph{sonic time}, i.e. the time required by the cavity to reach its projected distance $R$ at the speed of sound,
\begin{equation}t_s=R/c_s\end{equation} 
\noindent with $c_s=(\gamma kT/\mu m_H)^{1/2}$, $\mu$ the mean atomic weight of the plasma, and $m_H$ the proton mass;

\item the \emph{refill time} required by the gas to refill the displaced volume as the cavity rises upward,
\begin{equation}
t_{\rm ref} \approx \sqrt{r/g}
\end{equation} 
\noindent where $r$ is the radius of the cavity and $g=GM(<R)/R^2$ is the gravitational acceleration at the cavity position;

\item the \emph{buoyancy time}, i.e.  the time required for the cavity to rise buoyantly at its terminal velocity,
\begin{equation} 
t_{\rm buoy}=R/v_t \approx R\sqrt{SC/2gV}
\end{equation}
\noindent where $V$ and $S$ are the volume and the cross-section of the cavity, respectively, and $C=0.75$ is the drag coefficient \citep{Churazov:2001}

\end{enumerate} 

Cavity ages estimated using these three methods usually agree within a factor of two, with buoyancy times typically in between the shorter sonic time and the longer refill time \citep[][and references therein]{Gitti:2012}. Dividing the enthalpy $H$ by the characteristic timescale provides an observational estimate of the cavity power $P_{cav}$. The cavity power is a lower limit to the mechanical power of the AGN, given the paucity of detected shocks and other possible sources of energy feedback such as sound waves. 

Operationally, estimating $P_{cav}$ requires a measurement of the geometry and size of the cavity and the pressure of the surrounding ICM. The total cavity power can be compared with the gas luminosity inside the cooling radius, $L_{\rm cool}$, which needs to be balanced by the AGN mechanical feedback. $L_{\rm cool}$ is usually defined as the total luminosity inside the regions where the cooling time is less than $7.7 \times 10^{9}$ yrs \citep{Birzanetal08}, although different thresholds exist in the literature \citep[e.g. 3 Gyr,][]{Panagoulia:2014_cav}. $L_{\rm cool}$ is estimated by deprojecting the X-ray temperature and emissivity profiles and computing the corresponding bolometric luminosity \citep{Dunn:2006}. A number of possible biases and systematic errors can affect this apparently straightforward observational approach, as the detectability of cavities depends on the depth of the observation, position of the cavity with respect to the plane of the sky, and uncertainties in the assumed geometry \citep[][and references therein]{McNamara:2007,Gitti:2012,McNamara:2012}. The exact impact of these observational uncertainties on the statistics of cavities in clusters and groups is yet to be quantified.

\begin{figure}[t]
\centerline{\includegraphics[width=0.8\textwidth]{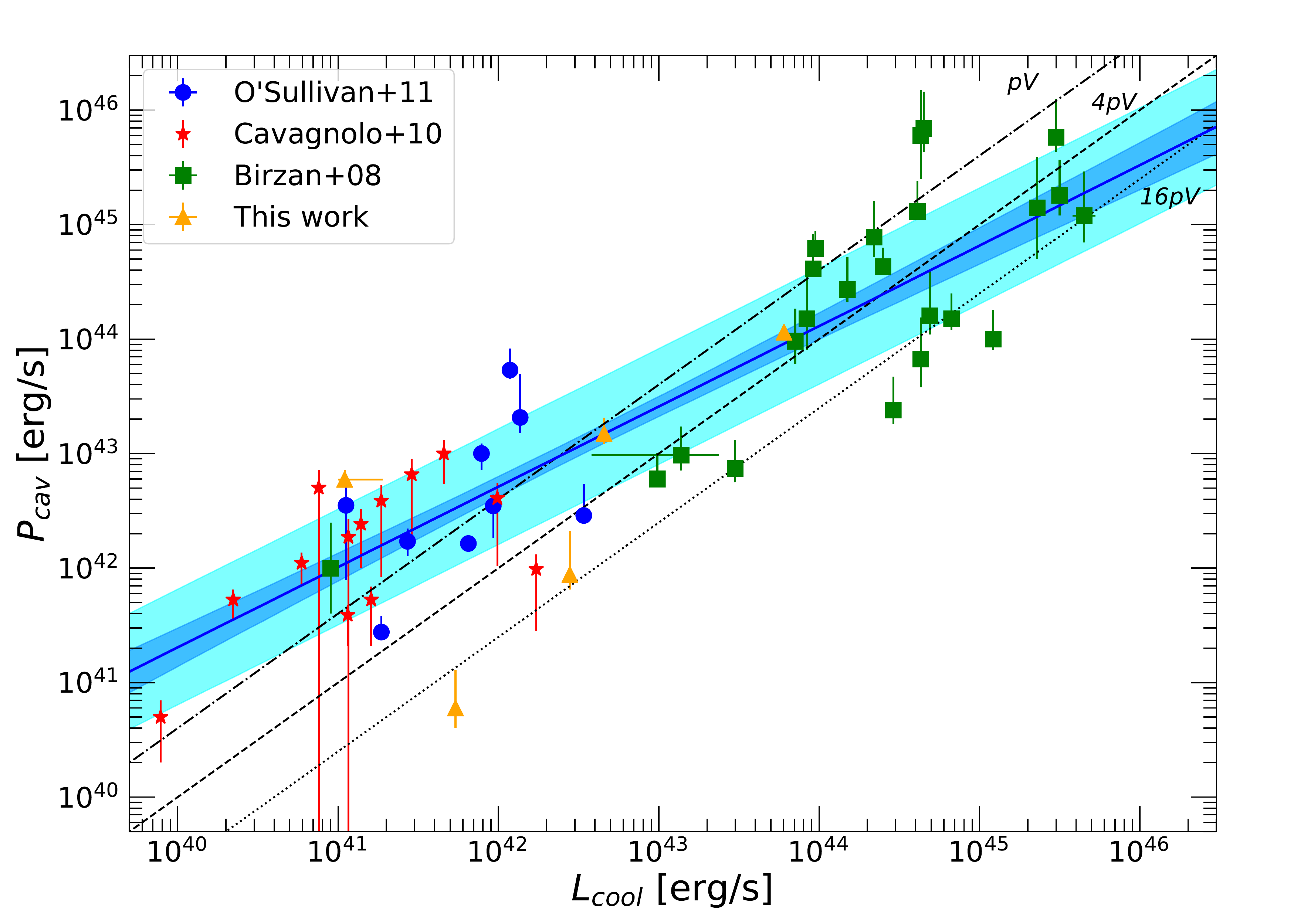}}
\caption{
\label{fig:pcav_lcool}
Relation between the luminosity within the cooling radius ($L_{cool}$) and the power injected by the cavities assuming $H=4pV$ ($P_{cav}$) for several literature samples. The orange points were either recomputed for this work or collected from papers on individual objects; we refer to Table \ref{tab:cavity_properties} for detailed references. The blue line shows a fit to the data using a power law with intrinsic scatter. The uncertainty on the fitted relation is indicated by the blue shaded area, whereas the cyan range indicates the intrinsic scatter around the relation. 
}
\end{figure}

The $P_{cav}-L_{cool}$ relation has been investigated through the years in an increasing number of objects ranging from ellipticals to groups and clusters 
\citep[][and references therein]{Gitti:2012,McNamara:2012}. 
The general consensus is that the cavity power is enough to offset cooling given an average $4pV$ injected energy per cavity and that the jet mechanical power correlates well with the cooling luminosity. In Fig. \ref{fig:pcav_lcool}, we show the relation between cooling luminosity and cavity power from a compilation of literature measurements \citep{Birzanetal08,OSullivan:2017,Cavagnoloetal10}; see Table \ref{tab:cavity_properties}. Here the total cavity power for each system was computed by summing up the power of each individual cavity. While at the high-mass end, the data are broadly consistent with an enthalpy $H=4pV$, typical of heating by a relativistic plasma, in the group regime, the cavity power is substantially higher. To quantify this effect, we fitted the $L_{\rm cool}-P_{\rm cav}$ relation with a power law using \texttt{PyMC3} \citep{pymc3}. The blue curve and shaded area show the best-fit relation, which reads

\begin{equation}
\log\left(\frac{P_{\rm cav}}{10^{43}\mbox{ erg/s}}\right) = (0.41\pm0.09) + (0.70\pm0.05)\log\left(\frac{L_{\rm cool}}{10^{43}\mbox{ erg/s}}\right)
\end{equation}

\noindent with an intrinsic scatter of $0.51\pm0.07$ dex. The slope of the fitted relation is significantly shallower than unity, which would be the expected slope if the feedback efficiency is independent of halo mass. At face value, this result implies that the feedback efficiency is higher in groups than in clusters, with the cavities injecting enough energy to overheat the cores and deplete the central regions from their gas content. We discuss this point in detail in \S\ref{s:baryon_content}.

\begin{figure}[t]
 \centering
     \includegraphics[width=1.0\textwidth]{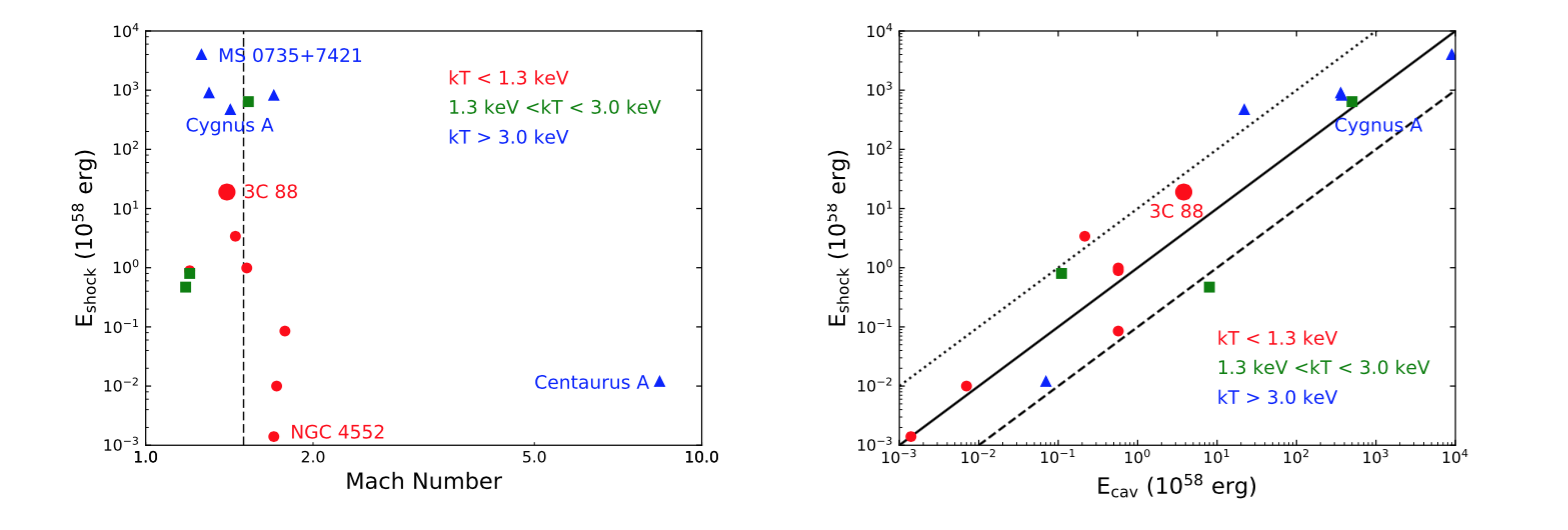}
     \caption{Shock energy versus Mach number (left panel) and shock energy versus cavity enthalpy (right panel) for groups and clusters with available shock energy in the literature. Figure reproduced with permission from \citet{Liuetal19}. Objects with a red circle have $kT < 1.3$ keV, green square: $1.3 < kT < 3.0$ keV and blue triangles $kT > 3.0$ keV. The dashed line in the left panel marks $\mathcal{M}=1.5$ and the dotted, solid, dashed  lines in the right panel represent $E_{shock}/E_{cav}$=10, 1, 0.1 respectively. The objects considered in the plots are: Hydra A \citep{Nulsen:2005_Hydra,Wise:2007}, MS 0735+7421 \citep{McNamara:2005}, Centaurus A \citep{Croston:2009}, Cygnus A \citep{Birzan:2004,Snios:2018}, 3C 444 \citep{Croston:2011}, M87 \citep{Forman:2017}, Abell 2052 \citep{Blanton:2009}, 3C 310 \citep{Kraft:2012}, NGC 4552 \citep{Machacek:2006}, NGC 4636 \citep{Baldi:2009}, HCG 62 \citep{Gitti:2010}, NGC 5813 \citep{Randall:2015}. } 
     \label{fig:shock}
 \end{figure}

Another clear observational hydrodynamical feature caused by AGN feedback is the presence of shock fronts. The passage of a shock front compresses and heats the gas, raising its entropy and providing an effective heat input 
\begin{equation}\Delta Q \approx T \Delta S= T \Delta \ln K\end{equation} 
\noindent with $K$ the entropy index usually quoted as entropy by X-ray astronomers (see \S\ref{s:entropy}). Because of their transient nature, single weak shocks fail to compensate for the radiative losses in a cool core, but the cumulative effect of multiple shocks can be relatively important \citep[see e.g. the discussion in ][]{McNamara:2012}. This is again highlighted by the exemplar case of NGC 5813 (see Fig.\ref{fig:N5813}), where each set of three cavities has been associated with an elliptical shock measured at 1 kpc, 10 kpc and 30 kpc, respectively, with Mach numbers $\mathcal{M}$ in the range 1.17-1.78. Generally speaking, the detected shock fronts are weak, i.e. their Mach number falls in the range $\mathcal{M}\sim1-2$ \citep{Liuetal19}. This range can be understood given the typical evolution of the Mach number as a function of the fundamental parameters, total energy and duration, of the AGN outbursts \citep[see for example the discussion in ][and section \S \ref{s:heating}]{Tang:2017}. The cumulative heating effect of these successive shock fronts is sufficient to offset cooling within the inner 30 kpc. The shock energy can be estimated as 

\begin{equation}
E_s = p_1 V_s (p_2/p_1 - 1)
\end{equation}
\noindent where $p_1$ and $p_2$ are the pre- and post-shock pressures, respectively, and $V_s$ is the volume enclosed by the shock \citep[e.g.,][]{Randall:2015}. \citet{Liuetal19} made an exhaustive search for groups and clusters with detected shocks and studied the dependence of the shock energies and related Mach numbers on the cavity enthalpies (see Fig.\ref{fig:shock}). The shock energies span almost 7 orders of magnitude from $4\times 10^{61}$~erg~s$^{-1}$ in the cluster MS 0735 + 7421 \citep{McNamara:2005} to $10^{55}$~erg~s$^{-1}$ in NGC 4552 \citep{Machacek:2006}, with group-scale objects in the range $10^{56}-10^{59}$~erg~s$^{-1}$ and Mach numbers all in the range 1-2 with the exception of Centaurus A. The shock energy is similar to the cavity energy (see the right-hand panel of Fig.\ref{fig:shock}), suggesting that shocks and cavities may play a comparable role in supplying mechanical energy provided by the AGN \citep[with the balance between the two mainly driven by the duration of the outburst, e.g.][]{Forman:2017,Tang:2017}. Other heating mechanisms discussed in \S\ref{s:heating}, such as turbulent heating, have yet to be explored observationally at the group scale.

\subsubsection{Non-gravitational feedback energy and entropy profiles}
\label{s:entropy}

One of the earliest pieces of evidence for non-gravitational feedback energy came from observed deviations
from the self-similar scaling relations driven only by gravity \citep{Kaiser:1986}, in particular the
deviation of the observed luminosity-temperature relation from the predicted $L \propto T^2$ 
\citep{Allen:1998,Arnaud:1999,Markevitch:1998}; see the companion review by \citet{Lovisari:2021} on the scaling relations of galaxy groups. 
One of the first attempts at a solution was to advocate a minimum entropy in the pre-collapse intergalactic medium to break self-similarity \citep{Kaiser:1991,Evrard:1991} with the result of bending the relation from self-similar at the scale of massive clusters to a steeper slope at the scale of groups. It was recognized that a given entropy level could be reached through different thermodynamic histories and that the key insight would have been given by the sequence of adiabats through which baryons evolve: the excess entropy could have been achieved prior accretion to the
collapsed halo (the external scenario) or in the higher density medium after accretion (internal scenario \citep[e.g.][]{Tozzi.ea:00,Tozzi:2001}. A third option was initially considered realizing that cooling alone could remove low entropy gas from the centers of halos producing a similar effect to non-gravitational heating \citep[e.g.,][]{Knight:1997,Bryan:2000}.

\begin{figure}[t]
 \centering
     \includegraphics[width=1.0\textwidth]{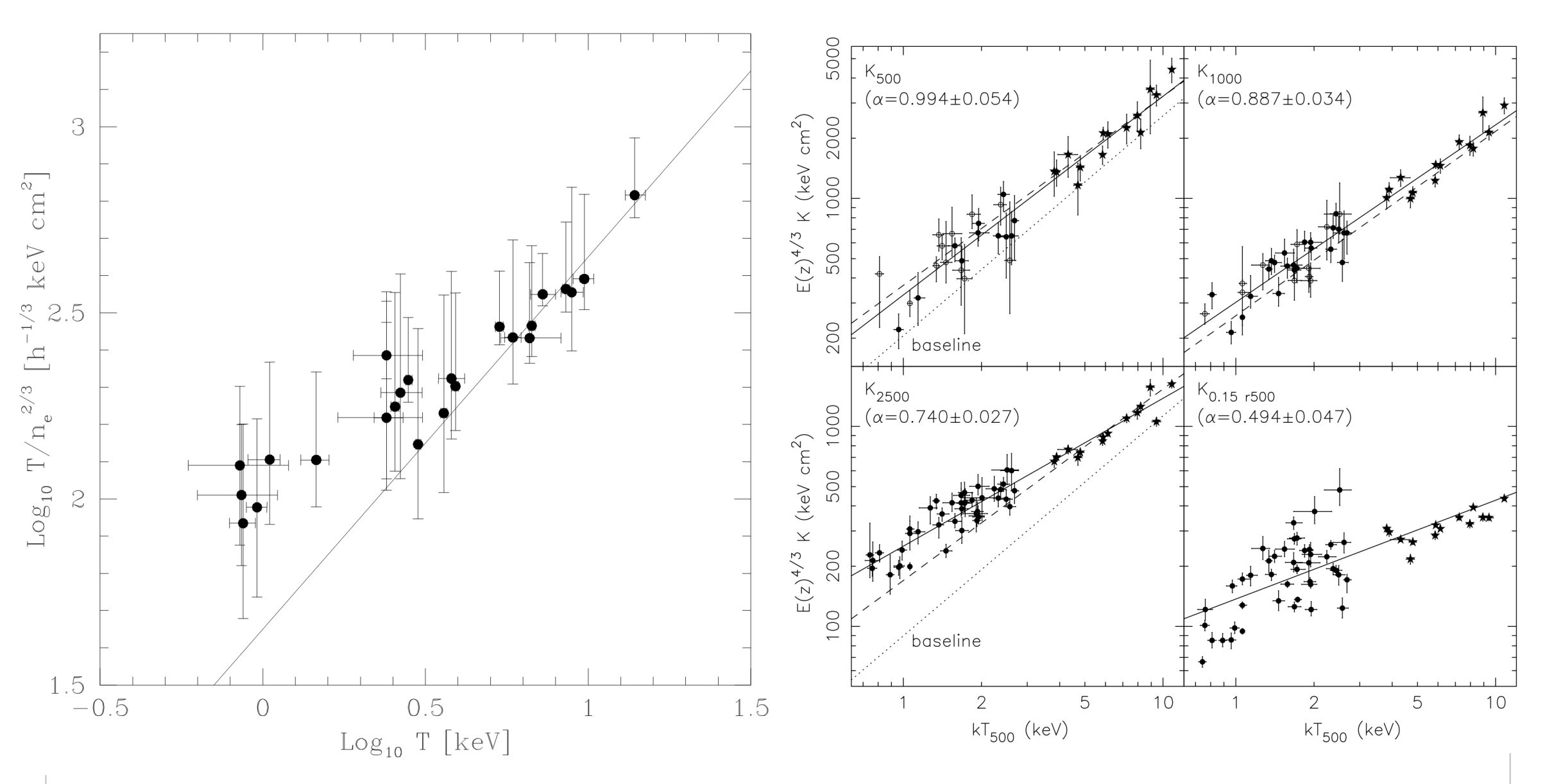}
     \caption{\textit{Left:} The gas entropy at the fiducial radius of $0.1 R_{virial}$ as a function of the temperature for the 25 systems in the sample of \citet{Ponman:1999}. The solid line shows the relation obtained from numerical non-radiative simulations \citep{Eke:1998}. Figure reproduced with permission from \citet{Ponman:1999}, arXiv author's version. \textit{Right:} The same relation at different radii for the groups analyzed in the sample of \citet{Sun:2009a} together with the sample of clusters in \citet{Vikhlinin:2009}, both analyzed with \Chandra\ data. Figure reproduced with permission from \citet{Sun:2009a}.}
     \label{fig:entropy}
 \end{figure}

\citet{Ponman:1999} provided  the first observational evidence by measuring the entropy ($S=T/n_e^{2/3}$) at a fixed scaled radius (0.1 $r_{virial}$) and showing that it does not follow the expected linear trend with temperature (see the left-hand panel of Fig.~\ref{fig:entropy}). \ROSAT\ surface brightness profiles were combined with \emph{Ginga} mean temperatures under the assumption of isothermality to derive entropy profiles for 25 objects, 6 objects with $T < 2$ keV (NGC4261, NGC2300, NGC533, HCG97, HCG94, HCG62), and objects like MKW4, MKW3, AWM4 and AWM7 all in the range $T$=2-4 keV. The authors concluded that the observational data were consistent with a scenario involving an external preheating mechanism through supernova winds, thereby raising the central entropy and enriching the medium in heavy elements. They also established that preheating would have a broader impact on the general picture of structure formation, as for example the level and timing of the heating required could not violate constraints from the Lyman-$\alpha$ forest \citep[see also ][]{Borgani:2009}.

A key improvement with respect to this early result was the ability to go beyond the assumption of isothermality by constraining the temperature profiles exploiting the combination of \ROSAT\ and \ASCA\ data \citep{Lloyd-Davies:2000,Ponman:2003}. These studies confirmed that low-mass systems exhibit higher scaled entropy profiles. However, they did not show the large isentropic cores predicted by simple preheating models \citep[e.g.,][]{Brighenti.ea:01} and the high entropy excess in galaxy groups was found to extend to large radii \citep[as also shown by the \ASCA\ analysis of][]{Finoguenov:2002}. 

The \Chandra\ and \XMM\ results have made the observational picture clearer and more solid. In the comprehensive work done by \cite{Sun:2009a}, an archival sample of 43 groups observed with \Chandra\ with a temperature range of $kT_{500}=0.7-2.7$~keV was analyzed, deriving detailed entropy profiles thanks to the superb spatial resolution of the satellite. The derived entropy-temperature scaling relations at six characteristic radii (30 kpc, 0.15 $R_{500}$, $R_{2500}$, $R_{1500}$, $R_{1000}$, $R_{500}$) show a large intrinsic scatter at small radii, but already at $R_{2500}$, the scatter reaches a value of 10\% and remains the same beyond this point. When combined with similar observations in the galaxy cluster regime \citep[the sample of][]{Vikhlinin:2009}, the slope of the relation is found to gradually approach the self similar value, steepening from $0.740 \pm 0.027$ at $R_{2500}$ to $0.994 \pm 0.054$ at $R_{500}$ (see the right-hand panel of Fig.\ref{fig:entropy}). The entropy ratios calculated with respect to the baseline entropy profile expected from purely gravitational processes \citep{Voit:2005c} confirm an excess entropy which is a function of mass and radius, with groups having higher ratios at small radii. The weighted mean ratio for groups decreases from 2.2 at $R_{2500}$ to 1.6 at $R_{500}$. In general the entropy profiles of groups have slopes (0.7-0.8) which are flatter than the self-similar expectation from pure gravitational processes of 1.1.  Deep observations of nearby poor clusters with \emph{Suzaku} (RX J1159, \citealt{Humprey:2012}; Virgo, \citealt{Simionescu:2017}; UGC 03957, \citealt{Tholken:2016}) have shown that the entropy excess can extend all the way to the system's virial radius. Similar observations on a larger sample of groups are needed to determine whether the high entropy of galaxy groups is a general feature linked to the AGN feedback phenomenon. 

The \XMM\ study performed by \cite{Johnson:2009} on a sample of 29 groups based on the two-dimensional \XMM\ Group Survey, 2DXGS \citep{Finoguenov:2006,Finoguenov:2007} supplemented by groups from the sample of \citet{Mahdavi:2005} divided the objects into cool core (CC) and non-cool core (NCC) objects on the basis of the presence of a temperature gradient in the core, the first time this had been done for galaxy groups. The slope of the scaling relations of the entropy with temperature, incorporating the cluster sample of \citet{Sanderson:2009}, at 0.1 $R_{500}$ has a slope of $0.79 \pm 0.06$ consistent with the results of \citet{Sun:2009a}. The entropy profiles of NCC groups show more scatter than the CC sub-sample, and they have higher central entropies, in qualitative agreement with the results at the cluster scale \citep[e.g.][]{Cavagnolo:2009}. The excess entropy with respect to the baseline expected from gravitational processes cannot be reproduced by simple theoretical models of entropy modification such as pure pre-heating or pure cooling and the required mechanism should provide increasingly large entropy shifts for higher entropy gas.

Another significant advance in the study of entropy profiles of group-scale objects has been provided by the work of \citet{Panagoulia:2014_prof}, which analyzed the entropy profiles of 66 nearby groups and clusters drawn from a volume-limited sample of 101 objects assembled from the NORAS and REFLEX catalogues. The study pointed out that the flattening of the entropy profiles at small radii found in previous studies \citep{Cavagnolo:2009} was mainly a matter of resolution and could be affected by the presence of multi-temperature gas. In particular, for nearby groups, a broken power law model provides the best description of the entropy profiles, with an inner slope of 0.64 within the central 20 kpc. This finding was confirmed in the recent studies of \citet{Hogan:2017} and \citet{Babyk:2018b}, who found that the behavior of the entropy profiles in the inner regions of relaxed clusters and groups can be well described by a broken power law with $K(r)\propto R^{2/3}$, a break around $100 $ kpc ($\sim0.1 R_{2500}$) and an outer slope of $\sim 1.1$ matching the predictions of gravitational collapse models \citep{Voit:2005c}. Since the cooling time is $\propto K^{3/2}$, the non-existence of an entropy floor affects the interpretation of the SMBH feeding and feedback processes, as we will discuss in \S\ref{s:ratios} and \S\ref{s:theory}.


\subsubsection{Thermal instability timescale profiles}
\label{s:ratios}
In the past decade, a substantial amount of work has been dedicated to understanding the triggering of AGN feedback in galaxy groups and clusters. As we will review in \S\ref{s:theory}, the energy injected by the central AGN and the cooling of the IGrM appear to be closely balanced over the long-term. This reflects a tight relation between cooling/feeding (\S\ref{s:cooling}) and heating/feedback (\S\ref{s:heating}) processes.
Initial works suggested that the onset of thermal instability in hot halos and the triggering of runaway cooling can be expressed in terms of the ratio of the cooling time to the free-fall time \citep{Gaspari:2012a,McCourt:2012,Sharma:2012,Voit:2015_nat,Voit:2015_SFR}. The cooling time is usually expressed as the ratio of the thermal energy to radiative cooling rate, 

\begin{equation} \label{e:tcool}
t_{\rm cool}=\frac{(3/2)n k_{\rm B} T}{n_{\rm e} n_{\rm i}\Lambda} \approx \frac{3 k_{\rm B} T}{n_{\rm e}\Lambda}
\end{equation}

\noindent with $\Lambda$ the cooling function (see Fig. \ref{f:Lambda}) and $n_{\rm i}\approx n_{\rm e}$ the IGrM ion number density. The free-fall time describes the timescale necessary for a gas particle to directly fall to the bottom of the potential well, 

\begin{equation} \label{e:tff}
t_{\rm ff}=[2R/g(R)]^{1/2} 
\end{equation}

\noindent with $g(R)$ the local gravitational acceleration. If $t_{\rm cool}\gg t_{\rm ff}$, the gas particle is losing internal energy too slowly and radial oscillations are eventually damped. Conversely, as $t_{\rm cool}$ becomes comparable to $t_{\rm ff}$, the gas rapidly loses pressure support, developing runaway thermal instability, and eventually sinking radially onto the central galaxy and SMBH. This `classical' thermal instability (and related TI-ratio) has been also referred to as `raining' (\citealt{Gaspari:2012a}) or `precipitation' \citep{Voit:2015_nat}, being analogous to early physics studies based on analytical approximations (e.g., \citealt{Field:1965}).

\begin{figure}[t]
\hspace{-0.3cm}
\includegraphics[width=0.43\textwidth]{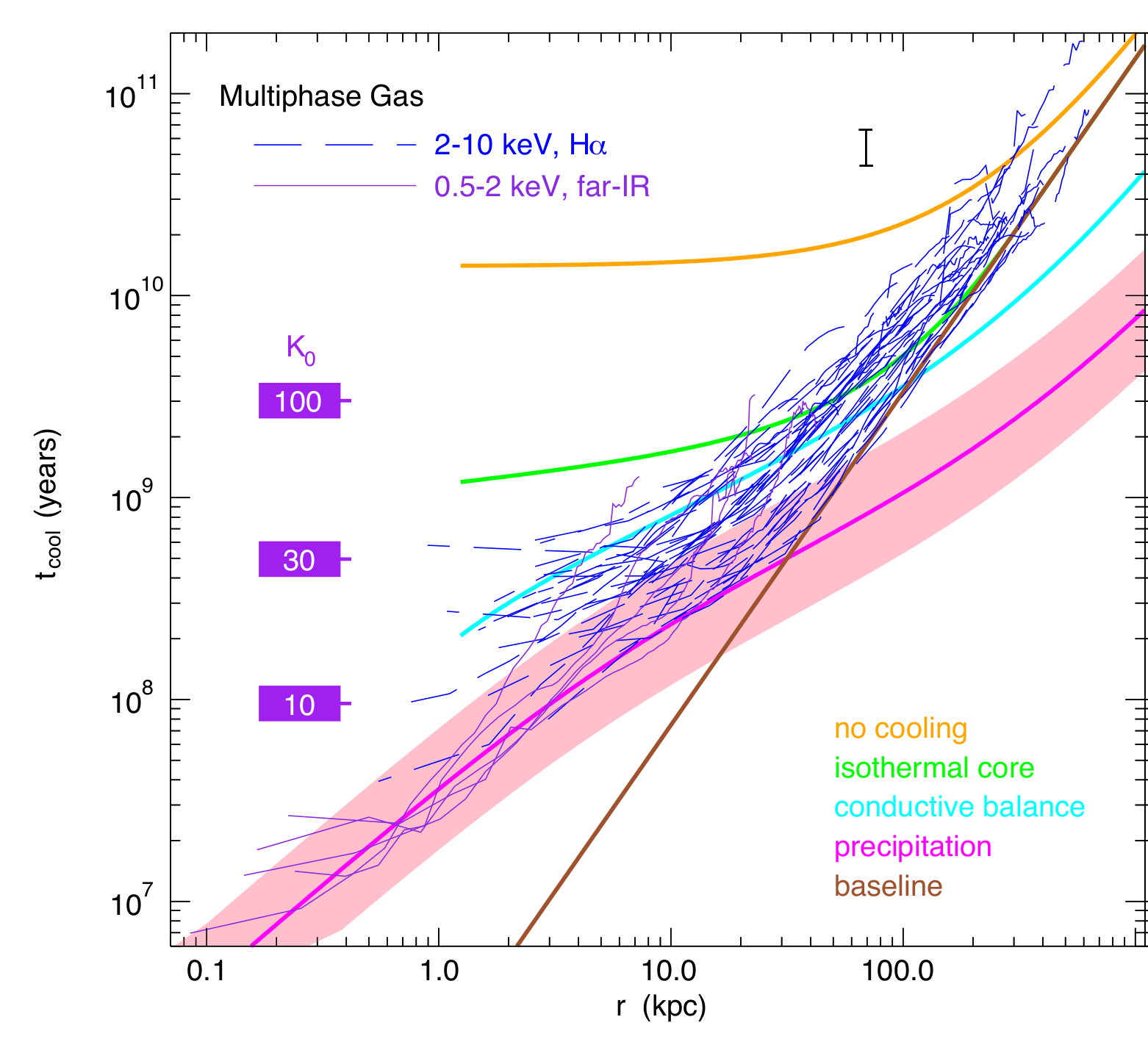}
\hspace{-0.15cm}
\raisebox{0.1cm}{\includegraphics[width=0.53\textwidth]{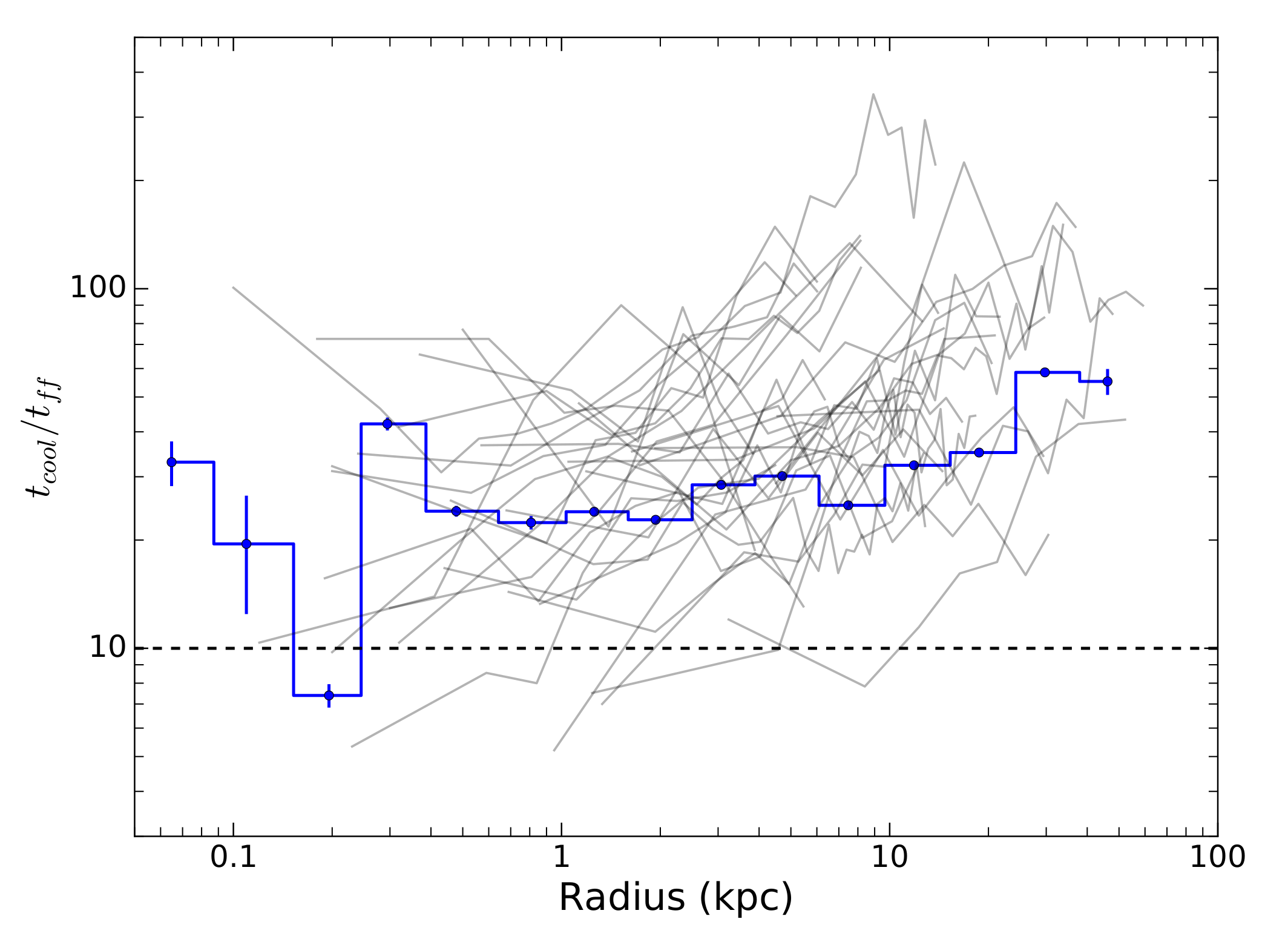}}
\caption{Observed timescales profiles in the IGrM (mainly related to TI). \emph{Left:} Cooling time profiles for a sample of cool-core clusters (blue) and groups (magenta) hosting multiphase gas (H$\alpha$, CO; figure reproduced with permission from \citealt{Voit:2015_nat}, arXiv author's version). The observed profiles are compared with the `baseline' structure formation profile (maroon) and the typical simulation threshold $t_{\rm cool}/t_{\rm ff}\sim 10$ (pink). \emph{Right:} The ratio of the cooling time to free-fall time in a sample of groups and ellipticals (figure reproduced with permission from \citealt{Babyk:2018b}; the average is shown with a blue line).
}
\label{fig:ratios}
\end{figure}

Observational constraints on thermal instability can be investigated by studying the radial profiles of such IGrM timescales. In the absence of cooling and non-gravitational heating ($t_{\rm cool}\gg1$ Gyr), the cooling time follows a baseline universal profile set by the structure formation process \citep{Borgani:2005,Voit:2005c}. Deviations from the baseline profile occur in the central regions where $t_{\rm cool}\ll1/H_0$ and radiative cooling losses can no longer be neglected. The left-hand panel of Fig. \ref{fig:ratios} shows the cooling time profiles for a large sample of galaxy clusters and groups from the ACCEPT database \citep{Voit:2015_nat,Cavagnolo:2008}. All the systems exhibiting evidence of multi-phase gas (e.g., H$\alpha$, CO) show a floor set by the threshold $t_{\rm cool}/t_{\rm ff}\sim10-30$, such that on average the IGrM does not experience runaway cooling. Figure \ref{fig:ratios} (right panel) includes 40 galaxy groups and massive early-type galaxies with deep \Chandra\ observations \citep{Babyk:2018b} showing that the ratio of cooling time to free-fall time reaches a floor at $t_{\rm cool}/t_{\rm ff}\sim10$. In the inner regions where the entropy rises approximately as $K\propto R^{2/3}$, $t_{\rm cool}/t_{\rm ff}$ is approximately constant with an average value of $t_{\rm cool}/t_{\rm ff}\sim30$ (blue line). 

On the one hand, the above results show that the TI-ratio is correlated with the presence of multiphase gas. On the other hand, it is clear that the threshold is puzzlingly not unity (as expected in classical thermal instability) and that there is a substantial intrinsic scatter even in multiphase systems. However, classical thermal instability starts from idealized linear fluctuations and does not account for key astrophysical processes, such as AGN feedback or mergers, both recurrently injecting substantial gas motions (e.g. turbulence) at small and large radii, respectively (e.g., \citealt{Lau:2017}; see \S\ref{s:heating}).
In this more realistic IGrM case of `turbulent' nonlinear TI, the key physical timescale is not the free-fall time but rather the turbulence eddy turn-over time (\citealt{Gaspari:2018}),
\begin{equation} \label{e:teddy}
t_{\rm eddy} = \frac{2\pi R^{2/3}L^{1/3}}{\sigma_{v,L}},
\end{equation}
where $\sigma_v$ is the turbulence velocity dispersion and $L$ is the related injection scale. 
The velocity dispersion of the \textit{ensemble} warm H$\alpha$-emitting gas should linearly correlate with $\sigma_v$ of the hot IGrM, allowing us to convert between the two, in particular by leveraging the higher spectral resolution of optical/IR telescopes. Rough estimates of the injection scale can be also obtained via the size of the ensemble warm gas filaments/nebulae, or via the AGN cavity diameter. In the presence of a turbulent halo, thermal instability develops chaotically and non-linearly in a very rapid way whenever $t_{\rm cool}/t_{\rm eddy} \sim 1$ \citep{Gaspari:2017_cca,Olivares:2019,Juranova:2019}. 
Future X-ray microcalorimeter missions like XRISM and Athena (see \S\ref{s:future}) will allow us to measure the eddy time directly and test in more depth the above scenarios. In \S\ref{s:theory}, we discuss the related processes from a theoretical perspective.

\subsubsection{Baryon content} \label{s:baryon_content}

A key quantity for AGN feedback models in cosmological simulations is the total integrated baryon budget and its dependence on halo mass. While galaxy cluster halos are massive enough to retain all of their baryons \citep[e.g.][]{Eckert:2019}, energy injection by AGN feedback can lead to an overall \emph{depletion} of baryons all the way out to the virial radius. Observational studies have found that the gas fraction within $R_{500}$ increases with halo mass \citep{Gonzalez:2007,Gasta:2007,Gonzalez:2013,Pratt:2009,Lovisari:2015,Eckert:2016,Ettori:2015,Nugent:2020}. Since an estimate of the gas density can be obtained from imaging data only, a lot of attention has been devoted to the study of gas density profiles \citep{Sun:2012,Croston:2008b,Eckert:2016}. Using a compilation of measurements from the literature, \citet{Sun:2012} showed that while the gas density of galaxy group cores is systematically lower than that of more massive systems, at $R_{500}$, the gas density is nearly independent of mass. \citet{Eckert:2016} studied the gas density profiles of the 100 brightest galaxy clusters and groups in the XMM-XXL survey. While at galaxy clusters scales, the measured profiles show a well-defined core and a relatively steep decline in the outskirts, the density profiles of the selected groups exhibit a power-law behavior with a very flat index $n_e(r)\propto r^{-1.2}$, indicating that the gas fraction in spherical shells increases steeply with radius. It is believed that most of the gas has been evacuated from the inner regions under the influence of AGN feedback and displaced to larger radii, thereby explaining the observed shallow slopes \citep{LeBrun:2014,Gaspari:2012a}. 

\begin{figure}[t]
 \centerline{
     \includegraphics[width=1.0\textwidth]{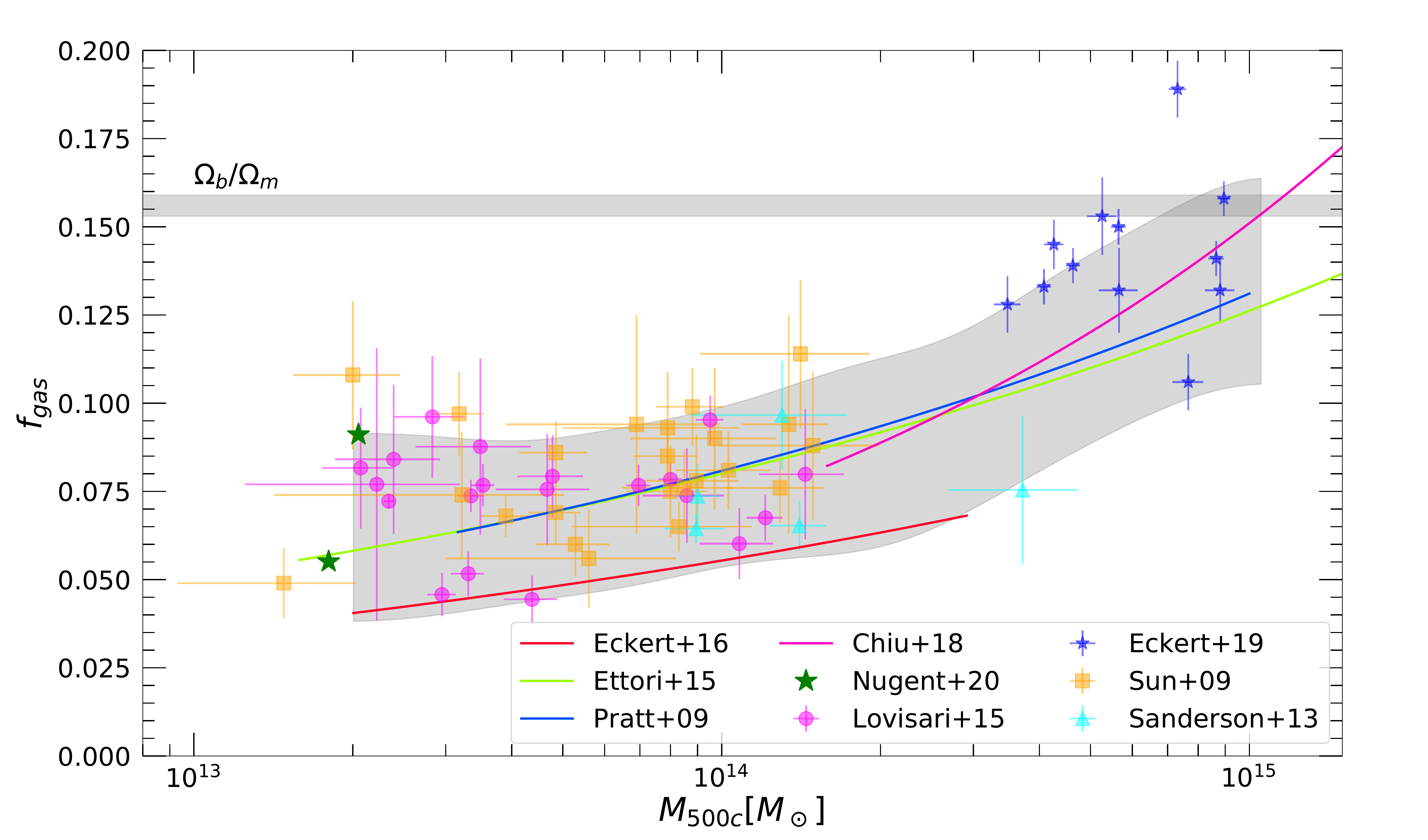}}
     \caption{Compilation of existing measurements of the hot gas fraction at $R_{500}$ in galaxy groups and clusters as a function of halo mass $M_{500}$. The data points show the galaxy group samples of \citet{Sun:2009a} (orange), \citet{Lovisari:2015} (magenta), \citet{Sanderson:2013} (cyan), and \citet{Nugent:2020} (green). The data from the X-COP sample \citep{Eckert:2019} at the high-mass end are shown as the blue points for comparison. The solid lines show the $f_{gas}-M$ relations derived from REXCESS \citep[blue,][]{Pratt:2009}, XMM-XXL \citep[red,][]{Eckert:2016}, SPT-SZ \citep[magenta,][]{Chiu:2018} and the literature sample of \citet{Ettori:2015}. The gray shaded area shows the 90\% confidence range encompassing the existing observational data and their corresponding uncertainties (see text).}
     \label{fig:fgas_obs}
 \end{figure}

In Fig. \ref{fig:fgas_obs}, we present a compilation of published measurements of the hot gas fraction at $R_{500}$ as a function of the corresponding halo mass. The gas fractions were derived from X-ray data under the assumption of hydrostatic equilibrium, with the exception of XMM-XXL and SPT-SZ. In the case of XMM-XXL \citep{Eckert:2016}, weak lensing measurements for a sub-sample of 35 clusters were used to calibrate the mass-temperature relation. SPT-SZ masses \citep{Chiu:2018} were derived as an ensemble from a joint fit to the cosmological parameters and the relation between SZ observable and halo mass. All studies find that the hot gas fraction contained within $R_{500}$ increases with halo mass roughly as $f_{\rm gas}\propto M_{500}^{0.2}$. Here we provide a conservative estimate of the $f_{\rm gas}-M_{500}$ relation which aims at encompassing all state-of-the-art observational studies and their uncertainties. To this aim, we collected the compilation of observational studies from Fig. \ref{fig:fgas_obs} and estimated in each mass bin the median and 90\% confidence range of the data points. The resulting relation is shown as the gray band in Fig. \ref{fig:fgas_obs}. The gray band can be approximated as a power law, which reads

\begin{equation}
	f_{\rm gas,500} = 0.079_{-0.025}^{+0.026} \times \left(\frac{M_{500}}{10^{14}M_\odot}\right)^{0.22_{-0.04}^{+0.06}}.
\end{equation}

While at the high-mass end, the gas fractions approach the cosmic baryon fraction, on galaxy group scales, the IGrM only contains about half of the baryons expected from the self-similar structure formation scenario. On the other hand, the stellar fraction $f_{\star}$ is a weak function of halo mass and decreases only slightly from $2-3\%$ at $10^{13}M_\odot$ to $1-1.5\%$ at $10^{15}M_\odot$ \citep{Andreon:2010,Leauthaud:2012,Gonzalez:2013,Chiu:2018,Coupon:2015,Eckert:2016,Kravtsov:2018}. The weak dependence of the stellar fraction on halo mass is insufficient to compensate for the steeper dependence of the gas fraction, which results in a deficit of baryons in galaxy groups with respect to the cosmic baryon fraction. We note here that this result is independent of the hydrostatic equilibrium assumption adopted by most authors. Indeed, an additional non-thermal pressure term would lead to a slight underestimation of the mass in these studies \citep[e.g.][]{Rasia:2006}, which in turn would result in the gas fraction being \emph{overestimated} \citep{Eckert:2019}. A high level of non-thermal pressure would thus render the lack of baryons in group-scale halos even more severe.

We caution here that the measurement of the gas fraction of group-scale halos is a difficult one and is hampered by numerous systematic uncertainties. While halo mass estimates definitely represent the leading source of systematics, several other sources introduce potential systematic errors. In the temperature range of galaxy groups, line cooling renders the X-ray emissivity highly dependent on gas metallicity, which is difficult to measure away from group cores (see the review by \citet{Gastaldello:2021} within this issue). This can introduce uncertainties as large as 20\% in the recovered gas mass \citep{Lovisari:2015}. Sample selection, usually based on \ROSAT~all-sky survey data, may bias the selected samples towards gas-rich systems, especially if the scatter at fixed mass is substantial \citep{Eckertetal11,Andreon:2017}. Finally, most studies do not detect the X-ray emission all the way out to $R_{500}$ \citep[see e.g. Fig. 8 of ][]{Sun:2009a} and must rely on extrapolation. For all these reasons, the question of what is the exact baryon fraction of galaxy groups within $R_{500}$ is still very much an open one, let alone within the virial radius.

\subsection{\textbf{Radio observations}} \label{s:radio}
\subsubsection{Interaction between radio sources and the IGrM}

Radio surveys have made clear that the centers of galaxy groups and clusters are special locations for AGN \citep[e.g.,][]{Best:2007,LinMohr07,Smolcicetal11}, with group-central galaxies twice as likely to host radio-mode activity than non-central galaxies of equal mass out to $z>1$. Deeper observations show that almost all central galaxies of X-ray luminous groups host some radio emission \citep{Dunnetal10,Kolokythasetal19}, though in the local universe some of these may be contaminated by emission from low-level star formation \citep{Kolokythasetal18}. Observations of nearby groups show a wide range of radio morphologies \citep[e.g.,][]{Giacintuccietal11}, with jet-mode feedback dominated by FR-I radio galaxies, as in clusters. Roughly one third of X-ray luminous groups appear to host currently or recently active jet sources in their central galaxies \citep{OSullivan:2017} with typical jet powers in the range 10$^{41}$-10$^{44}$~erg~s$^{-1}$ \citep{Kolokythasetal19}.

\begin{figure}[t]
\centering
\subfigure{
\includegraphics[width=0.72\textwidth]{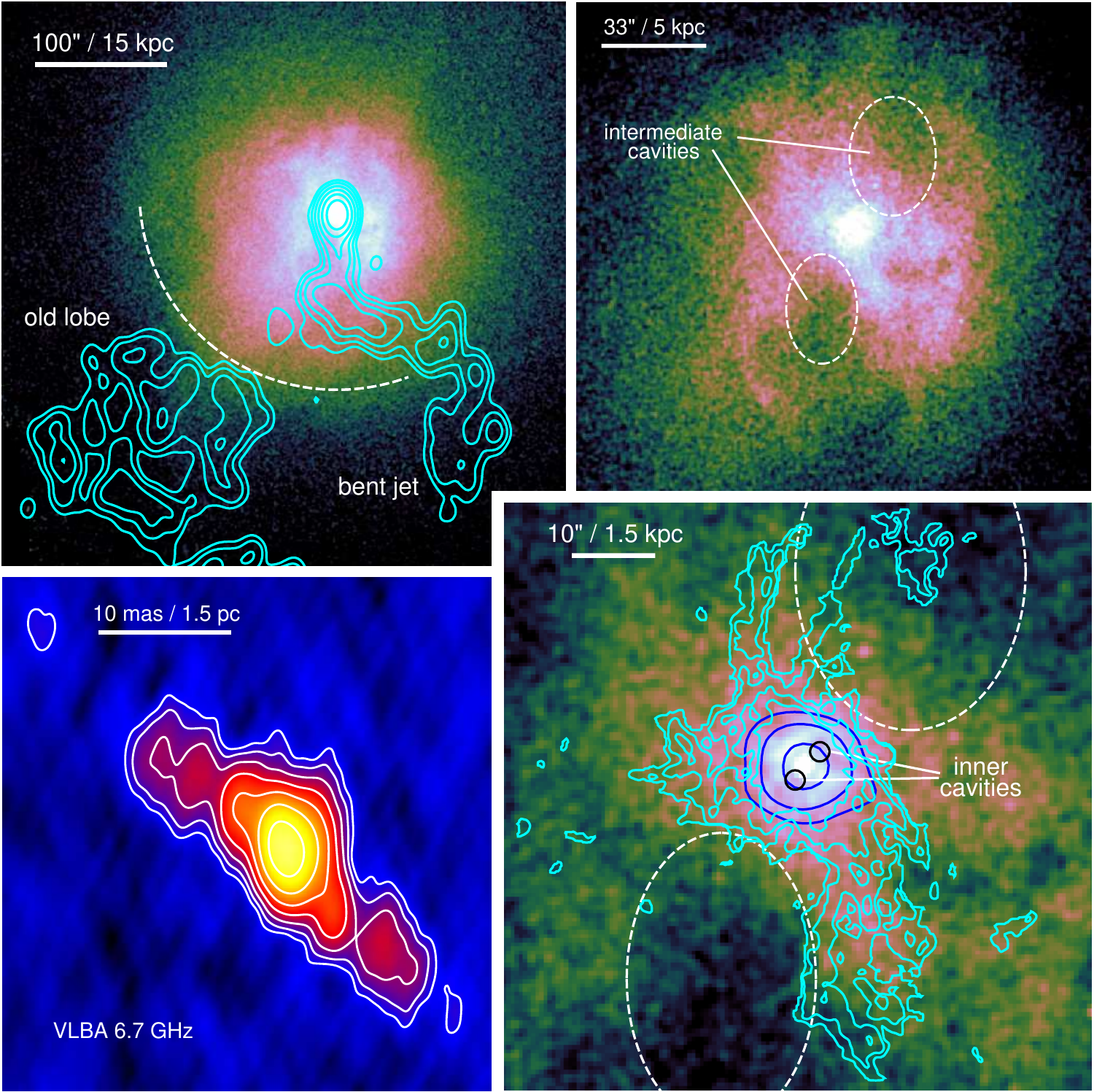}
}
\caption{\label{fig:N5044}Multiple cycles of AGN feedback in the NGC~5044 galaxy group. The \textit{upper left} panel shows the 0.5-2~keV \Chandra\ image with GMRT 235~MHz contours overlaid. These reveal an old, bent radio jet and detached lobe from a prior AGN outburst, whose structure has been affected by the sloshing front marked with a dashed line. The \textit{upper right} panel zooms in to show more detail of the complex of cavities and cool filaments in the group core. The \textit{lower right} panel zooms in further, with contours showing the MUSE H$\alpha$ (cyan) and ACA diffuse CO (blue) in the densest parts of the X-ray filaments and core. The \textit{lower left} panel shows parsec-scale VLBA 6.7~GHz radio emission in the nucleus of the galaxy, evidence of a new cycle of AGN jet activity \citep[adapted from][]{Schellenbergeretal21}.
}
\end{figure}

While cavities and shocks are the most accurate indicators of the impact of AGN feedback on the IGrM (see \S\ref{ss:X-ray}), current X-ray instruments have a limited ability to detect these features outside the high surface brightness cores of nearby groups. Radio studies offer an observationally cheaper way to measure feedback, particularly at higher redshifts. Radio galaxies are only periodically active, and once their AGN ceases to power them, their emission fades fastest at high frequencies. Low-frequency observations can therefore be particularly effective at identifying older, dying radio sources, and measuring their full extent and luminosity \citep[e.g.,][]{Kolokythasetal20,Birzanetal20}. The radio spectrum can also give an indication of the properties of the source, most notably its age and the lobe pressure, which for older sources is usually in equilibrium with the surrounding IGrM. Combining radio and X-ray observations, we can observe multiple cycles of outbursts in individual groups, e.g., NGC~5813 and NGC~5044 \citep[Figures~\ref{fig:N5813}, \ref{fig:N5044},][]{Randall:2015,Schellenbergeretal21}. In particular, in Fig. \ref{fig:N5044} we show the existing high-quality radio, X-ray and H$\alpha$ observations of NGC~5044 \citep{Schellenbergeretal21}. GMRT 235~MHz radio observations trace the oldest outburst via detached lobes and a bent, one-sided radio jet, while \Chandra\ detects cavities on $\sim$5~kpc and $\sim$150~pc scales. Interestingly, the current radio jets, traced by high-resolution VLBA observations (bottom-left panel), are not aligned with the X-ray cavities, possibly indicating precession of the jet axis with time.

The properties of group-central radio galaxies are closely linked to the IGrM. Both groups and clusters show a correlation between X-ray luminosity and the radio luminosity of the central source \citep{Dunnetal10,Inesonetal15,Pasinietal20}. In clusters, central radio source luminosity is observed to be higher in systems with cooling times $<$10$^9$~yr \citep{Birzanetal12}, and in groups it appears that radio jets are more common in the central galaxies of groups with short central cooling times, low t$_{\rm cool}$/t$_{\rm ff}$ ratios, and declining central temperature profiles \citep{OSullivan:2017}. But perhaps the most important correlation is that between jet power, as determined from the enthalpy of AGN-inflated cavities, and radio luminosity. This P$_{\rm cav}$-L$_{\rm radio}$ relation was first established for galaxy clusters by \citet{Birzan:2004,Birzanetal08} and later extended to early-type galaxies \citep{Cavagnoloetal10} and galaxy groups \citep{OSullivanetal11a}. Although there is significant scatter in the relation, it offers a mechanism for determining the energy available from AGN feedback in the many systems where direct determination in the X-ray is impossible.

Applying the P$_{\rm cav}$-L$_{\rm radio}$ relation to a large sample of SDSS groups and clusters with radio sources identified from the NVSS and FIRST surveys, \citet{Best:2007} showed that central radio galaxies dominate the heating of the IGrM within the cooling radius. They also found that the efficiency of AGN heating cannot be constant across the full mass range of groups and clusters; feedback must be less efficient in groups if they are not to be over-heated. Support for this result came from a study of groups observed in the COSMOS survey \citep{Giodinietal10} which found that, factoring in the likely duty cycle of the AGN population, group-central radio galaxies can inject energies comparable to the binding energy of the IGrM. These results suggest that group-central AGN have the potential to drive gas out of the group core, and perhaps out of the group altogether, unless some mechanism reduces their effectiveness in heating the gas. 

\subsubsection{Giant radio galaxies} 
Because feedback studies at the group scale are largely limited to the X-ray bright cores of nearby groups where cavities are most easily identified, they have tended to focus on relatively small radio sources, with jet sizes less than a few tens of kiloparsecs. However, the population of group-central radio galaxies includes much larger objects, some of which extend to very large radii, well beyond the cool core and even into the outskirts of their groups. \citet{Pasinietal20} show that radio galaxies larger than 200~kpc are more common in groups than clusters, and that the largest radio galaxies are located in groups, probably because the IGrM is less able to confine their growth than the ICM. There is also evidence from new radio surveys, with greater sensitivity to extended diffuse emission, that giant radio galaxies may be more common in general than previously believed \citep{Delhaizeetal21}.

Large radio sources pose a problem for AGN feedback models, in that they may inject a large fraction of their energy into the IGrM at large radii, rather than in the core, where it is needed to balance cooling. Even some medium sized sources appear to have jets which tunnel out of the cool core and inflate cavities outside it \citep[e.g., NGC~4261][]{OSullivanetal11b}. An extreme example in which at least one cavity is confirmed is IC~4296, which hosts an FR-I radio galaxy whose 160~kpc diameter lobes extend out to a projected radius of $\sim$230~kpc \citep{Grossovaetal19}. This is far beyond the cool core (20-30~kpc radius) and about half of $R_{500}$ for this $\sim$1~keV group. In such a system, while some of the energy involved in lobe inflation will likely have heated the core, the energy bound up in the relativistic particles and magnetic field of the radio lobes will likely be released at large radii, heating gas which is unlikely to contribute to fuelling the AGN. Other nearby examples of group-central giant FR-Is include NGC~315 and NGC~383 \citep{Giacintuccietal11} and NGC~6251 \citep{Cantwelletal20}. As in clusters, group-central FR-II galaxies are uncommon but not unknown \citep[see, e.g.,][]{Inesonetal15,Inesonetal17}. Their faster, more collimated jets likely provide feedback heating via shocks during expansion \citep[as in, e.g., 3C~88,][]{Liuetal19}, and lobe inflation will drive turbulence, but as with the giant FR-Is, it is less clear how they affect the cooling region once they grow beyond it. 

Giant radio galaxies therefore pose a number of important questions for feedback models of groups. Do they provide feedback that can balance the rapid cooling in group cores, and if so how? The large sizes of these systems, particularly the FR-Is, implies that their jets have been active for very long periods. How do these sources stay active for so long? 

\subsection{\textbf{Multiwavelength observations}} \label{s:multiw}

In the cool cores of galaxy clusters, many observations have shown evidence of material cooling from the hot atmosphere, in the form of highly multi-phase filamentary nebulae surrounding the central galaxy and containing gas and dust with temperatures ranging from $\sim$10$^6$~K to a few $\times$10~K. Some cool core galaxy groups show similar structures, though they are generally less luminous and are thus far less thoroughly explored. As yet, few studies have focused specifically on BGGs, but samples of giant ellipticals provide a window on the group regime.

H$\alpha$ emission from ionized gas with temperatures $\sim$10$^4$~K may be the most accessible tracer of cooled material. \citet{Lakhchaura:2018} find that, in giant ellipticals as in galaxy clusters \citep[c.f.][]{Pulidoetal18}, the presence of H$\alpha$ emission is associated with high IGrM densities, short cooling times, low values of the thermal instability criterion, t$_{cool}$/t$_{\rm ff}$ (see \S~\ref{s:ratios}) and disturbed X-ray morphologies, with the overlap between galaxies with and without detected H$\alpha$ suggesting that the transition between the two states can happen fairly easily. They also report a weak correlation between the mass of H$\alpha$-emitting gas and P$_{\rm cav}$, as expected if the H$\alpha$ traces cooling material, some of which will eventually fuel the AGN. Some of the best known X-ray bright groups contain examples of H$\alpha$ filaments similar to those seen in clusters \citep[e.g.,][]{Werneretal14,Randall:2015,Davidetal17,OSullivanetal19}. As in clusters, the filaments are closely correlated with feedback-related structures, showing signs of having been drawn out behind, or wrapping around, radio lobes and cavities. In some cases they are located in cool X-ray filaments that show signs of being thermally unstable \citep{Davidetal17}. Figure~\ref{fig:N5044} shows an example of this in the NGC~5044 group, where the H$\alpha$ nebula is correlated with the brightest cool X-ray emission and appears to wrap around the base of the intermediate-scale cavities. Spatially resolved spectroscopy shows that while the inner parts of these H$\alpha$ nebulae are generally cospatial with the stellar bodies of the BGGs, they do not rotate with the stars, supporting formation from the IGrM rather than stellar mass loss \citep[e.g.,][]{Gomesetal16,Dinizetal17,Schellenbergeretal20}. 
It should be noted that while BGGs do host some star formation (SF), their H$\alpha$ nebulae are not tracing SF. \citet{McDonald:2018} studied the relation between star formation rate inferred from infrared data and the X-ray cooling luminosity (\S\ref{ss:X-ray}) and found that the inferred star formation rates in BGGs are typically quenched by a factor 10-100 compared to the pure cooling scenario.

Molecular gas in groups has been observed via multiple tracers. \textit{Herschel} observations revealed [C\textsc{ii}] emission from $\sim$100~K gas with a similar distribution to the H$\alpha$, and [C\textsc{ii}]/H$\alpha$ flux ratios indicating that both phases are powered by the same source \citep{Werneretal14}. \textit{Spitzer} IRS spectra show rotational H$_2$ lines in the BGGs of some X-ray bright groups \citep{Kanedaetal08}, tracing gas at a few $\times$100~K, and CN has been detected in absorption in a handful of cases via the millimeter-wave band \citep{Rose:2019}. The forthcoming \textit{James Webb Space Telescope} will open an important observation window on H$_2$, which is likely the dominant mass component of the molecular phase. However, at present emission from CO is our best tracer for this phase, allowing us to examine the coolest, densest gas in the cooling regions of groups.

\citet{Babyketal19} examine CO in a large sample of local ellipticals (many of which are BGGs) and find that the molecular gas mass M$_{mol}$ is correlated with the density of the IGrM and its mass in the central 10~kpc, and that systems with t$_{cool}<1$~Gyr at 10~kpc are more likely to contain molecular gas. They also find that M$_{mol}$ is proportional to P$_{\rm cav}$, confirming that the molecular gas is the fuel source for the central AGN. However, cooling from a surrounding hot halo is not the only source of gas for ellipticals. \citet{Davisetal19} use a combination of the ATLAS3D and MASSIVE samples to show that gas-rich mergers are an important source of molecular gas in these galaxies. Observations of smaller samples of BGGs find some of the same trends, and show that BGGs of X-ray bright, cool core groups are not the CO-richest systems \citep{OSullivanetal15,OSullivanetal18}. BGGs of X-ray fainter groups can contain more CO (and H\textsc{i}) and it is more often located in disks rather than filaments. This again emphasizes the importance of gas-rich mergers in groups, though IGrM cooling is likely still the more important process in the cool core groups in which AGN feedback is most often observed.

The BGGs of X-ray bright cool core groups generally seem to contain only a few $\times$10$^6$ or $\times$10$^7$~M$_\odot$ of molecular gas \citep{OSullivanetal18}, making them challenging targets even for ALMA. However, being nearer than typical cool core clusters, groups offer an opportunity to study individual molecular cloud associations within the cool core, rather than the overall filamentary structures. Three well-known systems have been studied in detail by ALMA, NGC~4636, NGC~5846 \citep{Temietal18} and NGC~5044 \citep{Davidetal14,Davidetal17}. The velocity dispersions of the molecular clumps observed in these systems suggest that they are not gravitationally bound, and are likely collections of smaller, denser clouds, with more diffuse gas between them. Atacama Compact Array (ACA) observations of NGC~5044 show that a significant fraction of the molecular gas in the BGG is more diffuse than the clumps observed by ALMA \citep{Schellenbergeretal20}, and a similar argument can be made for the other two groups by comparing the CO masses derived from ALMA and IRAM~30m observations. The denser CO clumps are generally located within filamentary structures visible at other wavelengths \citep{Temietal18}, and the extent and velocity distribution of the diffuse CO in NGC~5044 is similar to that of the H$\alpha$ and [C\textsc{ii}] emission, supporting the idea that all the observed phases are material cooling from the IGrM. Figure~\ref{fig:N5044} shows the diffuse CO emission in the group core, collocated with the peak of the H$\alpha$ and X-ray emission. As with H$\alpha$, the CO in these ALMA-observed systems is cospatial with the stellar component, but shows little sign of rotation or velocity gradients, consistent with formation from the IGrM.

Intriguingly, ALMA studies of giant radio galaxies, some of them group-central systems, show a different CO morphology, with the molecular gas located in compact disks \citep{Ruffaetal20,Boizelleetal20}.
The difference in cold gas morphology may indicate a difference in the fuelling of the AGN. There are examples of group-dominant giant radio galaxies which appear to be fed by cold gas \citep[e.g., NGC~1167,][]{Shulevskietal12} and given the importance of galaxy interactions in groups, the potential for fuelling by gas rich mergers cannot be ignored. With only a handful of group-dominant galaxies mapped thus far in molecular gas, there is a significant opportunity for exploration of the mechanics of AGN fuelling in these important systems.

\section{Theoretical framework} \label{s:theory}

\begin{figure}[!t]
	\centering
	\subfigure{
		\includegraphics[width=0.82\textwidth]{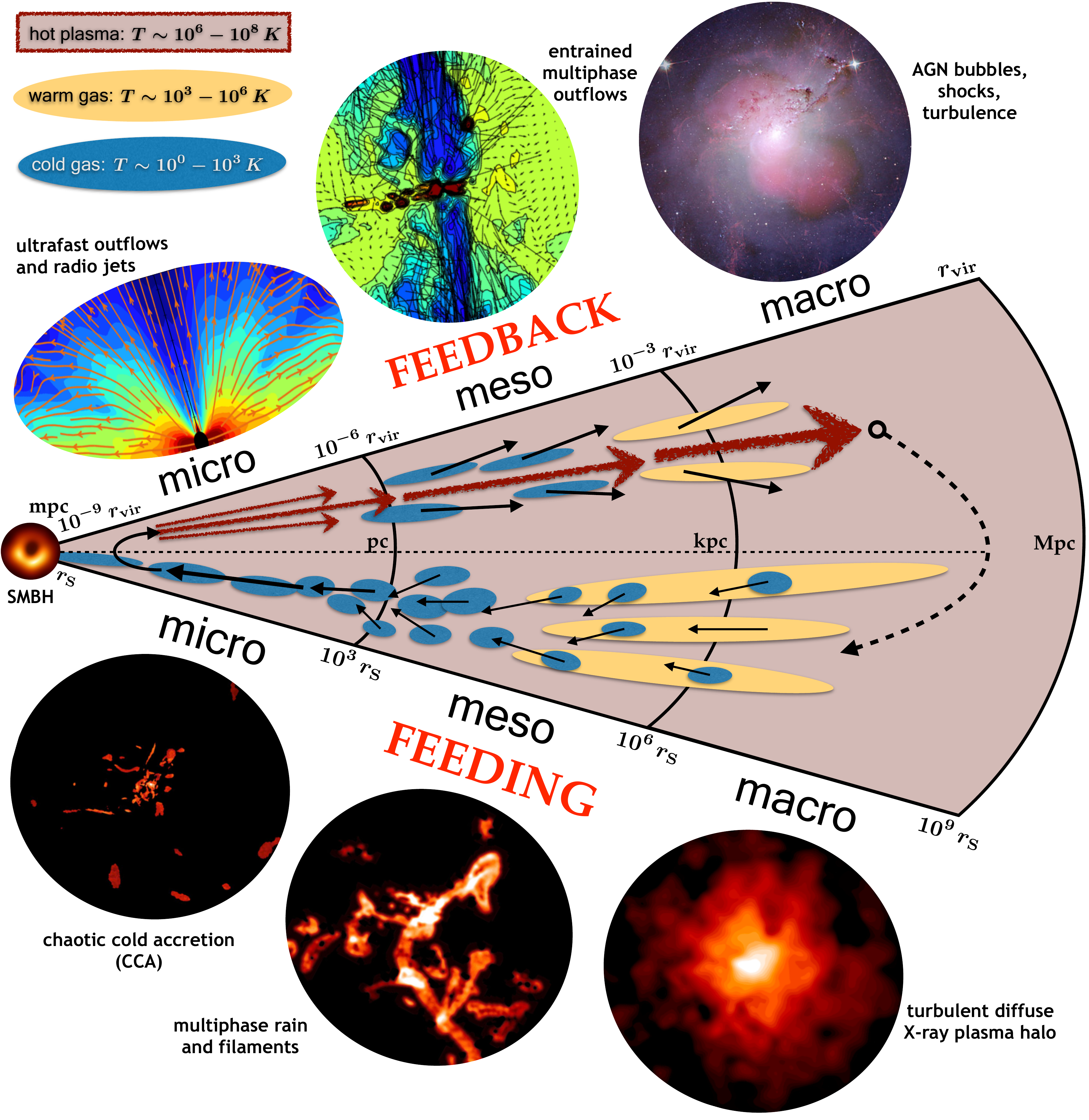}}
	\caption{Schematic representation of the self-regulation loop necessary to fully link feeding and feedback processes over nine orders of magnitude in space (and time) and over the multiphase/multiband cascade, including X-rays (hot halo/outflows), the IR/optical (warm filaments) and radio/mm (molecular clouds) bands -- reproduced from \citealt{Gaspari:2020} (arXiv authors' version). In particular, the IGrM experiences relatively stronger top-down condensation rain compared with clusters, due to the lower central cooling times. Tightly related to such an enhanced feeding is the higher frequency of the AGN outflow/jet feedback events, which gently self-regulate each galaxy group for several billion years during the cosmic evolution.}
	\label{f:loop}
\end{figure}

Hot halos are a fascinating and crucial element of virialized systems in the Universe, which have been unveiled to be a fundamental engine for the growth and triggering of SMBH, despite the large difference in spatial and temporal scales, which span over 9 orders of magnitude (commonly sub-divided in three major scales; see Fig.~\ref{f:loop}: micro -- meso -- macro). Here we review the AGN feedback process in terms of fundamental physics and why it is expected in the more theoretical framework of accretion out of the hot halo and onto SMBH, with a keen eye on galaxy groups. 
As shown in the summary diagram of Fig.~\ref{f:loop}, it is important to appreciate that AGN feedback is only half of the self-regulated cycle, which is bootstrapped via the AGN feeding, the key complementary mechanism on which we will also focus below.

While complex, non-linear thermo-hydrodynamical (THD) mechanisms are at play over the macro (kpc-Mpc), meso (pc-kpc) and micro scales (mpc-pc) -- thus requiring expensive numerical 3D Eulerian simulations -- it is useful here to understand the whole SMBH-halo system as a unified, co-evolving engine. In essence, such a global THD system can be described via the simple conservation of energy, or analogously via the (Lagrangian) entropy equation (\citealt{Peterson:2006,Gaspari:2015_xspec}):
\begin{equation}\label{e:entropy}
U\,\frac{\dd}{\dd t} \ln{K} = \mathcal{H} - \mathcal{L},
\end{equation}
where $K = k_{\rm b}T/n^{\gamma-1}$ is the astrophysical entropy (with $n=n_{\rm e} + n_{\rm i} \approx 2\,n_{\rm e}$ the sum of the electron and ion number densities), $U=P/(\gamma-1)$ is the internal/thermal gas energy per unit volume ($\gamma=5/3$ is the IGrM adiabatic index), $\mathcal{H}$ and $\mathcal{L}$ are the gas heating and cooling rates per unit volume ($\es\,{\rm cm}^{-3}$), respectively. A few immediate insights from Eq.~\ref{e:entropy}: the internal energy acts as an effective normalization knob (the larger the X-ray temperature times density, the stronger the required heating/AGN feedback, in absolute $\rm erg\, s^{-1}$ values); secondly, the macro entropy evolution is the sole result of the competition of heating and cooling processes, which translates in the competition between AGN feedback and feeding. Let us discuss first the (astro)physics and consequences of the cooling/feeding component of the cycle, i.e., $\mathcal{L}$.

\subsection{\textbf{AGN feeding \& cooling processes}} \label{s:cooling}

\begin{figure}[t]
	\centering
	\subfigure{
		\includegraphics[width=0.52\textwidth]{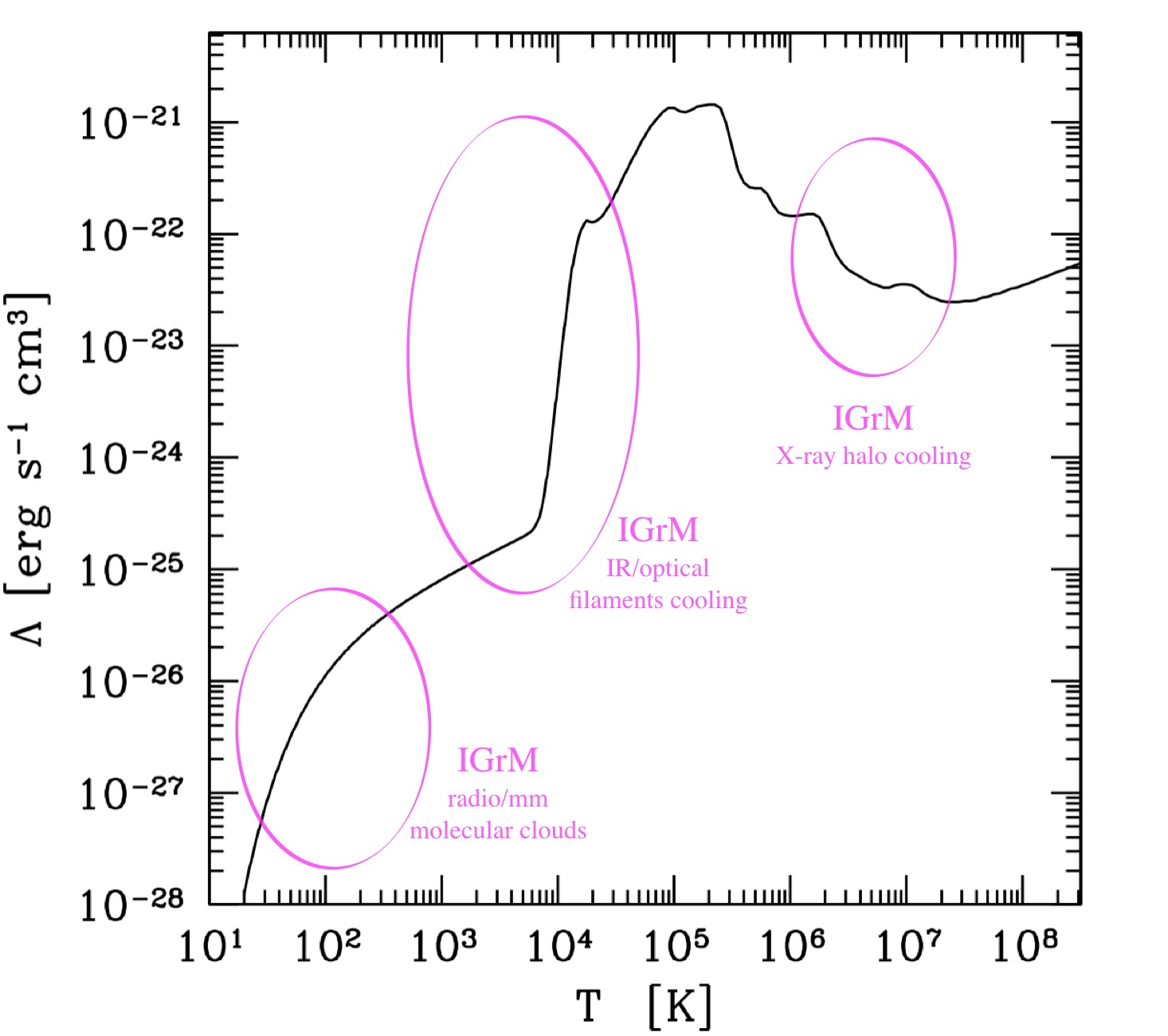}}
	\caption{Multiwavelength cooling function for the IGrM (adapted from \citealt{Gaspari:2017_cca}, unifying the atomic/plasma physics studies by \citealt{Sutherland:1993,Dalgarno:1972, Inoue:2008}; $Z=1\,Z_\odot$), which was specifically used for a typical galaxy group akin to NGC\,5044. 
		Above the neutral hydrogen recombination ($T\sim10^4$\,K) the gas is fully ionized and in collisional ionization equilibrium; below this threshold, the gas becomes progressively less ionized ($\lesssim 1\%$) leading to the formation of neutral filaments, and subsequently dense molecular clouds. The three magenta ellipses highlight the three key (semi)stable phases of the condensing IGrM, in particular during the feeding dominated part of the AGN cycle shown in Fig.~\ref{f:loop} (bottom insets).
	}
	\label{f:Lambda}
\end{figure}

The cooling process is very well understood from basic quantum physics and laboratory plasma/ionized gas experiments, with a radiative cooling loss \citep{Sutherland:1993} $\mathcal{L}=n_{\rm e} n_{\rm i}\, \Lambda(T,Z)$, where $\Lambda$ is a cooling function varying with gas temperature and metallicity $Z$ ($\sim0.6-1\,Z_\odot$ for group cores; \citealt{Mernier:2017}, cf.~the companion \citet{Gastaldello:2021} review). As shown in Fig.~\ref{f:Lambda}, the hot IGrM experiences a significantly enhanced $\Lambda$ due to the influence of line cooling (mostly recombination) taking over from the Bremsstrahlung/free-free emission ($\Lambda \propto T^{1/2}$), which instead shapes the more massive galaxy clusters ($T>2$\,keV).

Fig.~\ref{f:Lambda} depicts the three main (quasi)stable phases arising during the top-down condensation cascade \citep{Gaspari:2017_cca}, especially during the feeding dominated stage of the AGN cycle. 
Assuming relatively slow motions over the group macro gravitational potential (quasi pressure equilibrium), Eq.~\ref{e:entropy} can be approximated as (e.g., \citealt{Pope:2012})
\begin{equation}\label{e:dotM}
H - L \approx \frac{c^2_{\rm s}}{\gamma-1}\,\dot M_{net},
\end{equation}
where $c^2_{\rm s}=\gamma k_{\rm b}T/\mu m_{\rm p}$ is the (squared) IGrM sound speed (with $\mu\approx0.62$ the mean atomic weight of the IGrM) and $H$/$L$ is the gas heating/cooling power (or luminosity in $\es$). Notably, any net cooling or heating will induce a net mass inflow ($\dot M_{\rm cool}$) or outflow rate ($\dot M_{\rm OUT}$) in the macro-scale halo, denoted as $\dot M_{net}$; here uppercase subscripts denote macro properties, while lowercase subscripts denote micro properties. Therefore, the thermal evolution of the IGrM is deeply intertwined with feeding/feedback processes, as intuitively anticipated above.

In the absence of any heating, the entire hot atmosphere would rapidly condense and collapse, initiating from the inner denser radial regions 
(see \S\ref{ss:X-ray}). 
As noted in the previous sections, such massive cooling flows are not observed in our Universe, especially in galaxy groups.
On the other extreme of (idealized) feeding models, the IGrM halo might experience pure cooling while having significant angular momentum. In this regime, the gas would condense through helical paths onto the equatorial plane, and there forming a thin rotating multiphase disk \citep{Gaspari:2017_cca}.
While extended disks have been found in some BGGs (\citealt{Hamer:2016,Juranova:2020}, \citealt{Ruffaetal20}), such a scenario would induce both large (unobserved) cooling rates in X-ray spectra, as well as drastically reduced accretion rates onto the SMBH due to the preservation of high angular momentum and related centrifugal barrier. 
 
Realistic IGrM atmospheres instead often reside in an intermediate THD regime, neither strongly rotating nor in a spherical cooling flow (\citealt{Davidetal17}, \citealt{Lakhchaura:2018}, \citealt{OSullivan:2017}, \citealt{Temietal18} -- \S\ref{s:multiw}). Indeed, as shown by high-resolution HD/cosmological simulations (\citealt{Gaspari:2012b,Hillel:2017,Lau:2017,Weinberger:2018,Wittor:2020}) and X-ray spectroscopy (\citealt{Sanders:2013,Ogorzalek:2017,Hitomi:2018}) hot halos experience significant amount of turbulence, with an irreducible level of 3D turbulent velocity dispersion $\sigma_v \approx 100-300\,\kms$, due to both the previous AGN feedback outbursts and the secular cosmological flows (e.g., \citealt{Vazza:2011_turb,Valentini:2015}). While we review the kinematical features in a companion review \citep{Gastaldello:2021}, we focus here on its thermodynamical impact, namely the formation of \textit{chaotic cold accretion} (CCA) and related multiphase rain, a key process driving the bulk of AGN feeding and hence the recurrent AGN feedback triggering.
In a turbulent hot halo, chaotic multiscale eddies drive local perturbations in relative gas density proportionally to the turbulence sonic Mach number ($\delta n/n \propto \mathcal{M}_t$; \citealt{Gaspari:2013_coma,Zhuravleva:2014}). 
The relative increase in IGrM density produces in-situ enhanced radiative cooling ($\mathcal{L}\propto n^2$), thus leading to turbulent non-linear thermal instability (TI; \citealt{Gaspari:2013_cca}, \citealt{Voit:2018}). It is important to note that such a chaotic instability is different from classical TI (\citealt{Field:1965,Pizzolato:2005,McCourt:2012}), in the sense that direct non-linear fluctuations are seeded by chaotic motions, rather than growing from tiny linear amplitudes.
This triggers a quick \textit{top-down condensation} of localized (soft X-ray) patches to the first quasi-stable phase at $T\sim10^4$\,K, which is best traced via ionized line-emitting (e.g., H$\alpha$+[NII]; \citealt{Gastaldello:2009}, \citealt{McDonald:2011a}, \citealt{Werneretal14}) filaments or nebulae observed in optical/UV (e.g., see the synthetic image in the bottom middle inset of Fig.~\ref{f:loop}). Sustained turbulent perturbations lead to the further condensation cascade onto the last stable and compact gas phase, molecular gas clouds (bottom left inset of Fig.~\ref{f:loop}; see \S\ref{s:multiw}). Such cold clouds will then strongly and frequently collide inelastically within the meso/micro scale, cancelling angular momentum and thus feeding the central SMBH (hence the `CCA' nomenclature; \citealt{Gaspari:2017_cca}), with the consequent trigger of the next stage of AGN feedback (\S\ref{s:heating}). CCA feeding recurrently boosts the accretion rates over $100$ fold over the feeble and quiescent hot-mode (\citealt{Bondi:1952,Narayan:2011}) accretion, thereby overcoming the inefficiency of classical hot mode accretion.

\begin{figure}[t]
\centering
\includegraphics[width=0.45\textwidth]{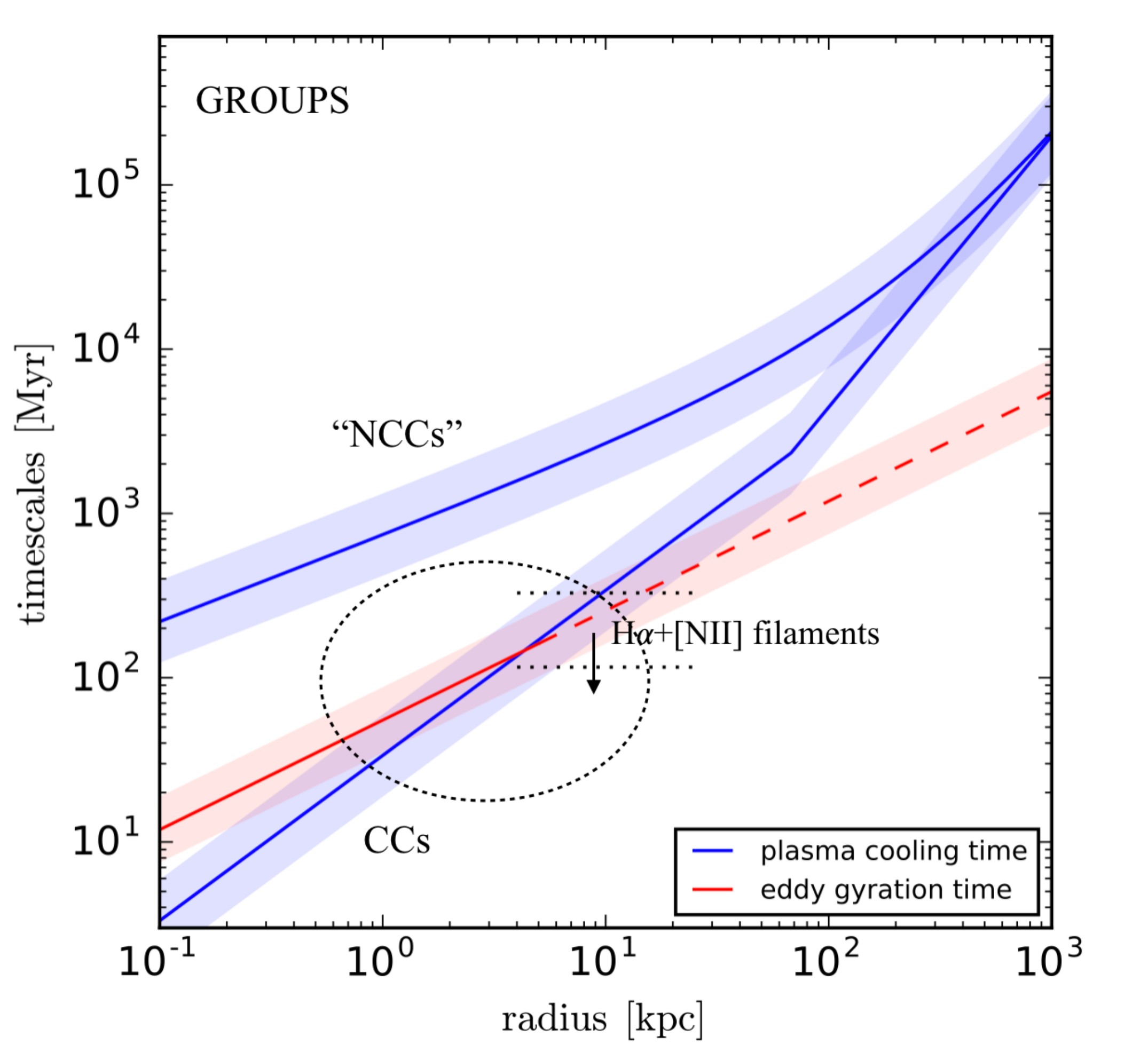}
\caption{Average cooling time and turbulence eddy time profiles in the IGrM (with 90\% confidence scatter bands), the latter constrained mainly via optical/IR telescopes (figure reproduced from \citealt{Gaspari:2018}; group sub-sample). The dotted circle marks the size of the condensed warm nebular emission, which matches the $C\equiv t_{\rm cool}/t_{\rm eddy}\sim 1$ turbulent TI threshold.
}
\label{fig:Cratio}
\end{figure}

On the macro scale, the hot halo can be assessed to reside or predicted to soon enter the CCA raining phase, whenever the ratio of the plasma cooling time and the turbulence eddy gyration/turnover time reaches unity. This reference dimensionless number is called $C$-ratio (from condensation or CCA; \citealt{Gaspari:2018,Olivares:2019}), 
\begin{equation}\label{e:C}
C \equiv \frac{t_{\rm cool}}{t_{\rm eddy}} \sim 1,
\end{equation}
where the cooling and turbulence timescales have been defined in Eq.~\ref{e:tcool} and \ref{e:teddy}.
A correlated ratio and thermal-instability threshold is the TI-ratio $\equiv t_{\rm cool}/t_{\rm ff} \lta 10-30$ (e.g., \citealt{Gaspari:2012a,Sharma:2012,Voit:2015_gE}), where the free-fall time is defined in Eq.~\ref{e:tff}.
As introduced in \S\ref{s:ratios}, all three IGrM timescales can be constrained from X-ray or optical/IR datasets.
While both the $C$-ratios and TI-ratios are valuable complementary tools, simulations show that the $C$-ratio is the more direct physical criterion to apply to probe the onset and extent of \textit{nonlinear} thermal instability (e.g., \citealt{Gaspari:2018}; see also Fig.~\ref{f:cond}). Indeed, unlike in classical linear TI, turbulence acts as an irreducible background of fluctuations over the whole IGrM (e.g., \citealt{Lau:2017}), . 
In particular, AGN bubbles are a key recurrent mechanism to induce such fluctuations in the IGrM (e.g., \citealt{McNamara:2016,Voit:2021}).
In this regard, it is not surprising that $t_{\rm cool}/t_{\rm ff}$ profiles show a large non-trivial deviation above unity, as well as a large intrinsic scatter (e.g., \citealt{Singh:2020}; see also Fig.~\ref{fig:ratios}).
Fig.~\ref{fig:Cratio} shows the average cooling and turbulence eddy time profiles (with scatter) for the galaxy group regime (\citealt{Gaspari:2018}). Evidently the crossing of $C\sim1$ well matches the dotted circle denoting the typical size of the condensed extended multiphase nebulae (e.g., \citealt{McDonald:2011a}).
%

Together with mild $C$-ratios, a CCA-driven atmosphere -- often found in the IGrM cores -- is described by a low turbulent Taylor number (\citealt{Gaspari:2015_cca,Juranova:2019}):
\begin{equation}\label{e:Tat}
{\rm Ta_t} \equiv \frac{v_{\rm rot}}{\sigma_v} \lta 1.
\end{equation}
Given that the dominant galaxies of hot gas rich galaxy groups tend to be early-type (as do those galaxies which host their own extended hot halos), they have a fairly weak coherent gas rotational velocity $v_{\rm rot}$ (e.g., \citealt{Caon:2000,Diehl:2007}), unlike lower-mass/spiral galaxies. Therefore, a median IGrM long-term evolution is to oscillate between stages of strong CCA rain (${\rm Ta_t}\sim0.3-1$, $C\sim0.5-1$) and mild rain superposed on a clumpy disk (${\rm Ta_t}\sim1-3$, $C\sim1-2$). Evidently, extremes of strong rotation (${\rm Ta_t \gg 1}$) or overheated quiescence ($C \gg 1$) can lead to periods of disk- or Bondi-driven accretion, both experiencing highly suppressed inflow rates and feedback (albeit such periods must be short-lived to avert the cooling flow catastrophe).
High-resolution HD simulations have shown that the chaotic behaviour of a CCA-driven halo is imprinted not only in the thermodynamical maps and kinematical properties, but also in the time-series spectra. Unlike quiescent and continuous hot modes (Bondi/ADAF), CCA drives a characteristic flicker/`pink' noise power spectrum (logarithmic slope of -1) in the Fourier space of frequencies $f$ (\citealt{Gaspari:2017_cca}), thus generating strong self-similar variability on all temporal scales (the integral over $\dd f$ of this spectrum yields constant variance), from several Myr down to years and minutes, as ubiquitously observed in multiwavelength AGN lightcurves (\citealt{Ulrich:1997,Peterson:2001_AGN_var}). 
This triggers the second key part of the self-regulated loop, AGN feedback, the focus of the next section.

\subsection{\textbf{AGN feedback \& heating processes}} \label{s:heating}
While pure cooling flows and catastrophic condensation are not detected in IGrM observations, strong overheating is equally ruled out, as virtually all galaxy groups exhibit central cooling times well below 1\,Gyr due to the high efficiency of radiative cooling in their characteristic temperature range (see Fig. \ref{f:Lambda}). We note that massive galaxy clusters instead show a dominant population of non-cool-core systems with central cooling times above the Hubble time \citep[e.g.][]{Hudson:2010,Ghirardini:2019}. 
As inferred from Eq.~\ref{e:entropy}, the system would be in perfect thermal equilibrium in the absence of any heating and cooling terms. However, analogous configuration can be achieved if a heating process (macro AGN feedback) balances the cooling rate (macro condensation). This configuration is the more realistic state of observed galaxy groups, with the characteristic feature that the self-regulation process is intrinsically \textit{chaotic} (from the macro down to micro scales; \S\ref{s:cooling}), hence leading only to a \textit{statistical} thermal balance $\mathcal{H} \sim \langle \mathcal{L} \rangle$.
Moreover, while Eq.~\ref{e:entropy} formally allows for the entropy to decrease (pure cooling), fundamental THD physics dictates that entropy shall always increase in real systems (even over the ensemble Universe). With such intuition, we can already expect that the heating rate is as essential -- if not eventually more vigorous -- than the cooling component (Fig.~\ref{f:loop}). 

\begin{figure}[t]
     \centering
     \subfigure{
     \includegraphics[width=0.55\textwidth]{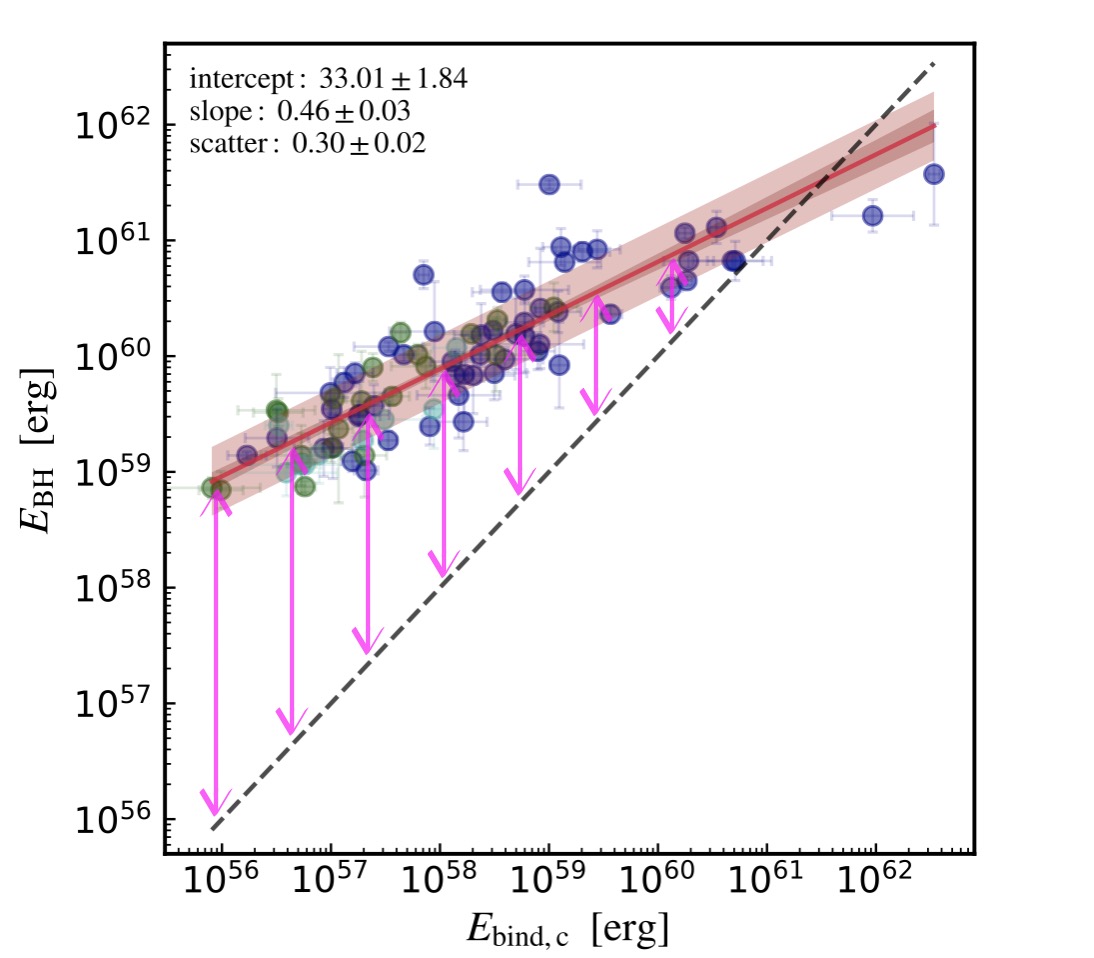}}
     \caption{Available mechanical feedback energy of the central SMBH versus gravitational binding energy of the hot gas within the core of the host halo ($R \lesssim 0.15\,\r500$). The SMBH energy is $E_{\rm BH} = 10^{-3}\,M_{\rm BH} c^2$, while
     the binding energy is related to the thermal energy via the virial theorem $E_{\rm bind} \approx 2\,E_{\rm th} \propto M_{\rm gas} \tx$.
     The 85 points are taken from \citet{Gaspari:2019}, which include the observed direct/dynamical SMBH mass with the X-ray halo detected in the host group or cluster. The solid red curve shows a fit to the relation with a power law, with the $16-84$ percentile interval indicated by the red shaded area. The 1-$\sigma$ intrinsic scatter is plotted as a light red band on top of the mean fit.
     The circle colors reflect the morphological type of the central galaxy: elliptical (blue), lenticular (green), spiral (cyan). 
     The black dashed line demarks the one-to-one energy equivalence, whereas the magenta arrows highlight the excess BH energy compared to the binding energy.
     }
     \label{f:Ebh_Ebind}
\end{figure}

Before tackling the AGN feedback physical sub-processes, we first discuss here the key difference between groups and clusters. 
In Fig.~\ref{f:Ebh_Ebind}, we show an analysis of the potential impact of the stored SMBH energy versus the gravitational binding energy of the hot halo cores (from small groups to clusters), by leveraging the large sample of \citet{Gaspari:2019} with both direct BH masses and extended hot halos detected.
The potentially available SMBH mechanical energy is $E_{\rm BH} = \varepsilon_{\rm M} M_{\rm BH} c^2$, where $\varepsilon_{\rm M}$ is the macro mechanical efficiency (\citealt{Gaspari:2017_uni}). 
We test for the now commonly used fixed $\varepsilon_{\rm M} \sim 10^{-3}$ (we will explore variations to this basic modeling further below).
The gravitational binding energy is tightly related to the thermal energy via the virial theorem, $E_{\rm bind} \approx 2\,E_{\rm th} \propto M_{\rm gas} \tx$. Here we consider the integration over a large scale, $R<0.15\,R_{500}$.
Evidently, the linear regression fit (including the intrinsic scatter band) is significantly shallower than the dashed line of the one-to-one balance. In particular, the mechanical feedback energy that a SMBH can release could potentially overcome the core binding energy if released in a very short period of time -- for instance, assuming a quasar-like/Sedov blast scenario.
This would drastically overheat and evacuate the gaseous core atmosphere, becoming more serious toward the galaxy group regime and lower mass halos ($E_{\rm bind,c}<10^{59}$\,erg), where the feedback energy might even evacuate the entire gas virial region (\citealt{Puchwein:2008,Gaspari:2014_scalings}, \S\ref{s:fgas_sim}). 
As observations almost ubiquitously detect hot atmospheres, this indicates that the AGN feedback in groups shall be well self-regulated and relatively gentler than in massive galaxy clusters, which can sustain much stronger and impulsive AGN feedback deviations over the cosmic evolution.

Reaching a gentle self-regulation, while avoiding strong overheating, implies two major features of AGN feedback in galaxy groups and related improvements compared with the above modeling. First, the conversion efficiency of accreted rest mass energy into feedback energy is expected to decline with lower halo mass: equating the macro AGN power to the gas X-ray radiative cooling rate requires to modify the above with $\varepsilon_{\rm M} \sim 10^{-3} (\tx/2\,\rm keV)$ (\citealt{Gaspari:2017_uni}; see also Eq.~\ref{e:balance}). This can be explained by the weaker macro-scale coupling of the AGN jets/outflows with the hot halo, as the IGrM atmospheres are more diffuse than the dense ICM counterparts. Second, in order to avoid the above evacuation outburst, such self-regulated AGN feedback has to be not only gentler but significantly more frequent, i.e.~with larger duty cycle (ratio of on/off activity). This is also naturally explained by the relatively lower cooling times (tens of Myr) in the inner IGrM regions (compared with the ICM counterparts), due to the lower $T_{\rm x}$ and the substantially enhanced $\Lambda$ via line emission (Fig.~\ref{f:Lambda}). Both of such key features have been extensively tested and retrieved by high-resolution HD simulations (\citealt{Gaspari:2011b}, \citeyear{Gaspari:2012b}, \citealt{Sharma:2012,Prasad:2015}), and found in observations (e.g., \citealt{Best:2007}, \citealt{OSullivan:2017}).

While we have discussed the key characteristics and requirements of AGN heating, the next major question is: how is the AGN feedback energy propagated and dissipated within the IGrM?
As shown in Fig.~\ref{f:loop} (top insets) the problem is challenging, as it entails a wide range of scales and phases, from the milliparsec up to at least the 100 kpc region.
General-relativistic, radiative-magnetohydrodynamical simulations (GR-rMHD; \citealt{Sadowski:2017}) resolving radial distance of $\sim500\,r_{\rm S}$ (Schwarzschild radii)  show that the AGN triggered via CCA is able to transform the inner gravitational energy into wide ultrafast outflows (UFOs) with velocities $\sim0.1 c$ (top-left inset in Fig.~\ref{f:loop}; \citealt{Fukumura:2010,Tombesi:2013}). Under strong magnetic field tower and spin conditions (\citealt{Tchekhovskoy:2011}), the AGN is also able to generate a very collimated relativistic (radio-emitting) jet, perpendicularly to the thick accretion torus. The above GR-rMHD simulations show that kinetic feedback appears to be present over both low and high Eddington ratios ($\dot M_{\rm BH}/\dot M_{\rm Edd} \equiv \dot M_{\rm BH} /[23\,\msun\,\rm yr^{-1}(M_{\rm BH}/10^9\msun)]$), with a retrieved \textit{micro} mechanical efficiency $\varepsilon_{\rm m} \simeq 0.03\pm 0.01$. At variance, the radiative efficiency declines dramatically below $\varepsilon_{\rm r} \ll 0.01$ at $\dot M_{\rm BH}/\dot M_{\rm Edd} < 1\%$, which is the typical regime of local AGN in massive galaxies (\citealt{Russell:2013}).
Further, in order to achieve an efficacious macro self-regulation, the AGN feedback has to satisfy energy conservation (\citealt{Costa:2014}) and related micro- to macro-scale power transfer (\citealt{Gaspari:2017_uni}), such as
\begin{equation}\label{e:balance}
(P_{\rm out} \equiv \varepsilon_{\rm m} \dot M_{\rm BH} c^2)\,=\,(P_{\rm OUT} \equiv \varepsilon_{\rm M} \dot M_{\rm cool} c^2)\,\sim\,L_{cool},     
\end{equation}
with $L_{cool}$ the cooling luminosity (see \S\ref{ss:X-ray}).
The discrepancy between the above \textit{macro} and the larger \textit{micro} efficiency is  crucial: it implies that most of the accreted matter ($\dot M_{\rm out}/\dot M_{\rm cool} =(1-\varepsilon_{\rm M}/\varepsilon_{\rm m}) >90\%$) is re-ejected back by the SMBH, as discussed above, driven mostly in the kinetic form of UFOs and relativistic jets. Such AGN outflows/jets propagate and percolate deeper into the meso-scale atmosphere and start to entrain progressively more IGrM, loading part of the surrounding gas mass and decreasing their velocity down to several 1000\,$\kms$ (e.g., \citealt{Giovannini:2004,Fiore:2017}).

\begin{figure}[t]
     \centering
     \subfigure{
     \includegraphics[width=0.9\textwidth]{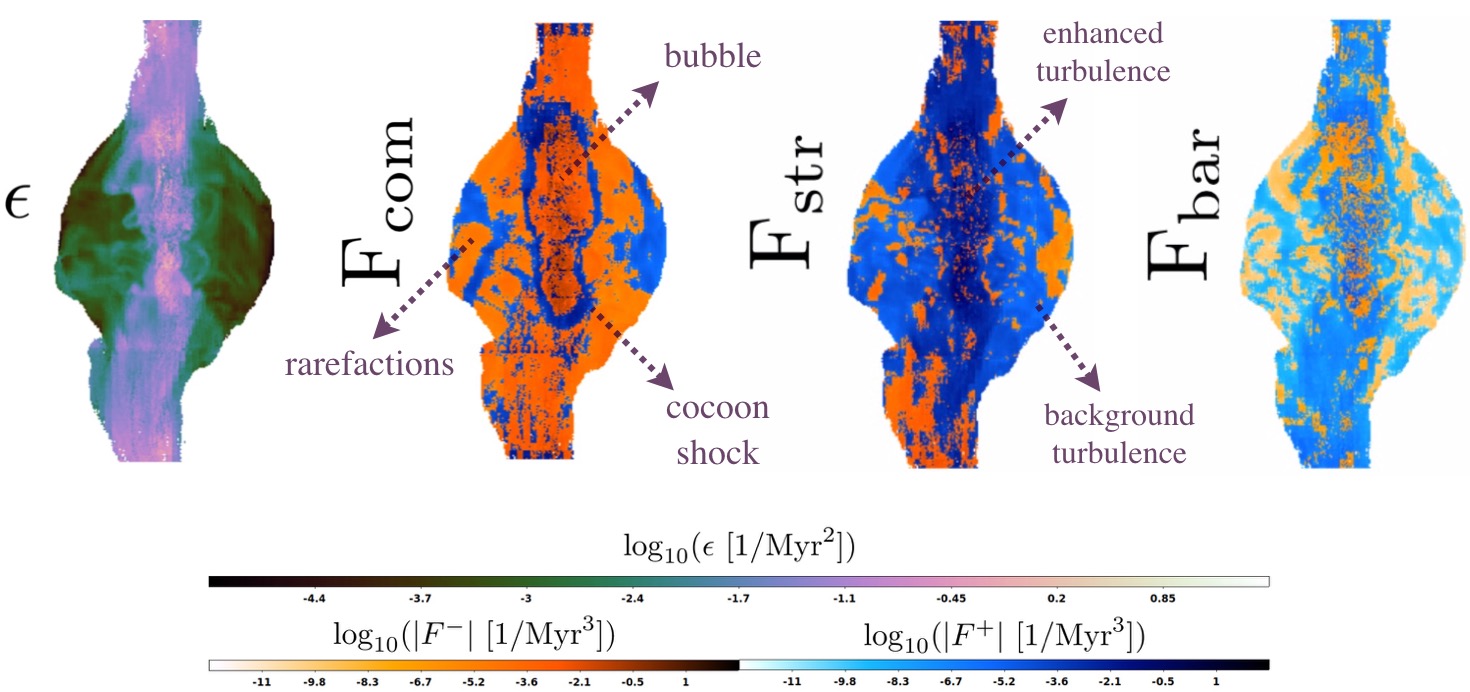}}
     \caption{Macro AGN feedback transfer and deposition highlighted via the enstrophy $\epsilon$ decomposition (Eq.~\ref{e:ens}) into its main positive/negative components: compressions/rarefactions (second panel), stretching/squeezing motions (third), baroclinicity (fourth) -- adapted from \citet{Wittor:2020}. This is achieved via Lagrangian tracer particles on top of an adaptive-mesh-refinement HD simulation of 
     self-regulated AGN outflows/jets in a central massive galaxy (\citealt{Gaspari:2012a}). The cycle of CCA rain, AGN outflow injection, bubble inflation, cocoon shock expansion, and turbulence cascade repeats self-similarly over several billion years, recurrently quenching the macro cooling flow.
     }
     \label{f:ens}
\end{figure}

The last missing tile of the self-regulated cycle is the macro AGN feedback deposition -- a strongly debated topic since the launch of \Chandra\ and \XMM\ telescopes, as their angular resolution is mostly limited to the macro scale (Fig.~\ref{f:loop}, top-right inset). 
While numerous physical mechanisms have been proposed to compensate macro cooling flows (e.g., \citealt{McNamara:2012}), 
we focus here on the physics of the three major mechanisms that have been firmly established to be present in the majority of hot halos, particularly the IGrM (see the  observational evidences in \S\ref{s:obs}), namely: buoyant bubbles, shocks, and turbulence.
While previous reviews tried to assess what is the dominant or sole driver of the AGN feedback, we show that the macro AGN feedback deposition is a strong nonlinear composition of at least three key processes. 
We can dissect such non-linearity and sub-processes via the local \textit{enstrophy} analysis, which we define as the squared magnitude of the flow vorticity $\epsilon = \frac{1}{2} |\boldsymbol{\omega}|^2 \equiv \frac{1}{2} |\boldsymbol{\nabla} \times \mathbf{v}|^2$. Neglecting the small dissipation term, the Lagrangian (tracer particle) framework leads to the following enstrophy evolution decomposition (\citealt{Wittor:2020}):
\begin{align} \label{e:ens}
 \frac{\dd \epsilon} {\dd t} = 
 \underbrace{-2\epsilon (\boldsymbol{\nabla}  \cdot \mathbf{v})}_{F_{\rm com}} +
 \underbrace{2\epsilon \left(\frac{\boldsymbol{\omega}}{|\boldsymbol{\omega}|} \cdot \boldsymbol{\nabla} \right) \mathbf{v} \cdot \frac{\boldsymbol{\omega}}{|\boldsymbol{\omega}|}}_{F_{\rm str}} +
 \underbrace{\frac{\boldsymbol{\omega}}{\rho^2} \cdot ( \boldsymbol{\nabla}  \rho \times \boldsymbol{\nabla}  P)}_{F_{\rm bar}},
\end{align}
where the three right-hand-side (positive/negative) terms are compressions/rarefactions, stretching/squeezing motions, and baroclinicity, respectively.

The more we want to zoom into the detailed processes of the AGN heating/weather, the more we need to rely on nonlinear HD simulations. Fig.~\ref{f:ens} shows a typical AGN-heating dominated period taken from a self-regulated AGN feedback simulation of meso AGN outflows/jets that consistently balance the macro cooling flow down to $1-10\%\,\dot M_{\rm cool, pure}$ (\citealt{Gaspari:2012a}). The 30 million tracer particles injected on top of the Eulerian grid can dissect the major components of the macro AGN feedback. 
First, the progressively slower and entrained outflow/jet inflates a pair of underdense cavities/bubbles, as its ram pressure is balanced by the surrounding IGrM thermal pressure (e.g., \citealt{Brighenti:2015}). 
Such bubbles are often -- albeit not universally -- traced by radio synchrotron emission spilling from the micro/meso jets (see \S\ref{s:radio}). Within their ellipsoidal volumes $V_{\rm b}$ (\citealt{Shin:2016}), they contain a substantial amount of enthalpy $E_{\rm cav} \simeq 4 P_{\rm b} V_{\rm b}$ in the purely relativistic case (see \S\ref{ss:X-ray}); dividing by the buoyancy time (\citealt{Churazov:2001}), the related cavity power/heating rate is $H_{\rm cav} = E_{\rm cav}/t_{\rm buoy}$.
Second, the bubbles are often encased within a cocoon shock, which is the result of the strong compressional motions of the expanding outflow and bubbles (see the thick blue contours in the second panel of Fig.~\ref{f:ens}). At this stage, the shock Mach number has become already weakly transonic, $\mathcal{M} \sim 1-2$ (see Fig. \ref{fig:shock}); as the AGN outflow recurrently ignites, they generate a series of weak shock ripples in the IGrM (\citealt{Randall:2015}, \citealt{Liuetal19}) that heat the gas non-adiabatically via cumulative entropy jumps, with heating rate $H_{\rm shock}=(e_{\rm th}\,\Delta\ln K)/t_{\rm age}$ (where $e_{\rm th}$ is the specific thermal energy and $t_{\rm age}^{-1}$ is the frequency of shocks).
Fig.~\ref{f:ens} shows that both processes are indeed present, although the relative heating ratio varies as a function of time, with shock heating initially more vigorous toward the inner regions, while cavity deposition more effective at 10-100 kpc radii.

Without the third component -- subsonic turbulence (see also \S\ref{s:cooling}) -- the final macro AGN feedback deposition would be either highly anisotropic (bubble pairs) or localized (thin shock jumps). 
Fig.~\ref{f:ens} (third panel) shows that the AGN feedback induces major turbulence/vorticity in a quasi isotropic manner. While the jet direction is a continuous source of enhanced turbulence, the whole IGrM core experiences a quasi irreducible level of turbulent motions ($\sigma_v\sim100-300\,\kms$; $\mathcal{M}_t\lta 0.5$). At the same time, the simulation shows that the (negative) rarefactions avoid the runaway accumulation of large vorticity, by balancing the (positive) stretching term in a volume-filling way. The final panel finally shows that baroclinicity is negligible during the macro AGN feedback deposition, as subsonic turbulence is able to preserve the alignment of density and pressure gradients. It is important to note that, while turbulence provides a key source of isotropic \textit{mixing} (with characteristic scale $t_{\rm eddy}$; \S\ref{s:cooling}), its subsonic nature implies that the heating rate ($H_{\rm turb} = \frac{1}{2} M_{\rm gas} \sigma_v^2/t_{\rm turb} \propto \sigma_v^3$, where $t_{\rm turb}=t_{\rm eddy}/{\mathcal{M}_t^2}$; \citealt{Gaspari:2014_coma2}) is not only a fraction of the global cooling rate, but also has a substantially delayed deposition time $t_{\rm turb} \gg t_{\rm cool}$ \citep{Hillel:2017,Fabian:2017}. Alternatively, reorienting jets similar to the case of NGC 5044 (see Fig. \ref{fig:N5044}) may provide an alternative way of heating the gas in a quasi-isotropic way. \citet{Cielo:2018} presented simulations of AGN/IGrM interaction in the case of precessing jets and claimed that the distributed energy is sufficient to offset cooling and reproduce features seen in real cool-core clusters.

In closing this theoretical section,  we remark a few remaining important differences between galaxy groups and the more massive clusters. 
While we have discussed above that the IGrM shall be strongly self-regulated to avoid overheating/overcooling, this does not imply that groups are less variable than clusters. Indeed, the tails of the chaotic feeding/feedback loop can generate \textit{relatively} more disruptive imprints in the less bound IGrM (e.g., \citealt{Voit:2020}). This is reflected in the increased morphological diversity of groups (\citealt{Sun:2009a}) and larger intrinsic scatter of the scaling relations toward the low-mass regime, as found in the fundamental $L_{\rm x}-\tx$ (\citealt{Goulding:2016}; see the companion \citet{Lovisari:2021}~review) and $M_{\rm BH}-\tx$ (see \S\ref{s:Mbh}) relations. While in absolute values the AGN deposition radius is significantly larger in galaxy clusters (up to several 100 kpc), normalized to function of $R_{500}$ the AGN feedback outliers can pierce through \textit{relatively} larger regions of the less bound IGrM (e.g., \citealt{Grossovaetal19}). 
Interestingly, many elliptical galaxies (including non-centrals) show the presence of
a mini-cool core 
with a size of $\sim 1$\,kpc, 
which could represent the irreducible inner CCA condensation region enabling the more frequent self-regulated AGN feedback discussed above for galaxy groups. 

%
%
%

\section{Impact of AGN feedback on large scales} \label{s:large_scale}

\subsection{\textbf{AGN feedback in cosmological simulations}}
\label{s:cosmo_sims}

Hydrodynamical cosmological simulations are paramount for self-consistently modelling the highly non-linear formation of large-scale structure. They can simultaneously precisely solve for the gravitational and hydrodynamical aspects of structure formation. Yet, as these simulations have limited spatial and mass resolutions, one needs to implement simplified `sub-grid’ prescriptions for including crucial physical processes such as cooling, star formation, and the feedback from supernovae and AGN, since these phenomena take place at scales which cannot be resolved by the simulations. For a complete review on numerical simulations of galaxy groups, we refer the reader to \citet{Oppenheimer:2021} within this issue.  Modern simulations of galaxy groups broadly fall into three categories:
\begin{enumerate}
\item High resolution zoom simulations (a few hundred pc spatial resolution and $\sim10^{5}~M_{\odot}$ particles) of a few groups or of a small volume. These types of simulations are typically used to develop new baryonic physics and study the details of its impact on the IGrM (e.g. ROMULUS \citep{Tremmel2017}; NewHorizon \citep{Dubois2020}; FABLE \citep{Henden2018}).
\item Moderate resolution simulations (spatial resolution of the order of $\sim1$kpc and $\sim10^{6}~M_{\odot}$ particles) of volumes that are large enough (about 100 Mpc on a side) to contain a sizable sample of groups, but not many clusters (e.g. Horizon-AGN \citep{Dubois2014}; EAGLE \citep{Schaye2015}; Illustris(TNG) \citep{Vogelsberger2014,Springel2018}; SIMBA \citep{Dave2019,Robson2020}; MassiveBlack-II \citep{Khandai2015}).
\item Low resolution simulations (about 5 kpc spatial resolution and $\sim10^{9}~M_{\odot}$ particles) of much larger volumes ($\sim300-1,000$ Mpc on a side with the most common value being around 500 Mpc) to contain a large sample of groups and clusters (e.g. IllustrisTNG-300 \citep{Springel2018}; cosmo-OWLS \citep{LeBrun:2014}; BAHAMAS \citep{McCarthy:2017}; Magneticum \citep{Hirschmann2014}; Horizon Run 5 \citep{Lee2021}).
\end{enumerate}

These simulations have been run with codes that use different methods for solving the equations of hydrodynamics in a cosmological context. Namely, the Tree Particle Mesh (TreePM) and smoothed particle hydrodynamics (SPH) code \textsc{GADGET} \citep{Springel:2005} in various versions for the majority (EAGLE, MassiveBlack-II, cosmo-OWLS, BAHAMAS, Magneticum), the moving mesh codes \textsc{AREPO} \citep{Springel2010,Weinberger2020} and \textsc{GIZMO} \citep{Hopkins2015} for a smaller number (FABLE, Illustris(TNG), SIMBA) and, finally, the adaptive mesh refinement (AMR) code \textsc{RAMSES} \citep{Teyssier2002} for an even smaller fraction of them (Horizon-AGN, Horizon Run 5, and NewHorizon). Note that ROMULUS was run with the Tree+SPH code \textsc{ChaNGa} \citep{Menon2015}. All these simulations include a sophisticated modelization of the non-gravitational processes of galaxy formation such as metal-dependent radiative cooling, star formation, chemical evolution, accretion onto supermassive black holes and feedback processes from supernovae, asymptotic giant branch stars as well as AGN. Some of them have even calibrated the free parameters of these models on observations (e.g., FABLE, EAGLE, Illustris(TNG), and BAHAMAS). Note that the value of these parameters are often at least informed by higher-resolution simulations.

In the majority of cases, cosmological simulations (type 2 and 3) implement some variation of the \citet{Booth2009} AGN feedback model (hereafter BS09), which is itself largely based upon the \citet{Springel:2005} model (hereafter S05). In the BS09 model, halos are seeded with BH seeds in their center when their mass as evaluated by an on-the-fly halo finder first reaches $M_{\rm h,min}=100m_{DM}$, where $m_{DM}$ is the mass of a dark matter particle \citep[as in][]{DiMatteo2008}. At that point, BH seeds are introduced at the bottom of the potential well, with masses $M_{seed}=0.001m_{g}$ where $m_{g}$ is the mass of a gas particle. BH can then grow either by gas accretion or mergers. Specifically, BH accrete from the surrounding gas at a rate proportional to that given by the Eddington-limited Bondi–Hoyle–Lyttleton \citep{Hoyle1939,Bondi1944} formula,
\begin{equation} \dot{M}_{acc}=\alpha\dot{M}_{\rm Bondi}=\alpha\frac{4\pi G^{2}M_{\rm BH}^{2}\rho}{(c_s^2+v^2)^{3/2}}, \label{e:mdotBondi} \end{equation}
where $v$ is the velocity of the BH relative to the ambient gas. 
The dimensionless `boosting' factor $\alpha$ was introduced by S05 as a numerical correction factor that attempts to correct for the limitations of numerical simulations (see also \S\ref{s:cooling}). 
It was independent of density and had a constant value of $\sim 100$ \citep[e.g.][]{Springel:2005,DiMatteo2005,Sijacki2007,Bhattacharya2008,Puchwein:2008}. BS09 introduced a density-dependent efficiency that varies as a power law of the density with a power-law index $\beta=2$ when the density is above $n^{*}_{H}=0.1~\textrm{cm}^{-3}$ and $\beta$=1 otherwise. Bondi-Hoyle accretion is spatially resolved when the local gas density $n^{\star}_{H}<0.1~\textrm{cm}^{-3}$, which corresponds to the threshold for the formation of a cold ($T\lesssim10^{4}$ K) phase,  and when the simulations resolve the Jeans length (see BS09 and e.g. \cite{Schaye2010} for a detailed discussion). The BH growth rate can then be determined from the mass accretion rate by assuming a given radiative efficiency $\epsilon_r$, $\dot{M}_{\rm BH}=\dot{M}_{acc}(1-\epsilon_r)$. The total radiative efficiency is always assumed to be 10 per cent, which is the mean value for the radiatively efficient \cite{Shakura1973} accretion on to a Schwarzschild BH. 
As discussed in \S\ref{s:cooling}, Bondi accretion -- albeit very simple to implement in subgrid models -- is far from a realistic representation of the feeding processes (such as CCA and multiphase precipitation), thus we advocate for fundamental updates of subgrid models in future works.

BH inject a fixed fraction of the rest-mass energy of the gas they accrete into the surrounding medium. The feedback is implemented only thermally. In that case, the energy is deposited into the surrounding gas by increasing its internal energy, as opposed to kinetic feedback, which deposits energy by kicking the gas. The fraction of the accreted rest-mass energy that is injected is assumed to be independent of both the environment and the accretion rate \citep[i.e. no distinction between `quasar mode’ and `radio mode’ feedback as in the models of e.g.][which still injected energy thermally in both cases]{Sijacki2007}. The amount of energy returned by a BH to its surrounding medium in a time-step $\Delta t$ is given by

\begin{equation} E_{\rm feed}=\epsilon_f\epsilon_r\dot{M}_{\rm BH}c^2\Delta t \end{equation}

\noindent where $\epsilon_f$ is the efficiency with which a BH couples the radiated energy into its surroundings (a free parameter) and $c$ is the speed of light. In order to ensure that the thermal feedback from BHs is efficient, and that it is not immediately radiated away, BS09 introduced a minimum heating temperature $\Delta T_{\rm min}$. BHs store feedback energy until they have accumulated an energy $E_{\rm crit}$ that is large enough to increase the temperature of a number $n_{\rm heat}$ of their neighbours by an amount of $\Delta T_{\rm min}$, that is,
\begin{equation}
\label{eq:DTmin}E_{\rm crit}=\frac{n_{\rm heat}m_{g}k_{B}\Delta T_{\rm min}}{(\gamma-1)m_H}.\end{equation} 

\noindent The internal energy of the heated gas is instantaneously increased by $E_{\rm crit}$. If $\Delta T_{\rm min}$ is set too low, the cooling time of the heated gas remains very short and the energy is efficiently radiated away. If $n_{\rm heat}\Delta T_{\rm min}$ is set too high, the energy threshold and the time period between AGN heating events become very large. Thus, $n_{\rm heat}\Delta T_{\rm min}$ is connected to the AGN duty cycle. The energy is then deposited isotropically into the gas.

The recent increase in resolution (from category 3 to category 2) led to improvements of the AGN feedback modeling, as more physical processes could be taken into account. For instance, the EAGLE team modified the Bondi acrretion rate formula for $\dot{M}_{acc}$ differently to what had been previously done by BS09 to take into account the angular momentum of the gas accreted by the BH, such that the Bondi accretion rate defined in equation \ref{e:mdotBondi} was multiplied by a factor
$\alpha=\min(1,C_{visc}^{-1}(c_s/V_\phi)^3)$,
where $V_{\phi}$ is the rotation speed of the gas around the BH \citep{Rosas-Guevara2015} and $C_{visc}$ is a free parameter related to the viscosity of the accretion disc. Improvements to the \cite{Sijacki2007} model used by \cite{Puchwein:2008} have also been made as part of the Illustris, IllustrisTNG and FABLE projects for the same reasons. Specifically, the authors added a third mode of AGN feedback for Illustris \citep[i.e. the feedback is now thermal, mechanical and radiative as described in][]{Vogelsberger2013}. The kinetic AGN feedback model in the low accretion rate regime was updated for IllustrisTNG\citep{Weinberger2017} \citep[see][for an exhaustive discussion of the changes between Illustris and IllustrisTNG]{Pillepich2018}. In parallel, the FABLE team also modified the Illustris model to alleviate some of its shortcomings such as the underestimation of the gas fractions of groups and clusters (see \S\ref{s:fgas_sim} and in particular the discussion of Fig.~\ref{fig:fgas_modern} below). The parameters of the feedback model were calibrated on the gas mass fractions using a strategy similar to the one employed for BAHAMAS (\cite{McCarthy:2017}; see \cite{Henden2018} for detailed discussion). In particular, inspired by the BS09 model, they introduced a 25 Myr duty cycle for the AGN feedback to reduce artificial overcooling. We note that both Horizon-AGN \citep{Dubois2014} and NewHorizon \citep{Dubois2020} use two modes of feedback as originally introduced by \cite{Sijacki2007} but that the low accretion rate or `radio' mode is kinetic instead of thermal \citep[the detailed modelling can be found in][]{Dubois2012}.   
Indeed, as discussed in \S\ref{s:heating}, the observed and realistic astrophysical deposition of heating in hot halos is most of the time carried out via the AGN outflows and jets. 

\subsection{\textbf{The hot gas fraction and the AGN feedback model}}\label{s:fgas_sim}

\begin{figure}[t]
 \centering
     \includegraphics[width=1.0\textwidth]{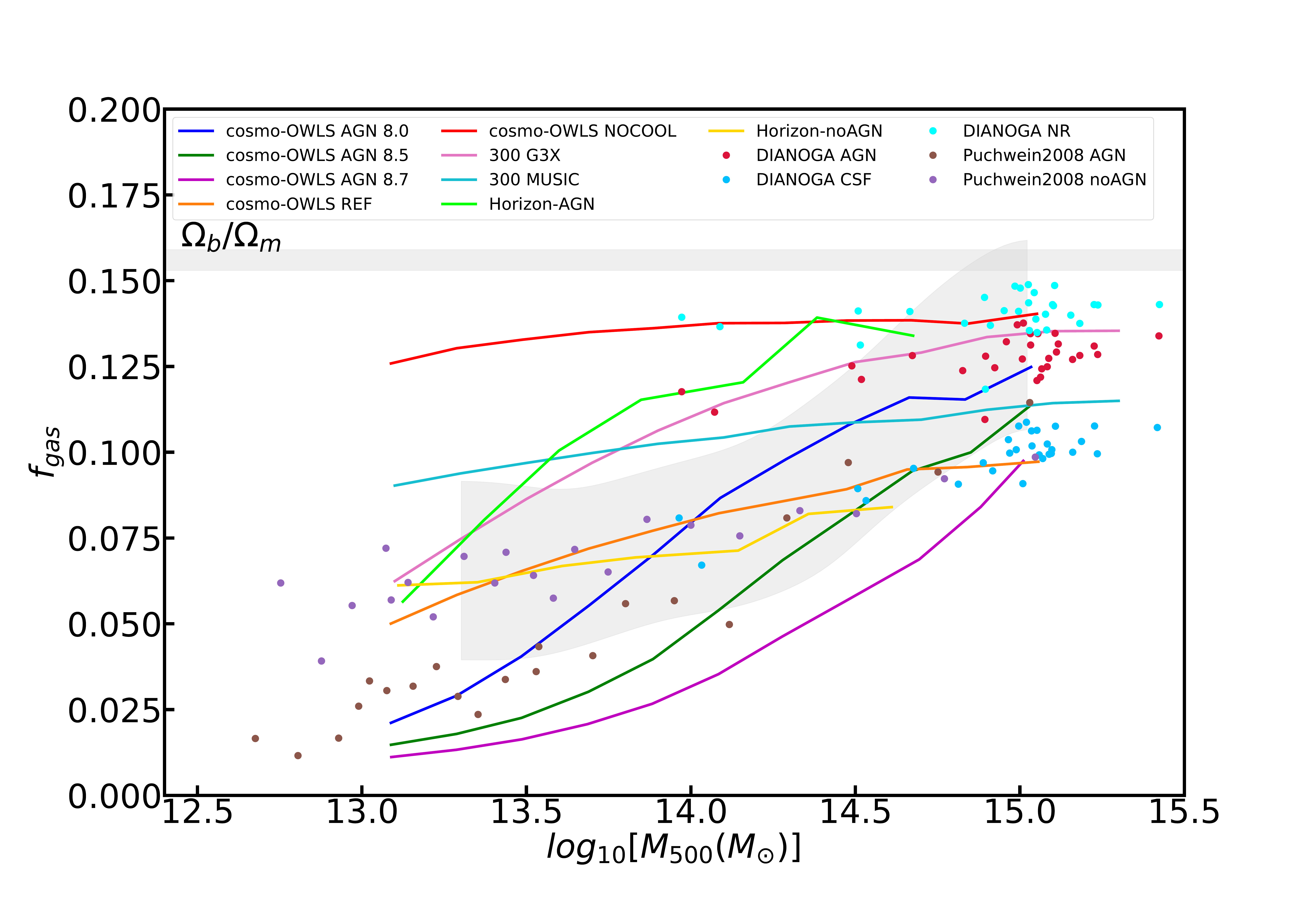}
     \caption{Compilation of historical simulation results for the gas fraction within $R_{500}$ as a function of $M_{500}$. The red, orange, magenta, green and dark blue solid lines correspond to the different sub-grid models of cosmo-OWLS \citep{LeBrun:2014}, the pink and cyan ones to the 300 clusters run with the GADGET3X \citep{Beck2016} and MUSIC \citep{Sembolini2013} codes respectively as part of The Three Hundred Project \citep{Cui2018}, the lime and gold ones correspond to Horizon-AGN and, its counterpart without AGN, Horizon-noAGN \citep{Dubois2014}, the cyan, blue and crimson ones to the various physical models of the DIANOGA suite \citep{Rasia2015} and finally, the purple and brown symbols correspond to the simulations of \cite{Puchwein:2008} without and with AGN, respectively. The compilation of observations presented in Fig.~\ref{fig:fgas_obs} is shown as a gray band.}
     \label{fig:fgas_historical}
 \end{figure}

\begin{figure}[t]
 \centering
     \includegraphics[width=0.95\textwidth]{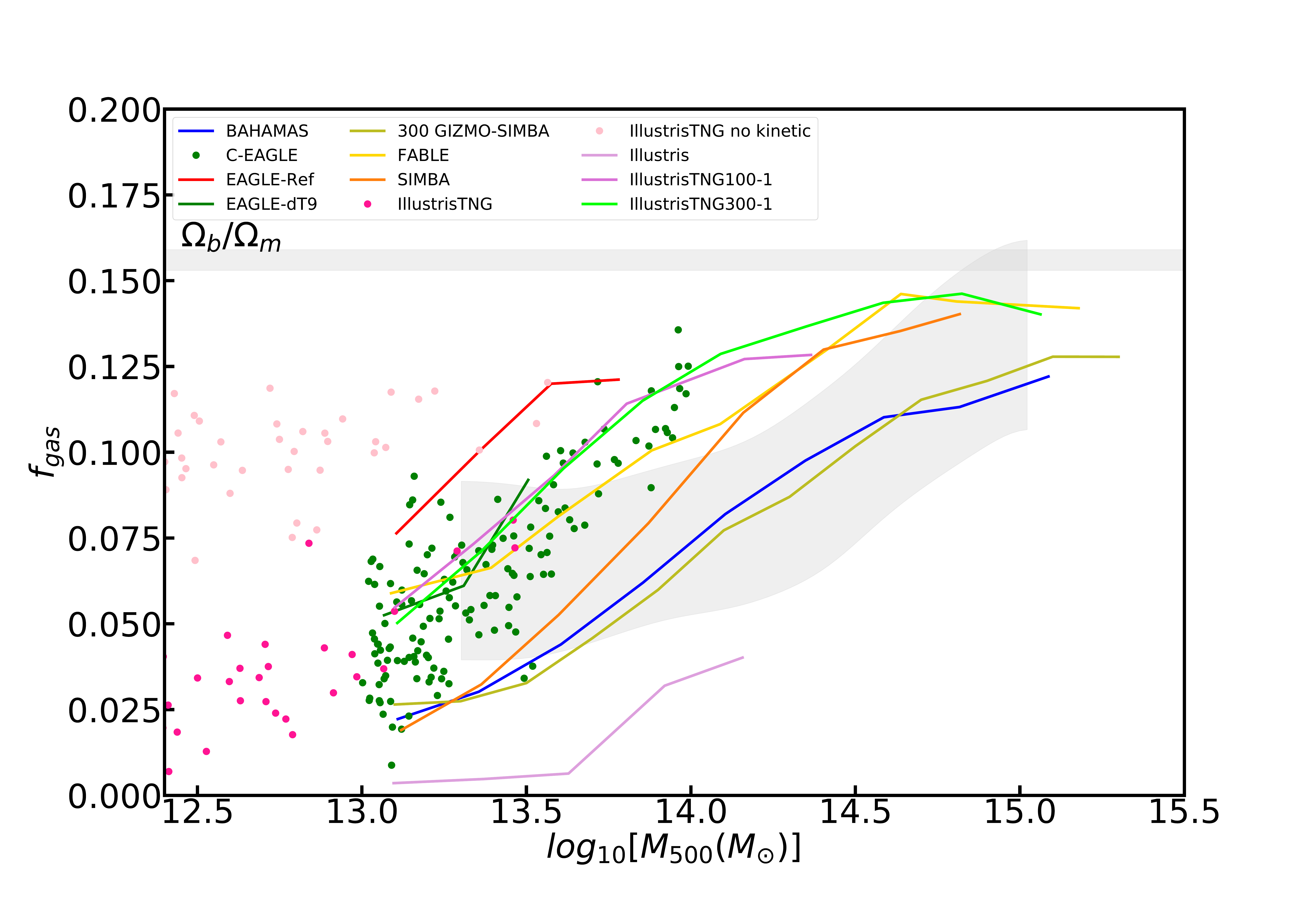}
     \caption{Compilation of gas fractions within $R_{500}$ from modern simulations as a function of $M_{500}$. The blue line corresponds to BAHAMAS \citep{McCarthy:2017}, the red and green lines to the Reference and dT9 models of EAGLE \citep{Schaye2015}, the green symbols to C-EAGLE and Hydrangea \citep{Barnes2017,Bahe2017} as it uses the same sub-grid model as EAGLE-dT9, the olive line to the 300 clusters run with the GIZMO-SIMBA code (Cui et al. in preparation) as part of The Three Hundred Project \citep{Cui2018}, the salmon line to Horizon-AGN \citep{Dubois2014}, the gold one to FABLE \citep{Henden2018}, the orange one to SIMBA \citep{Dave2019,Robson2020} and the pink and deep pink symbols as well as the plum, orchid and lime lines correspond to various models from the Illustris and IllustrisTNG suites \citep{Vogelsberger2014,Springel2018}. The compilation of observations presented in Fig.~\ref{fig:fgas_obs} is shown as a gray band.}
     \label{fig:fgas_modern}
 \end{figure}

As stated in \S\ref{s:baryon_content}, the total baryon content and its partition between the various gas and stellar phases put fundamental constraints on galaxy formation models and, in particular, on the strength of AGN feedback. We thus present here the gas mass fraction-$M_{500}$ relation at $z=0$ for two compilations of massive galaxies, groups and cluster simulations that include different sorts of baryonic physics: \emph{i)} historical simulations, that is, simulations run before 2014-2015 in Fig.~\ref{fig:fgas_historical}; and \emph{ii)} modern simulations run as from 2014-2015 that calibrated the free parameters of their subgrid models to reproduce at least the galaxy stellar mass function at $z=0$ in Fig.~\ref{fig:fgas_modern}. The various simulation sets will be compared with the compilation of observations we presented in \S\ref{s:baryon_content} and especially Fig.~\ref{fig:fgas_obs}, which is shown as a gray band on both figures.

In Fig.~\ref{fig:fgas_historical}, we present the results of a compilation of historical simulations for the gas fraction within $R_{500}$ as a function of $M_{500}$. 
The NOCOOL model of cosmo-OWLS and the NR model of DIANOGA both correspond to classical non-radiative simulations, where one includes hydrodynamics but do not allow the gas to cool through radiative processes. The REF model of cosmo-OWLS, the CSF model of DIANOGA, the 300 clusters run with the MUSIC code, as well as the noAGN models of \cite{Puchwein:2008} and the Horizon suite, all include prescriptions for radiative cooling, star formation and stellar feedback, but not for AGN feedback. As first noted by \cite{Puchwein:2008} and \cite{McCarthy2010}, the inclusion of AGN feedback substantially lowers the gas fractions of both groups and clusters.
The intensity and thus the duty cycle of the AGN feedback as parameterized by $\Delta T_{\rm min}$ in BS09 (see Eq. \ref{eq:DTmin}) can be used to eject more or less gas from the potential well, as can be seen by comparing the models AGN 8.0, 8.5 and 8.7 of cosmo-OWLS. Here the number corresponds to the logarithm of the value of $\Delta T_{\rm min}$ chosen, i.e. 8.0 corresponds to $\Delta T_{\rm min}=10^8$ K. The REF, CSF and noAGN models also yield reasonable gas mass fractions, but the relation with mass is flatter than observed, because the SF efficiency does not depend strongly on halo mass. The low gas fractions in these models are achieved by overly efficient star formation \citep[e.g.][and Fig.~\ref{fig:GSMF}]{McCarthy2010,LeBrun:2014}. Note that while the cosmo-OWLS models that include AGN feedback use the BS09 AGN feedback model summarized in \S\ref{s:cosmo_sims} which is fully thermal, Horizon-AGN, DIANOGA and 300 G3X resort to a mixture of thermal and kinetic feedback as originally developed by \cite{Dubois2012} for the former and \cite{Steinborn2015} for the latter two.

In Fig.~\ref{fig:fgas_modern}, we present the results of a compilation of modern simulations for the gas fraction within $R_{500}$ as a function of $M_{500}$. Despite the fact that most modern simulations have been calibrated to reproduce the local galaxy stellar mass function (see Fig. \ref{fig:GSMF}), the predictions on the hot gas fraction are vastly different. For instance, Illustris (plum line) vastly underpredicts the observed hot gas fractions, whereas the reference EAGLE model (red line) clearly overpredicts them. Therefore, a setup that broadly reproduces the stellar content of galaxies in the Universe may simultaneously fail at reproducing the properties of the hot gas phase. Note that in the case of BAHAMAS (blue line) and FABLE (gold line) the free parameters of the stellar and AGN feedback have been adjusted to reproduce both the $z=0$ galaxy stellar mass function and the gas content of groups and clusters \citep[see discussions in][]{McCarthy:2017,Henden2018}. IllustrisTNG and EAGLE-dT9 (and associated simulations) are versions of Illustris and EAGLE in which the AGN feedback parameters were slightly adjusted to reduce the discrepancies with the gas content of massive groups and clusters. It is worth noting that SIMBA uses a fully kinetic AGN feedback model \citep{Dave2019} while the simulations from the Illustris series and FABLE include a mix of thermal, kinetic/mechanical and radiative feedback \citep{Vogelsberger2014,Pillepich2018,Henden2018} in the vein of the one first developed by \cite{Sijacki2007}.

Generally speaking, we stress that the hot gas fraction of galaxy groups is an extremely sensitive probe of the feedback scheme implemented in cosmological simulations. Modern simulation suites have little predictive power on the baryon content of groups, even when the properties of the galaxy population are accurately reproduced (see Fig. \ref{fig:GSMF}). Some simulations are actually \emph{calibrated} on the gas mass fractions, i.e. the parameters governing the feedback model were tuned to produce reasonable gas fractions in the group regime. Major observational (\S\ref{s:baryon_content}) and theoretical advances (\S\ref{s:theory}) are required to understand the ejection of baryons from halos by AGN feedback and inform the mainstream galaxy evolution models.


\subsection{\textbf{Co-evolution between the IGrM and the central AGN}} \label{s:Mbh}

\begin{figure}[t]
	\centering
	\subfigure{
		\includegraphics[width=0.95\textwidth]{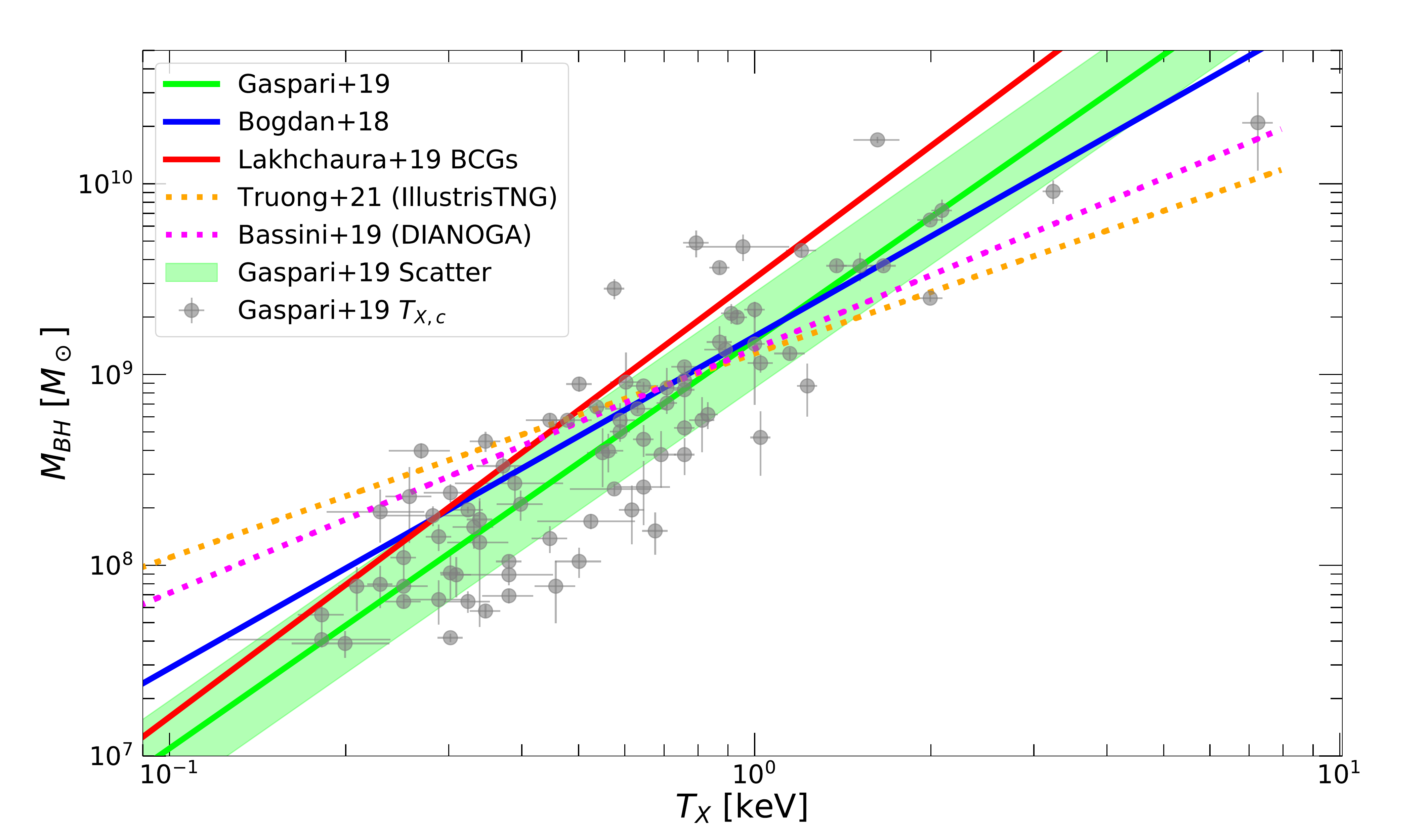}}
	\caption{Relation between BH mass and IGrM X-ray temperature. The data points show dynamical measurements of BH mass plotted against the spectroscopic X-ray temperature of the host halo. The solid curves show the fitted observational relations from \citet{Gaspari:2019} (green; together with the quoted log-normal scatter), \citet{Bogdan:2018} (blue), and the ``BCG'' subsample of \citet{Lakhchaura:2019} (red). The dashed lines show the predictions of cosmological simulations with AGN feedback (Illustris TNG, \citet{Truong:2021}; DIANOGA, \citet{Bassini:2019}). The gray data points are taken from the sample of \citet{Gaspari:2019}, which already included the smaller samples used by \citet{Bogdan:2018} and \citet{Lakhchaura:2019}.}
	\label{f:mbh_tx}
\end{figure}

As discussed in \S\ref{s:coevol}, SMBH masses are known to correlate with the properties of their host galaxy, in particular the integrated K-band luminosity $L_K$ and the velocity dispersion of the stars in the bulge, $\sige$ \citep[see][for a review]{Kormendy:2013}. However, it is still unclear whether the optical scaling relations of SMBH are fundamental or derive from correlations with other key quantities. Recent findings have instead unveiled that the SMBH masses are more tightly correlated with the properties of the host X-ray gaseous halos, especially in the IGrM regime \citep{Bogdan:2018,Gaspari:2019,Lakhchaura:2019}. In Fig.~\ref{f:mbh_tx}, we summarize our current knowledge of the relation between SMBH mass and X-ray temperature within the core of galaxy groups ($R\lta0.15\,\r500$). It is important to note that the SMBH masses shall be directly observed via dynamical measurements to properly unveil intrinsic scaling relations. 
The largest existing study is provided by \citet{Gaspari:2019} with 85 systems with measured SMBH masses, most of which with temperatures $\sim0.5-1$ keV typical of galaxy groups. A Bayesian fit to the relation finds slopes $M_{\rm BH} \propto \tx^{2.1}$ (Fig.~\ref{f:mbh_tx}, green) and $M_{\rm BH} \propto \lx^{0.4}$. At the high-mass end, \citet{Bogdan:2018} measure a somewhat flatter slope, $M_{\rm BH}\propto \tx^{1.7}$. Notably, the intrinsic scatter goes down to $\sim0.2$ dex, with very high correlation coefficient at or above the 0.9 level, compared to $\sim0.5$ dex for the K-band luminosity. The correlations hold regardless of the large diversity of systems, from BGGs and ETGs to non-central lenticular/spiral galaxies. Varying the extraction radius by slightly enlarging or decreasing it (group outskirts or CGM) does not vary significantly such conclusions (see the companion \citet{Lovisari:2021}~review for the complementary $R_{500}$ scalings, such as $M_{\rm BH}-L_{500}$ and $M_{\rm BH}-M_{\rm tot}$). We note that multi-variate X-ray correlations (aka `fundamental planes') do not further improve the intrinsic scatter.
In sum, by comparing the different X-ray/optical scaling relations, it has emerged that the extended plasma (collisional) atmospheres seem to play a more fundamental role than small-scale (collisionless) stellar properties in the co-evolution of SMBH and groups. This is further supported by zoom-in cosmological simulations \citep{Bassini:2019,Martin-Navarro:2020,Truong:2021}. On the other hand, the slope (and scatter) of the current cosmological simulations still remain too low compared with the observations (dotted lines in Fig.~\ref{f:mbh_tx}), indicating the need to model more realistic feeding and feedback physics (see \S\ref{s:theory}) into the coarse subgrid numerical modules.

\begin{figure}[t]
     \centering
     \subfigure{\hspace{-0.5cm}
     \includegraphics[width=0.95\textwidth]{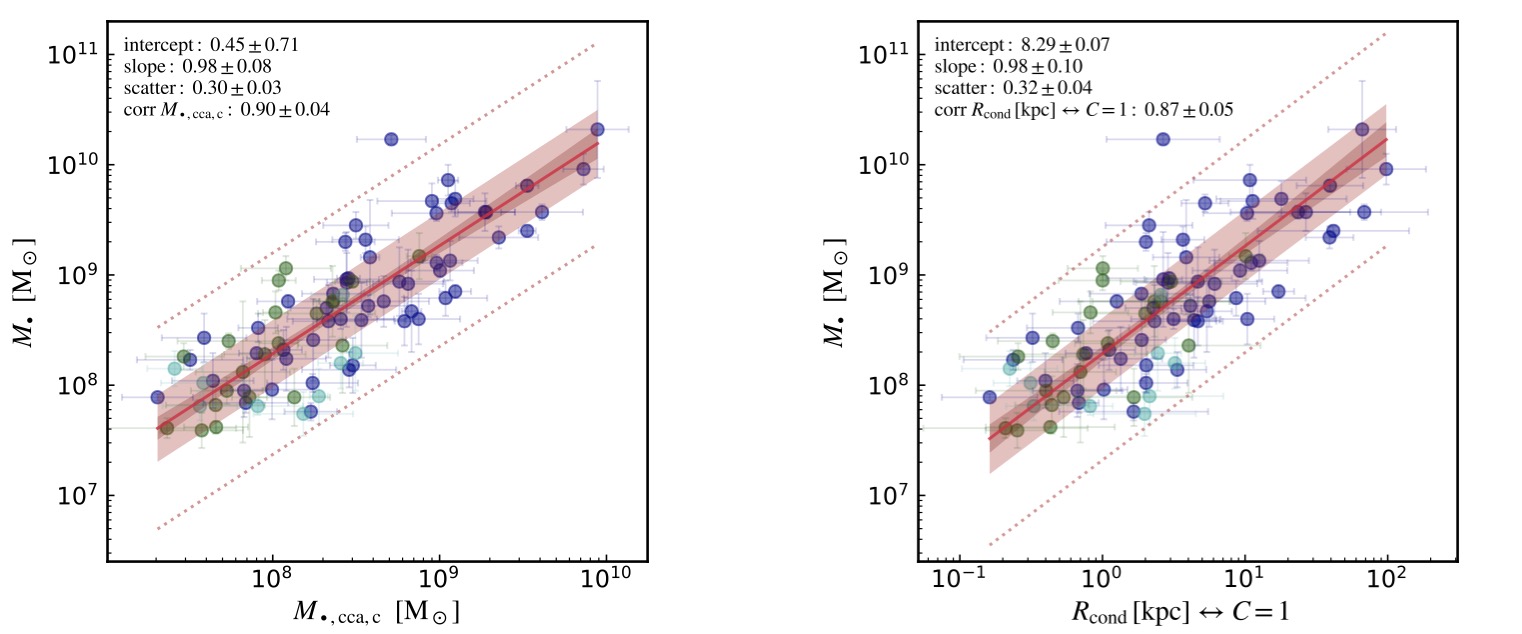}}
     \caption{X-ray scaling relations are key to constrain different baryonic physics of the IGrM, here in terms of feeding models.
     \emph{Left:} Direct SMBH masses plotted against the mass derived from theoretical/numerical predictions of chaotic cold accretion (CCA) via the X-ray core properties -- adapted from \citet{Gaspari:2019}. 
     \emph{Right:} Hot-halo condensation radius as a function of BH mass and thus IGrM halo mass (the locus where $C\equiv t_{\rm cool}/t_{\rm eddy}=1$; see \S\ref{s:ratios} and \S\ref{s:cooling}). 
     See Fig.~\ref{f:Ebh_Ebind} for the description of the analysis, color coding, and sample. Figure reproduced from \citet{Gaspari:2019}.
     }
     \label{f:cond}
\end{figure}

The above SMBH versus X-ray correlations are crucial to test models of galaxy/group formation and evolution. 
As discussed in \S\ref{s:cooling}, accretion/feeding models can be broadly divided into cold and hot accretion modes. Besides the cosmic dawn, hierarchical binary BH mergers are a present, but subdominant growth channel over most of cosmic time \citep{Bassini:2019,Gaspari:2019}. In hot accretion (usually Bondi or Advection Dominated Accretion Flow -- ADAF; \citealt{Bondi:1952,Narayan:2011}), the larger the thermal entropy of the gas, the more strongly feeding is stifled, since the inflowing gas has to overcome the outward thermal pressure of the hot halo. This would induce negative correlations with the IGrM properties, which are ruled out by the strongly positive correlations shown in Fig.~\ref{f:mbh_tx}. Conversely, cold-mode accretion (\citealt{Gaspari:2013_cca,Voit:2018}; \S\ref{s:cooling}) -- typically in chaotic form (due to the turbulent condensation out of the IGrM generating randomly colliding clouds) -- is linearly and tightly correlated with the X-ray luminosity and gas mass. 
Fig.~\ref{f:cond} (left) shows the CCA final mass growth via theoretical/numerical predictions \citep{Gaspari:2019} compared to direct measurements of SMBH masses. During the Gyr evolution, the turbulent IGrM locally condenses into extended warm filaments and cold molecular clouds via nonlinear thermal instability. The clouds inelastic collisions at the meso/kpc scale boost the micro accretion rate down to the Schwarzschild radius, 
hence triggering strong AGN feedback heating. 
Such recurrent SMBH growth drives the $M_{\rm BH}$ shown in Fig.~\ref{f:cond}, with excellent agreement with the SMBH mass observed in our local universe. 

Scaling relations allow us to predict other baryonic physics of the IGrM, such as the extent of the IGrM multiphase condensation radius, which is shown in the right panel of Fig.~\ref{f:cond}. As introduced in \S\ref{s:cooling}, such a radius is the locus at which the IGrM cooling time and the turbulent eddy-turnover time match ($C\equiv t_{\rm cool}/t_{\rm eddy} = 1$), both of which can be retrieved via the X-ray scaling relations as a function of \Tx\ and \Lx\ (\citealt{Gaspari:2019}).
Evidently, lower mass groups have condensation radii of less than a few kpc, 
while massive groups can reach $R_{\rm cond}$ of a few 10 kpc (e.g., \citealt{Davidetal17,Lakhchaura:2018,Olivares:2019}).
Overall, scaling relations between X-ray macro-scale properties translates into scaling relations of micro-scale properties ($M_{\rm BH}$), corroborating a tight co-evolution between multi-scale processes in the IGrM (as depicted in Fig.~\ref{f:loop}).

\subsection{\textbf{Impact on cosmological probes}}\label{s:cosmo}

During the past few years, it has become clear that AGN feedback will play an important role as a leading source of systematic uncertainties for upcoming high-profile cosmology experiments. Indeed, the energy injected by the central AGN affects the global distribution of baryons (see \S\ref{s:fgas_sim}), leading to local depletion or excesses of matter with respect to the expectations of models including dark matter only. This effect is most important in galaxy groups, since these systems correspond to the peak of the local halo mass density and their baryonic properties are highly sensitive to feedback. As a result, the matter power spectrum at $z=0$ can be substantially altered by baryonic physics, and in particular AGN feedback \citep[e.g.][]{VanDaalen:2011,Semboloni:2011}. The shape of the power spectrum is strongly affected by baryonic processes on scales $k\gtrsim 10^{-1}$ $h$/Mpc. For $10^{-1}<k<10$ $h$/Mpc most simulations predict a \emph{deficit} of power with respect to the N-body case, although the actual amplitude of the effect is highly uncertain \citep{Chisari:2019OJAp}. The evacuation of baryons from the central regions of galaxy groups under the influence of feedback is responsible for the deficit of power on scales of $\sim1$ Mpc, i.e. roughly the typical size of galaxy groups. On smaller scales ($k\gtrsim10$ $h$/Mpc), cooling and condensation of baryons in the central regions lead to a rapidly increasing power. 

\begin{figure}[t]
     \centering
     \subfigure{
     \includegraphics[width=0.5\textwidth]{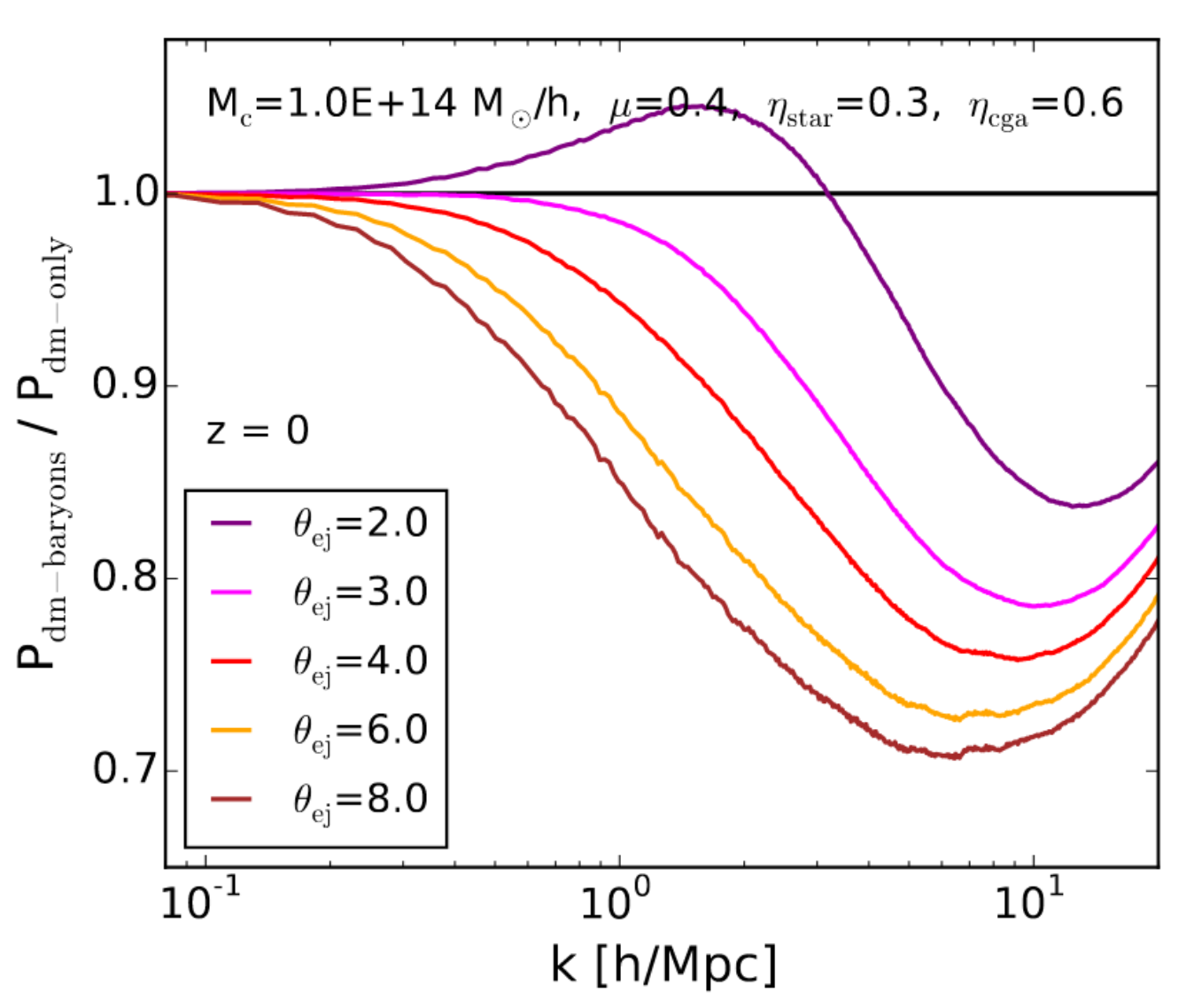}}
     \caption{Modification of the local matter power spectrum with respect to pure N-body simulations in the presence of AGN feedback at the scale of galaxy groups in the semi-analytic model of \citet{Schneider:2019}. The various curves show how the power spectrum depends on the parameter $\theta_{ej}$ governing gas ejection from the central regions of groups under the influence of AGN feedback. A strong ejection of gas from the core of halos implies substantial modifications to the matter power spectrum on scales of $\sim1$ Mpc. Figure reproduced with permission from \citet{Schneider:2019}.}
     \label{f:ps}
\end{figure}

Accurately predicting the shape of the matter power spectrum is crucial for the success of future cosmic shear experiments such as \emph{Euclid}, which aim at determining the growth of structures by measuring the matter power spectrum and its evolution \citep{Euclid:2018}. \citet{Semboloni:2011} showed that neglecting baryonic effects would imply important systematics on the determination of cosmological parameters. Systematic effects can be mitigated by excluding the small scales ($k\gtrsim10^{-1}$ $h$/Mpc) when fitting the measured power spectra, although that comes at the price of greatly increased uncertainties in the resulting cosmological parameters. Constraints on extended cosmologies such as massive neutrinos, variable dark energy equation of state, or chameleon gravity, require sensitivity on smaller scales and their effect is strongly degenerate with that of baryonic physics \citep{Schneider:2020a,Schneider:2020b}. 

\citet{Chisari:2019OJAp} showed that numerical simulations have not yet converged on the actual impact of feedback on the power spectrum (see their Fig. 3). For instance, the very strong feedback implemented in the original Illustris simulation, which was sufficient to completely evacuate the gas content from most groups (see Fig. \ref{fig:fgas_modern}), leads to a very strong suppression of power ($>30\%$) on scales of a few hundred kpc. Conversely, simulations implementing a more gentle feedback scheme (EAGLE, Horizon-AGN, MassiveBlackII) predict small corrections with respect to the fiducial DM-only case for $k\lesssim 10$ $h$/Mpc. These simulations also predict a high gas fraction in galaxy groups \citep[see \S\ref{s:fgas_sim} and][for an extensive discussion]{VanDaalen:2020}. Recently, \citet{Schneider:2019} used a semi-analytic model to predict the impact of baryons on the matter power spectrum based on the observed gas properties of groups. The authors modified the mass profiles of halos in large N-body simulations to account for star formation and AGN feedback. In particular, the semi-analytic model of \citet{Schneider:2019} is highly sensitive to the parameter $\theta_{ej}$ which governs the ejection of gas from the central regions of the halo by AGN feedback. In Fig. \ref{f:ps}, we show how the predicted matter power spectrum depends on $\theta_{ej}$. With increasing feedback, a progressively larger fraction of the gas is ejected from the halo, and thus the expected power gets more strongly suppressed. Calibrating their semi-analytic model on the observed gas fraction and gas density profiles of group-scale halos, \citet{Schneider:2019} provided a range of predictions matching the existing observational constraints. High-precision measurements of the gas density profiles in a representative sample of galaxy groups would allow us to determine precisely the expected shape of the power spectrum \citep{Schneider:2020b}, thereby providing a key input for upcoming cosmology experiments. 

In addition to the matter power spectrum, AGN feedback on the scales of galaxy groups also affects several other cosmological observables such as the thermal SZ power spectrum \citep[e.g.][]{Battaglia:2012,McCarthy2014}. Indeed, AGN feedback affects the pressure profiles of halos and thus modifies the amplitude of the power spectrum on small scales  \citep[$\ell\gtrsim1,000$,][]{McCarthy:2018}. \citet{Ramos-Ceja:2015} showed that tSZ models based on the universal pressure profile \citep{Arnaud:2010} overpredict the power measured by SPT and ACT on small scales. A strong feedback scenario and a low gas fraction on group scales are needed to fit the measured power. The effect of feedback also implies modifications to the cross-correlations between the tSZ and other observables \citep[e.g.][]{Battaglia:2015,Hojjati2015}. Finally, the choice of the feedback scheme affects the shape of the halo mass function \citep[e.g.][]{Velliscig2014,Cui2014,Bocquet:2016} and the structure of dark-matter halos \citep[e.g.][]{Velliscig2014,Schaller:2015}. We refer to the companion \citet{Oppenheimer:2021} review for a more general discussion of the topic.

\section{Future observatories}
\label{s:future}

\subsection{\textbf{eROSITA}}
At present, our knowledge of the population of galaxy groups in the local Universe comes largely from the \ROSAT\ All-Sky Survey (RASS). Groups identified in the RASS (or even the \textit{Einstein} slew survey) form the basis of most studies of the mechanics of AGN feedback at this mass scale, but unfortunately groups are at the lower limit of sensitivity for these surveys. RASS is therefore biased toward the detection of relaxed, centrally-concentrated, cool-core systems, with the strength of the bias increasing as mass decreases from poor clusters to groups \citep{Eckertetal11}. Searches tailored to the detection of more extended sources in RASS reveal a population of low surface brightness groups undetected by the original survey \citep{Xu:2018} confirming the expected bias, while \XMM\ observations of optically-selected groups identify both low luminosity and disturbed systems previously not detected or not recognised as groups in RASS \citep{OSullivan:2017}.

\begin{figure}[t]
\centerline{\includegraphics[width=0.85\textwidth]{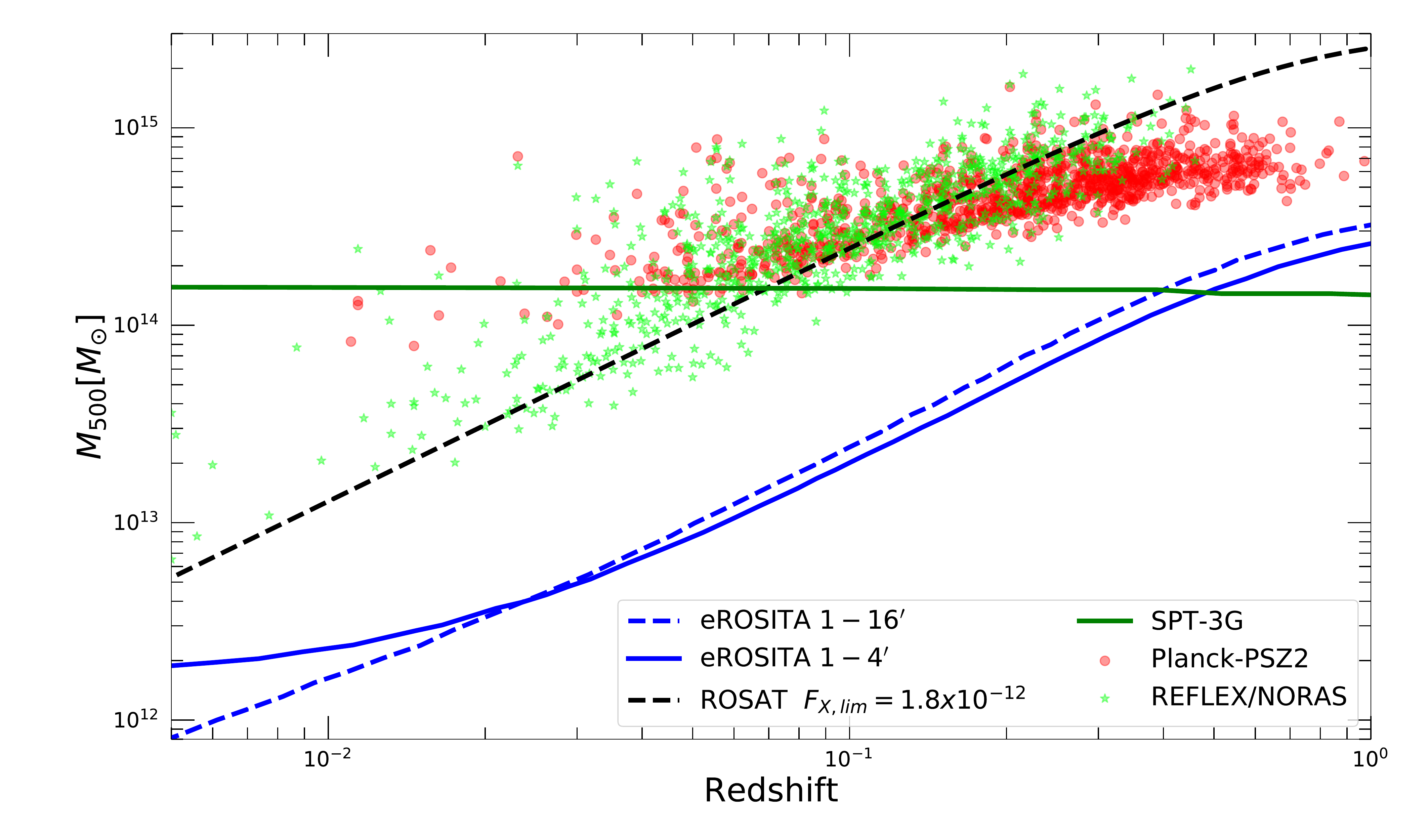}}
\caption{\label{fig:erosita} Expected sensitivity curve of the final eROSITA survey (eRASS:8) compared to the sensitivity of existing and upcoming X-ray and SZ surveys. The eROSITA sensitivity curve was computed from synthetic data using a wavelet decomposition algorithm \citep{Kaefer:2020} sensitive to scales of $1-4^\prime$ (solid blue curve) and $1-16^\prime$ (dashed blue curve). For comparison, the  dashed black curve shows the sensitivity of the ROSAT all-sky survey assuming a fixed soft X-ray flux threshold of $1.8\times 10^{-12}$ erg~s$^{-1}$, which is the typical sensitivity of the REFLEX and NORAS samples \citep[green asterisks, ][]{Boehringer:2013}. The red points show the systems selected from the second \emph{Planck} SZ catalogue \citep{Planck:PSZ2}. The green curve shows the expected sensitivity of the SPT-3G experiment \citep{Benson:2014}.}
\end{figure}

\textit{Spectrum Roentgen Gamma} (SRG, launched in 2019) hosts the eROSITA instrument, a set of seven co-aligned soft X-ray telescopes covering the 0.2-10~keV band, with a field of view of 1$^\circ$ and $\sim$15$^{\prime\prime}$ spatial resolution. SRG will spend four years surveying the whole sky once every 6 months, with eROSITA building up a map $\sim$20$\times$ deeper than RASS in the 0.5-2~keV band \citep{Merlonietal12}. This is sufficient to detect essentially every galaxy group with a virialized halo in the local universe \citep{OSullivan:2017}. More massive groups with luminosities $\sim$10$^{42}$~erg~s$^{-1}$ should be detectable to $z\sim0.1$ in the final eRASS:8 survey. \citet{Kaefer:2020} performed detailed simulations to evaluate the sensitivity of eRASS:8 to galaxy groups. The authors used a wavelet decomposition algorithm sensitive to large-scale diffuse emission. In Fig. \ref{fig:erosita}, we show the corresponding sensitivity curves for two possible source detection setups: a decomposition over wavelets of scales $1-4^\prime$ optimized for relatively compact sources, and the other for scales in the range $1-16^\prime$ sensitive to the most extended nearby sources. Using these setups, the authors predict eRASS:8 will detect all the galaxy groups with $M_{500}>10^{13}M_\odot$ out to $z=0.05$. The most massive groups ($\sim10^{14}M_\odot$) will be detected out to $z=0.5$. While the survey observations will typically only provide luminosity and morphology information for individual halos, the group samples derived from them will be a solid base from which to investigate the impact of cooling and AGN feedback in groups, particularly when combined with radio surveys.

Pointed observations with eROSITA, possible once the survey phase is complete, may also prove useful for studies of groups. The combination of a large field of view and soft band effective area (roughly double that of the \XMM\ EPIC-pn) is well-suited to observations of the outskirts of nearby groups, and the search for cavities or other structures associated with giant group-central radio galaxies. Thanks to its short focal length and very stable background \citep{Freyberg:2020}, eROSITA is very sensitive to diffuse X-ray emission below 2 keV, meaning it is well suited to study the diffuse X-ray emission of galaxy groups. 

\subsection{\textbf{XRISM}}
The X-ray Imaging and Spectroscopy Mission (XRISM), expected to launch by April 2023, will open a new era of high spectral-resolution observations of galaxy groups. Its X-ray microcalorimeter \citep[\textit{Resolve},][]{XRISM2020} will have a constant 7~eV energy resolution across its 0.3-12~keV band, resolving the forest of emission lines which characterizes emission from the IGrM. This offers opportunities in a number of important areas, including measurements of bulk flows and turbulence in the hot gas. At present, grating spectra only provide upper limits on the turbulence of the IGrM \citep{Sanders:2013,Pinto:2015}, with possible hints of asymmetries associated with sloshing motions \citep{Ahorantaetal16}. \textit{Hitomi} demonstrated the capabilities of microcalorimeters, but was only able to make a single turbulence measurement in the Perseus cluster \citep{Hitomi:2016}. XRISM should be able to measure turbulent velocities down to tens of km~s$^{-1}$, providing a measure of the kinetic energy stored in the IGrM and allowing us to determine how much of the energy of AGN outbursts can be diffused out into the IGrM by these gas motions. Spectra from the \textit{Resolve} microcalorimeter will also provide a detailed view of shock heating and cooling, with individual emission lines accurately tracing gas at different temperatures. Performance verification targets for the mission include NGC~5044 and NGC~4636, and while the spatial resolution ($>$1$^{\prime}$) and effective area of the observatory may limit its use to relatively bright nearby groups, its results are likely to be ground-breaking.

\subsection{\textbf{Athena}}
The Advanced Telescope for High Energy Astrophysics (Athena), expected to launch in the early 2030s, represents the next generation of major X-ray observatories, with 5$^{\prime\prime}$ spatial resolution and an effective area of 1.4~m$^2$ at 1~keV (roughly 45$\times$ that of XRISM's \textit{Resolve} instrument). It will carry a 40$^\prime\times$40$^\prime$ active pixel detector (the wide field imager, WFI) providing CCD-like spectral imaging, and a 5$^\prime$-diameter microcalorimeter array (the X-ray Integral Field Unit, X-IFU) with $\leq$5$^{\prime\prime}$ pixels, capable of 2.5~eV spectral resolution \citep[see, e.g.,][]{Barretetal20}. The combination of the very large collecting area with these two instruments will open up several new fields of study for galaxy groups. The WFI survey, performed over the first 4 years of operations, is expected to find $>$10,000 groups and clusters at $z\geq$0.5, including $\sim$20 groups with M$_{500}\geq$5$\times$10$^{13}$~M$_\odot$ at $z\sim2$, and measure their temperature to better than 25\% accuracy \citep{Zhangetal20}. This will provide a clear view of the evolution of AGN feedback in groups back to the era of peak star formation and black hole growth. Identification of cavities and spectral mapping will be possible out to moderate redshift, showing us the impact of feedback over the past few gigayears.

X-IFU offers capabilities similar to that of XRISM's \textit{Resolve}, but with greatly improved spatial resolution and the ability to examine even low luminosity systems in the local universe. It will allow mapping of turbulence and bulk flows in the IGrM, tracing gas motions associated with mergers, sloshing, uplift behind rising radio bubbles or AGN-driven outflows. By mapping the kinetic and thermal energy content of the IGrM on spatial scales similar to those at which energy is injected by radio galaxies, it will allow us to quantify how much energy is injected into the hot gas by outbursts, determine where and when energy is transferred out of the radio jets and lobes, and see how it is then transported out into the surrounding halo \citep{Crostonetal13}. It will also provide a clear view of the location of the coolest gas and allow us to trace the process by which it cools out of the hot phase.   

\subsection{\textbf{Lynx}}
The Lynx mission concept \citep[\url{https://www.lynxobservatory.com/report}, ][]{Gaskinetal19} will, if approved, go beyond Athena, with sub-arcsecond resolution over a 22$^\prime\times$22$^\prime$ field of view and an effective area of 2~m$^2$ at 1~keV, giving 50$\times$ the throughput of \Chandra. As with Athena, an active pixel array (the high-definition X-ray imager, HDXI) would provide wide-field CCD-like spectral imaging, but with 0.3$^{\prime\prime}$ pixels to take advantage of the exquisite spatial resolution. The Lynx X-ray Microcalorimeter (LXM) would bring 3~eV spectral resolution on 1$^{\prime\prime}$ spatial scales over a 5$^\prime$ field of view, with sub-arrays offering 0.5$^{\prime\prime}$ spatial resolution or 0.3~eV spectral resolution in 1$^\prime$ fields. Much of Lynx's proposed science relates to the detailed physics of accretion and galaxy evolution, and the X-ray universe at high redshift; survey observations with the HDXI would be capable of detecting groups with masses as low as 2$\times$10$^{13}$~M$_\odot$ out to a $z\sim3$. In low-redshift systems, LXM could examine the conditions within individual cooling filaments in group cores, and measure the velocities of the weak shocks and sound waves produced during AGN outbursts. As with Athena, the observatory would be sensitive enough to trace the IGrM out to the virial radius in a large sample of groups, but with fine spatial resolution making identification of structure easier, and rejection of background sources cleaner. It is notable that Lynx would be the first mission after \Chandra\ to be able to provide the finely detailed images that have proved so useful in the study of AGN feedback, reaching scales comparable to those of optical and radio observations.

\subsection{\textbf{The Square Kilometer Array and its precursors}}
Radio astronomy has undergone something of a renaissance in recent years, with new and improved capabilities coming online, e.g., the upgraded Jansky Very Large Array (JVLA) and Giant Metrewave Radio Telescope (uGMRT), the Low-Frequency Array (LOFAR) and the Atacama Large Milimeter Array (ALMA). These are providing improved radio continuum surveys in the northern hemisphere and equatorial sky. However, new telescopes in southern Africa and Australia are opening up new opportunities, as they begin to survey the relatively poorly-explored southern sky. These include the Murchison Widefield Array (MWA), the Australian Square Kilometer Array Pathfinder (ASKAP), and MeerKAT in South Africa's Karoo region. The Galactic and Extragalactic All-sky MWA Survey \citep[GLEAM,][]{HurleyWalkeretal17} provides an early example, covering the entire southern sky below Declination +30$^\circ$ at frequencies 72-231~MHz. While its spatial resolution is modest ($\sim$100$^{\prime\prime}$) its high sensitivity at low frequency and provision of fluxes in multiple bands makes it a powerful tool for studying radio galaxies, particularly old, fading sources. Higher frequency ($\sim$1~GHz), higher spatial resolution (8-30$^{\prime\prime}$) surveys of continuum emission and H\textsc{i} are becoming available from ASKAP \citep[e.g., RACS, WALLABY,][]{McConnelletal20, Koribalskietal20} and MeerKAT \citep[e.g., MIGHTEE,][]{Jarvisetal16}, and these observatories are beginning to produce interesting findings on, e.g., group dynamics and evolution \citep{Oosterlooetal18,Foretal19} and the population of giant radio galaxies \citep{Delhaizeetal21,Pasinietal20}.

These telescopes are the precursors of the Square Kilometer Array (SKA) a set of next-generation telescopes expected to begin operations in the late 2020s, combining wide frequency coverage with unprecedented sensitivity. The SKA will be built in phases, with phase 1 consisting of two components: the SKA1-Low, covering the 50-350~MHz band, with baselines up to 65~km providing spatial resolution of $\sim$4$^{\prime\prime}$ at 300~MHz and sensitivity a factor of 5-10 better than LOFAR or GMRT; and the SKA1-Mid, covering 350~MHz to 15~GHz with resolution 0.4$^{\prime\prime}$ at 1.4~GHz and a sensitivity up to an order of magnitude better than JVLA \citep{Braunetal19}. The proposed phase 2 SKA would improve sensitivity by another order of magnitude. Much of the SKA's proposed science relates to the early universe, but it will be an extraordinary tool for studies of feedback in groups and clusters, tracing the entire AGN population to high redshift, not merely the radio-loud systems that dominate current samples \citep{PrandoniSeymour15}. SKA surveys are likely to be sensitive to sources down to 10$^{22}$~W~Hz$^{-1}$ out to $z$=3-4 \citep{McAlpineetal15}. This offers an opportunity to detect the radio counterparts of most galaxies identified in current optical surveys, including essentially all group and cluster-dominant galaxies, with sufficient resolution to allow AGN and star formation emission to be disentangled, and with the wide frequency coverage necessary to determine the state and age of jets and lobes. Given the very large numbers of groups and clusters likely to be detected in the southern sky by, e.g., eROSITA, such surveys will play an important role in identifying systems with active cooling and feedback. H\textsc{i} observations reaching low column densities may provide another window on cooling from the IGrM. SKA will also open up the study of diffuse radio structures and magnetic fields in groups \citep[e.g.,][]{Healdetal20}, providing constraints on rates of energy transport and conduction in the IGrM, as well as information on gas motions \citep{Kaleetal16} and perhaps even on turbulence and shocks.

\subsection{\textbf{Upcoming SZ facilities}}

Upcoming surveys of the cosmic microwave background (CMB) such as CCAT-prime \citep{CCAT:2018}, Simons Observatory \citep{SO:2019}, and CMB-S4 \citep{CMBS4:2016} will likely also play a role in advancing our understanding of AGN feedback at group scale by providing complementary information to X-ray surveys. While the thermal SZ effect (hereafter tSZ) is a steep function of mass ($Y_{SZ}\propto M_{500}^{5/3}$), stacking of the tSZ effect over large samples (either X-ray or optically selected) can lead to a detection down to $M_{500}\sim10^{13}M_\odot$. As a pilot study, \citet{Planck:2013_YM} presented the stacked tSZ signal from a large sample of SDSS galaxies selected to be central to their halo, and found that the tSZ-to-mass scaling relation extends with no break all the way down to $M_{500}\sim10^{12.5}M_\odot$. However, the interpretation of the result is rendered difficult by the large \emph{Planck} beam ($\sim 8^\prime$), which dilutes the signal \citep{LeBrun:2015}. While detecting the tSZ signal from individual galaxy groups will be challenging for the new generation of CMB survey instruments, the angular resolution of the foreseen facilities ($\lesssim1^{\prime}$) will be sufficient to study the distribution of the stacked tSZ signal and determine the origin of the signal identified by \citet{Planck:2013_YM}. On the other hand, large single-dish facilities such as AtLAST \citep{AtLAST} will be sensitive enough to detect the tSZ effect from galaxy groups \citep{Mroczkowski:2019}.

Recently, the kinetic SZ effect (kSZ), i.e. the Doppler shift of the CMB spectrum induced by moving electron clouds, has emerged as a promising tool for studying the baryon content of galaxy groups \citep{Hill:2016,ferraro:2016}. The kSZ signal is independent of the gas temperature, which makes it in principle more suitable than the tSZ for the study of low-mass systems. The kSZ signal cannot be detected directly by stacking CMB observations, given that the average velocity of structures with respect to the CMB rest frame vanishes. However, the kSZ signal can be measured by cross-correlating CMB maps with spectroscopic galaxy surveys, thereby fixing each system's velocity; this technique is known as the \emph{pairwise} kSZ. Several recent studies reported low-significance detections of the kSZ with this technique \citep{DeBernardis:2017,Calafut:2021,Planck:kSZ}. These early results may indicate that the flat gas density profiles inferred from X-ray data (see \S\ref{s:baryon_content}) extend far beyond the halo's virial radius, which, if confirmed, provides a detection of the gas expelled from the central regions of halos by AGN feedback. Cross-correlating the kSZ data from future CMB experiments with large spectroscopic surveys like DESI may yield a detection of the pairwise kSZ at high significance \citep{Battaglia:2017} and possibly out to high redshifts \citep{Chaves-Montero:2019}.

\vspace{6pt} 



\authorcontributions{DE: lead author, Sect. 1, 2, 3.1.3, 3.1.4, 5.2, 5.3, 5.4, 6.1, 6.6; MG: Sect. 3.1.3, 4, 5.3; FG: Sect. 3.1.1, 3.1.2; AMCLB: Sect. 2, 5.1, 5.2; EOS: Sect. 3.2, 3.3, 6}

\funding{MG acknowledges partial support by NASA Chandra GO8-19104X/GO9-20114X and HST GO-15890.020-A grants. AMCLB is supported by a fellowship of PSL University at the Paris Observatory. EOS acknowledges support from NASA through \textit{XMM-Newton} award number 80NSSC19K1056 and \Chandra\ award number GO8-19112A.}

\acknowledgments{
We deeply thank Yannick Bahé, Weigang Cui, Marco De Petris, Yohan Dubois, Scott Kay, Ian McCarthy, Alisson Pellissier, Ewald Puchwein, Elena Rasia, Dylan Robson, Marcel van Daalen,  Federico Sembolini, Mark Vogelsberger and Rainer Weinberger for providing data from their simulations for inclusion in Fig. \ref{fig:fgas_historical} and \ref{fig:fgas_modern}, even if some could not be included in the end. We also thank Paul Nulsen for providing cavity parameters for inclusion in Table~\ref{tab:cavity_properties}, and the anonymous referees for useful comments. We thank Ming Sun, Mark Voit, Iurii Babyk, Trevor Ponman and Aurel Schneider for granting us permission to reprint some of their figures. EOS thanks G. Schellenberger and K. Kolokythas for useful conversations. DE thanks Alexis Finoguenov for useful discussions. DE and AMCLB thank Benjamin Oppenheimer for useful discussions. We are also grateful to Tony Mroczkowski, Joop Schaye and Ming Sun for sending comments on the accepted version while we were checking the proofs, which were taken into account for the published version. 
}

\conflictsofinterest{The authors declare no conflict of interest.} 

\appendixtitles{no} 
\appendix
\section{List of the properties of detected AGN cavities in galaxy groups. Only quantities available from the literature are included, thus for some systems the listing will be incomplete.}

\begin{center}
\begin{longtable}{lccccccc}
    \hline
    \hline
   Source      &  $a^{1}$ & $b^{2}$ & $R^{3}$ & $pV$  & Age$^4$ & $L_{cool}^{5}$ & Ref.\\
               & (kpc) & (kpc) & (kpc) & ($10^{56}$ ergs) & (Myr) & ($10^{42}$ ergs/s) & \\
   \hline
    HCG 62~N        & 5.0 & 4.3 & 8.4 & $2.9^{+4.1}_{-1.5}$ & 18-15-31 & $1.8\pm0.2$ & 1,2,3 \\ 
    HCG 62~S       & 4.0 & 4.0 & 8.6 & $2.1^{+3.7}_{-1.3}$ & 19-16-29 &             &   \\ 
    3C88~E         & 23 & 23 & 28 & 95 & 60 &  & 4 \\ 
    3C449~S        & 13 & 13 & 39 & 14.6 & 70 &    & 5 \\
    IC1262~N       & 2.2 & 1.5 & 6.5 & 58.0 & 17-24-52 & $3.3^{+0.2}_{-0.3}$ & 6 \\
    IC1262~S       & 4   &  2  & 6.1 & 50.1 & 12-21-42 & & \\
    NGC~5813~in~SW & 0.95 & 0.95 & 1.3 & 0.11 & 1.2 & & 7 \\ 
    NGC~5813~in~NE & 1.03 & 0.93 & 1.4 & 0.15 & 1.4 & &  \\
    NGC~5813~mid~SW & 3.9 & 3.9 & 7.7 & 1.53 & 7.2 & &  \\
    NGC~5813~mid-1~NE & 2.9 & 2.2 & 4.9 & 0.93 & 4.6 & &  \\
    NGC~5813~mid-2~NE & 2.8 & 2.4 & 9.3 & 0.41 & 8.8 & &  \\
    NGC~5813~out~SW & 5.2 & 3.0 & 22.2 & 0.6 & 20.8 & &  \\
    NGC~5813~out~NE & 8.0 & 4.4 & 18.0 & 2.6 & 17.0 & &  \\
    IC~4296~NW      & 80  & 80 & 230   & 920 & 220 &  & 8 \\ 
    NGC~741~W       & 8   & 8  & 16    & $12.2\pm1.2$ & 30 &  & 9 \\ 
    NGC~193 & 63.4 & 47.2 & 0.0 & 22.7$^{+17.3}_{-17.7}$ & 44.2-20.4-76.9 & 0.11$\pm$0.01 & 10 \\
    NGC~507~E & 21.7 & 8.7 & 22.1 & 308$^{+494}_{-63}$ & 48.4-229-38.9 & 1.37$\pm$0.02 & 10 \\
    NGC~507~W & 13.4 & 5.0 & 11.7 & 90$^{+254}_{-30}$ & 22.2-17.4-38.9 & & \\
    NGC~1550~E & 5.96 & 2.31 & 9.0 & 6.70$^{+10.7}_{-1.98}$ & 12.8-27.9-37.0 & 2.79$^{+0.03}_{-0.01}$ & 11 \\
    NGC~1550~W & 4.09 & 1.88 & 14.65 & 2.00$^{+2.45}_{-0.68}$ & 19.8-52.6-33.0 & & \\
    NGC~4261~E & 20.94 & 15.51 & 24.82 & 31.02$^{+11.98}_{-6.15}$ & 40.5-36.6-105.3 & 0.11$^{+0.08}_{-0.01}$ & 12 \\
    NGC~4261~W & 18.62 & 16.91 & 21.72 & 32.78$^{+6.30}_{-6.52}$ & 35.4-31.8-99.4 & & \\
    NGC~4636~NE & 2.67 & 1.11 & 3.25 & 0.28$^{+0.40}_{-0.06}$ & 8.8-20.2-46.8 & 0.18$\pm$0.01 & 10 \\
    NGC~4636~SE & 2.40 & 1.52 & 2.80 & 0.47$^{+0.28}_{-0.10}$ & 7.6-17.0-47.2 & & \\
    NGC~4636~SW & 2.78 & 1.88 & 4.62 & 0.88$^{+0.48}_{-0.31}$ & 11.7-31.2-62.1 & & \\
    NGC~4636~NW & 2.53 & 1.32 & 3.33 & 0.38$^{+0.36}_{-0.08}$ & 9.1-21.5-49.7 & & \\
    NGC~4782 & 10.7 & 10.7 & 23.0 & 11.0 & 35-54- & & 13 \\
    NGC~5044~SW & 6.54 & 2.84 & 8.60 & 2.17$^{+2.87}_{-0.49}$ & 21.1-14.4-35.5 & 4.72$\pm$0.01 & 12 \\
    NGC~5044~NW & 3.04 & 2.34 & 4.85 & 0.62$^{+0.22}_{-0.13}$ &  11.4-16.4-35.5 & & \\
    NGC~5044~in~SW & 0.15 & 0.15 & 0.45 & 0.0007 & 1 & & 21 \\
    NGC~5044~in~NE & 0.15 & 0.15 & 0.45 & 0.0007 & 1 & & 21 \\
    NGC~5098~N  & 3.0  & 1.6 & 2.97 & 7.0 & 18 &  & 14 \\
    NGC~5098~S  & 3.0  & 1.6 & 2.97 & 7.0 & 18 &  & \\
    NGC~5846~N & 0.74 & 0.58 & 0.64 & 0.35$^{+0.15}_{-0.13}$ & 1.7-1.2-4.6 & 0.27$\pm$0.01 & 10 \\
    NGC~5846~S & 0.74 & 0.58 & 0.68 & 0.35$^{+0.15}_{-0.13}$ & 1.8-1.4-4.8 & & \\
    NGC~5903    & 16.0 & 13.0 & 24.6 & 2.3$\pm$0.10 & 82.5 & 0.0047$\pm$0.0005 & 15 \\
    NGC~6269~N & 5.2 & 5.2 & 10.7 & 22.4$^{+7.2}_{-9.0}$ & 14.0-14.3-26.2 & 0.78$^{+0.04}_{-0.03}$ & 10 \\ 
    NGC~6269~S & 5.5 & 5.5 & 12.3 & 27.2$^{+8.2}_{-10.5}$ & 16.1-17.1-28.9 & & \\
    NGC~6338~in~NE & 4.60 & 3.22 & 3.96 & 20.38$^{+9.65}_{-4.83}$ & 7.3-5.6-19.3 & 4.56$^{+0.08}_{-0.06}$ & 16 \\
    NGC~6338~in~SW & 4.22 & 3.22 & 6.34 & 10.20$^{+3.85}_{-2.53}$ & 11.4-14.7-30.8 & & \\
    NGC~6338~outer & 6.49 & 4.01 & 18.21 & 12.98$^{+8.38}_{-2.88}$ & 25.8-33.5-32.7 & & \\
    VII~Zw~700~NE & 3.96 & 2.43 & 5.54 & 0.24$^{+0.17}_{-0.11}$ & 44.0-41.7-95.5 & 0.54$\pm$0.03 & 16 \\
    VII~Zw~700~SW & 4.70 & 1.90 & 3.43 & 0.43$^{+0.65}_{-0.19}$ & 27.3-29.3-82.8 & & \\
    NGC~6868~NW & 11.7 & 11.7 & 38.7 & 1.48 & 88-107-119 & & 17 \\
    NGC~6868~SE & 8.14 & 8.14 & 25.3 & 1.0  & 55 & & \\
    A~1991~N & 16.8 & 5.5 & 12.4 & 496 & 18-28-68 & 60.4 & 18 \\
    A~1991~S & 13.3 & 6.2 & 11.5 & 535 & 18-29-59 &      & \\
    A~3581~1 & 7.9 & 3.8 & 3.1 & & & & 19 \\
    A~3581~2 & 3.4 & 3.7 & 3.1 & & & & 19 \\
    NGC~533~1 & 2.2 & 1.3 & 1.2 & & & & 19 \\
    NGC~533~2 & 3.1 & 1.6 & 1.6 & & & & 19 \\
    NGC~4104  & 1.9 & 1.5 & 0.0 & & & & 19 \\
    RXC~J0352.9+1941~1 & 8.0 & 5.9 & 9.3 & & & & 20 \\
    RXC~J0352.9+1941~2 & 7.9 & 4.3 & 10.8 & & & &  \\
    RX~J0419+0225 & 1.2 & 0.9 & 1.7 &  &  &  & 20\\
    A~2550~1 & 18.9 & 9.3 & 10.3 &  &  &  & 19\\
    A~2550~2 & 10.7 & 5.9 & 7.8  &  &  &  & 19\\
    A~2717~1 & 11.2 & 6.3 & 7.9  &  &  &  & 19\\
    A~2717~2 & 13.4 & 5.8 & 8.4  &  &  &  & 19\\
    AS1101~1 & 21.0 & 14.7 & 24.2  &  &  &  & 19\\
    AS1101~2 & 24.1 & 15.7 & 23.6  &  &  &  & 19\\
    ESO~351-021 & 12.2 & 8.6 & 14.8  &  &  &  & 19\\
    RX ~J1159+5531~1 & 7.7 & 3.9 & 7.5  &  &  &  & 19\\
    RX~J1159+5531~2 & 6.7 & 4.3 & 9.7  &  &  &  & 19\\
    RX~J1206-0744 & 27.6 & 21.4 & 29.1  &  &  &  & 19\\
    NGC~2300 & 1.3 & 1.0 & 1.5 &  &  &  & 19\\
    UGC~5088~1 & 7.3 & 5.4 & 8.4 &  &  &  & 19\\
    UGC~5088~2 & 6.5 & 3.6 & 5.4 &  &  &  & 19\\
    NGC~777~E & 1.9 & 2.3 & 4.6 &  &  & 0.99 & 22,23 \\
    NGC~777~W & 2.1 & 2.4 & 4.0 &  &  &  & 22 \\
    NGC~4235~E & 2.4 & 4.6 & 11.8 &  &  &   &  22 \\
    NGC~4235~W & 2.1 & 2.4 & 4.0 &  &  &   & 22\\
    NGC~1553 1 & 4.1 & 3.5 & 4.6 & 0.42$^{+0.49}_{-0.22}$ & 13.5-10.7-33.0 & 1.72 & 23,24 \\
    NGC~1553 2 & 3.5 & 2.7 & 3.3 & 0.28$^{+0.23}_{-0.13}$ & 9.8-7.4-25.2 & 1.72 & 23,24 \\
    NGC~1600 1 & 0.87 & 0.82 & 1.21 & 0.15$^{+0.24}_{-0.09}$ & 2.2-1.6-4.2 & 0.12 & 23,24 \\
    NGC~1600 2 & 0.83 & 0.72 & 1.42 & 0.11$^{+0.13}_{-0.06}$ & 2.5-2.1-4.3 & 0.12 & 23,24 \\
    NGC~3608 1 & 3.2 & 2.2 & 6.0 & 0.04$^{+0.03}_{-0.01}$ & 18.7-15.9-26.4 & 0.008 & 23,24 \\
    NGC~3608 2 & 2.5 & 1.7 & 5.3 & 0.02$^{+0.02}_{-0.01}$ & 16.3-14.8-21.5 & 0.008 & 23,24 \\
    NGC~7626 1 & 5.6 & 3.1 & 14.4 & 0.37$^{+0.21}_{-0.14}$ & 33.1-36.5-39.9 & 0.12 & 23,24 \\
    NGC~7626 2 & 1.4 & 0.7 & 3.0 & 0.03$^{+0.03}_{-0.01}$ & 6.8-7.0-8.7 & 0.12 & 23,24 \\
    NGC~7626 3 & 1.6 & 1.1 & 3.8 & 0.06$^{+0.04}_{-0.02}$ & 8.7-8.8-11.5 & 0.12 & 23,24 \\
    NGC~7626 4 & 4.7 & 4.0 & 16.2 & 0.36$^{+0.48}_{-0.21}$ & 37.2-42.5-43.2 & 0.12 & 23,24 \\
    A~262~E    & 2.6 & 2.6 & 6.2  & 1.7$^{+3.2}_{-1.1}$    & 11-13-20       &      & 1,25 \\
    \hline
    \caption{\label{tab:cavity_properties}Properties of the cavities detected in groups. \\
    $^1$ Projected semi-major axis of the cavity \\
    $^2$ Projected semi-minor axis of the cavity \\
    $^3$ Projected distance from the cavity center to the core \\
    $^4$ Ages are reported as $t_s$-$t_{buoy}$-$t_{refill}$. Where only a single value is reported, this is $t_s$. \\
    $^{5}$ Bolometric luminosity between 0.001 and 100 keV inside $r_{cool}$, where the cooling time is less than 7.7 Gyr \\
    References: 1 \citet{Birzan:2004};  2 \citet{Rafferty:2006}; 3 \citet{Gitti:2010}; 4 \citet{Liuetal19}; 5 \citet{Lal:2013}; 6 \citet{Pandge:2019}; 7 \citet{Randall:2011}; 8 \citet{Grossovaetal19}; 9 \citet{Schellenberger:2017}; 10 \citet{OSullivanetal11a}; 11 \citet{Kolokythasetal20}; 12 \citet{OSullivan:2017,Kolokythasetal18};  13 \citet{Machaceketal07};  14 \citet{Randalletal09};  15 \citet{OSullivanetal18b};  16 \citet{OSullivanetal19};  17 \citet{Machaceketal10}; 18 \citet{Pandgeetal13}; 19 \citet{Dong:2010}; 20  \citet{Shin:2016};  21 \citet{Davidetal17,Schellenbergeretal21}; 22 \citet{Panagoulia:2014_cav}; 23 \citet{Cavagnoloetal10}; 24 Nulsen, P., priv. comm.; 25 \citet{Blanton:2004}
    }
    \end{longtable}
    
\end{center}



\externalbibliography{yes}
\bibliographystyle{Definitions/aa}
\bibliography{biblio,biblio2}

\end{document}